\documentclass[10pt]{article}

\usepackage{color}
\usepackage{amsmath}
\usepackage{graphicx}
\usepackage{footnote}
\usepackage{subscript}
\usepackage{amsfonts,color}
\usepackage{amsmath,amssymb}
\usepackage{epsfig}
\usepackage{amssymb} 
\usepackage{mathtools}
\usepackage{amsthm,wasysym}
\usepackage{framed}

\usepackage[english]{babel}
\usepackage[utf8]{inputenc}
\makeatletter
\usepackage{float}
\usepackage{wrapfig}
\restylefloat{figure}

\makeatletter
\let\@fnsymbol\@arabic
\makeatother

\usepackage{bbm}
\newcommand{\id}{{\boldsymbol{\mathbbm{1}}}}
\newcommand{\U}{\mathrm{U_e}}

\newcommand{\tr}{{\rm tr}}
\newcommand{\dev}{{\rm dev}}
\newcommand{\sym}{{\rm sym}}
\newcommand{\skw}{{\rm skew}}
\newcommand{\curl}{{\rm curl}}

\newcommand{\norm}[1]{\|#1\|}

\def\dd{\displaystyle}

\def\dd{\displaystyle}

\newtheorem{theorem}{Theorem}[section]

\newtheorem{definition}[theorem]{Definition}
\def\barr{\begin{array}}
\usepackage{lscape}

	\def\earr{\end{array}}
\def\bec#1{\begin{equation}\label{#1}}
\def\becn{\begin{equation*}}
\def\endec{\end{equation}}
\def\endecn{\end{equation*}}
\def\dd{\displaystyle}
\def\bfm#1{\mbox{\boldmat}}

\usepackage{color}
\usepackage{amsmath}
\usepackage{graphicx}
\usepackage{footnote}
\usepackage{subscript}
\usepackage{amsfonts,color}
\usepackage{amsmath,amssymb}
\usepackage{epsfig}
\usepackage{amssymb} 
\usepackage{mathtools}
\usepackage{amsthm,wasysym}
\usepackage{framed}
\usepackage{cancel}

\usepackage[english]{babel}
\usepackage[utf8]{inputenc}
\makeatletter
\usepackage{float}
\usepackage{wrapfig}

\makeatletter
\let\@fnsymbol\@arabic
\makeatother

\usepackage{bbm}

\usepackage{multicol}

\usepackage{nicefrac,enumitem}

\usepackage{nicefrac, xfrac}

\def\dd{\displaystyle}

\theoremstyle{definition}

\def\barr{\begin{array}}
	\usepackage{lscape}

	\def\earr{\end{array}}
\def\bec#1{\begin{equation}\label{#1}}
\def\becn{\begin{equation*}}
\def\endec{\end{equation}}
\def\endecn{\end{equation*}}
\def\dd{\displaystyle}
\def\bfm#1{\mbox{\boldmat}}

\allowdisplaybreaks

\makeatletter
\renewcommand*\env@matrix[1][*\c@MaxMatrixCols c]{%
	\hskip -\arraycolsep
	\let\@ifnextchar\new@ifnextchar
	\array{#1}}
\makeatother
\begin{document}

	\title{	\vspace*{-1.5cm}An essay on deformation measures in isotropic thin shell theories. Bending versus curvature.
	}
\author{ 
	  Ionel-Dumitrel Ghiba\thanks{Corresponding author:  Ionel-Dumitrel Ghiba,  \ Department of Mathematics,  Alexandru Ioan Cuza University of Ia\c si,  Blvd.
	 	Carol I, no. 11, 700506 Ia\c si,
	 	Romania; and  Octav Mayer Institute of Mathematics of the
	 	Romanian Academy, Ia\c si Branch,  700505 Ia\c si, email:  dumitrel.ghiba@uaic.ro}\,\,, \quad     Peter Lewintan\!\,\thanks{Peter Lewintan,  \ \  Lehrstuhl f\"{u}r Nichtlineare Analysis und Modellierung, Fakult\"{a}t f\"{u}r
 	Mathematik, Universit\"{a}t Duisburg-Essen,  Thea-Leymann Str. 9, 45127 Essen, Germany, email: peter.lewintan@uni-due.de}\,\,, \quad   Adam Sky\,\thanks{Adam Sky,\ \ Institute of Computational Engineering and Sciences, Department of Engineering, Faculty of Science, Technology and Medicine, University of Luxembourg, 6 Avenue de la Fonte, L-4362 Esch-sur-Alzette, Luxembourg, email: adam.sky@uni.lu}\,\,,   \\ Patrizio Neff\,\thanks{Patrizio Neff,  \ \ Head of Lehrstuhl f\"{u}r Nichtlineare Analysis und Modellierung, Fakult\"{a}t f\"{u}r
		Mathematik, Universit\"{a}t Duisburg-Essen,  Thea-Leymann Str. 9, 45127 Essen, Germany, email: patrizio.neff@uni-due.de} 
}

\maketitle
\begin{abstract}
It has become commonplace for the stored energy function of any realistic shell model to align ``within first order" with the classical Koiter membrane-bending (flexural) shell model. In this paper, we assess whether certain extended Cosserat shell models are consistent with the classical linear Koiter model. In doing this, we observe that there are numerous reasons why a modified version of the classical Koiter model should be considered, a consensus reached not only by Koiter himself but also by Sanders and Budiansky, who independently developed the same theory during the same period.
To provide a comprehensive overview of the strain measures employed in our Cosserat shell models, this paper presents them in a unified manner and compares them with the strain measures previously utilized in the literature. We show that all our new strain tensors either generalize (in the case of nonlinear constrained or unconstrained models) or coincide (in the case of the linear constrained model) with the strain tensors recognized as the ``best" or those possessing a well-defined geometric interpretation connected to bending or curvature.

\smallskip

\end{abstract}

\begin{footnotesize}
	\setcounter{tocdepth}{2} 
\tableofcontents
\end{footnotesize}
\allowdisplaybreaks

\section{Introduction}

Recent papers published by {\v{S}}ilhav{\`y} \cite{vsilhavycurvature}  and Virga  \cite{virga2023pure}  together with the novel shell strain tensors obtained by us using various dimensional reduction methods  \cite{GhibaNeffPartI,GhibaNeffPartII,GhibaNeffPartIII,GhibaNeffPartIV,GhibaNeffPartV,saem2023geometrically,saem2023explicit,gastel2022regularity}  have prompted us to undertake a more detailed examination of the physical and geometric significance of various strain tensors used in shell models. Anicic and L{\'{e}}ger also explored this question, some time ago  in the context of linear models, as documented in \cite{anicic1999formulation}, along with references  \cite{anicic2002mesure} and \cite{anicic2003shell}. The issue of appropriate nonlinear shell bending strain measures was also already addressed by Acharya in \cite{acharya2000nonlinear}.

In the classical shell-models of order $O(h^3)$ the total energy is given in the form \begin{equation}\label{energieshell}
\begin{array}{l}
\dd\int_\omega \left(h\,W_1(\mathcal{E}) +\dd\frac{h^3}{12}W_2(\mathcal{F})\right)\, {\rm d}a,
\end{array}
\end{equation}
where $\omega\subset \mathbb{R}^2$ is the planar reference domain, $\mathcal{E}$ and $\mathcal{F}$ are strain tensors measuring the deformation of the shell, and $W_1$  and $W_2$ are  energy densities. The strain tensor $\mathcal{E}$, through the $O(h)$ energy $W_1(\mathcal{E})$, measures the change of metric and it is called {\it a membrane shell term}. There  is complete agreement  which strain tensor should be used as measure of the change of metric in the linearised models, while in the nonlinear models there still  are some differences. For instance,  the strain tensors for measuring the change of the metric  after the derivation approach \cite{Neff_plate04_cmt,GhibaNeffPartI} and in the expression of the Gamma-limit \cite{neff2007geometrically,saem2023geometrically} starting from the  Biot-type quadratic parental 3D energy\footnote{In this paper, 
	for $a,b\in\mathbb{R}^n$ we let $\bigl\langle {a},{b} \bigr\rangle _{\mathbb{R}^n}$  denote the scalar product on $\mathbb{R}^n$ with
	associated vector norm $\lVert a\rVert _{\mathbb{R}^n}^2=\bigl\langle {a},{a} \bigr\rangle _{\mathbb{R}^n}$. 
	The standard Euclidean scalar product on  the set of real $n\times  {m}$ second order tensors $\mathbb{R}^{n\times  {m}}$ is given by
	$\bigl\langle  {X},{Y} \bigr\rangle _{\mathbb{R}^{n\times  {m}}}={\rm tr}(X\, Y^T)$, and thus the  {(squared)} Frobenius tensor norm is
	$\lVert {X}\rVert ^2_{\mathbb{R}^{n\times  {m}}}=\bigl\langle  {X},{X} \bigr\rangle _{\mathbb{R}^{n\times  {m}}}$. The identity tensor on $\mathbb{R}^{n \times n}$ will be denoted by $\id_n$, so that
	${\rm tr}({X})=\bigl\langle {X},{\id}_n \bigr\rangle $, and the zero matrix is denoted by $0_n$.} $W_{\rm Biot}:=\dd\mu\,\lVert \sqrt{F^T F}-\id_3\rVert^2+
\dd\frac{\lambda}{2}\,[{\rm tr}(\sqrt{F^T F}-\id_3)]^2$ is\footnote{For a given matrix $M\in \mathbb{R}^{2\times 2}$ we define the lifted quantities $
	M^\flat =\begin{footnotesize}\left(\begin{array}{cc|c}
	M_{11}& M_{12}&0 \\
	M_{21}&M_{22}&0 \\
	\hline
	0&0&0
	\end{array}\right)\end{footnotesize}
	\in \mathbb{R}^{3\times 3}$ and $
	\widehat{M} =\begin{footnotesize}\left(\begin{array}{cc|c}
	M_{11}& M_{12}&0 \\
	M_{21}&M_{22}&0 \\
	\hline
	0&0&1
	\end{array}\right)\end{footnotesize}
	\in \mathbb{R}^{3\times 3}$. Here, as specified in the next section, $[\nabla \Theta]$ means the value for $x_3=0$ of the gradient of 
	the diffeomorphism   $\Theta:\mathbb{R}^3\rightarrow\mathbb{R}^3$, describing the reference configuration (i.e., the curved surface of the shell),
	$
	\Theta(x_1,x_2,x_3)\,=\,y_0(x_1,x_2)+x_3\ n_0(x_1,x_2), \  n_0\,=\,\dd\frac{\partial_{x_1}y_0\times \partial_{x_2}y_0}{\lVert \partial_{x_1}y_0\times \partial_{x_2}y_0\rVert}\, ,
	$
	where $y_0:\omega\to \mathbb{R}^3$ is a function of class $C^2(\omega)$. This specific form of the diffeomorphism 
	$\Theta$ maps the midsurface $\omega$ of the fictitious Cartesian configuration  parameter space $\Omega_h$ onto the midsurface $\omega_\xi=y_0(\omega)$ of $\Omega_\xi$ and $n_0$ is the unit normal vector to $\omega_\xi$. } $\mathcal{G}_{\infty}^\flat:=[\nabla\Theta ]^T\left(\sqrt{[\nabla\Theta ]^{-T}\,\widehat{\rm I}_m\,\id_2^{\flat }\,[\nabla\Theta ]^{-1}}-
\sqrt{[\nabla\Theta ]^{-T}\,\widehat{\rm I}_{y_0}\,\id_2^{\flat }\,[\nabla\Theta ]^{-1}})\right)[\nabla\Theta ]$,  other classical models which are derived starting from the Saint-Venant-Kirchhoff energy $W_{\rm SVK}:=\dd\mu\,\lVert \dd\nicefrac{1}{2}\,({F^T F}-\id_3)\rVert^2+
\dd\frac{\lambda}{2}\,[{\rm tr}(\nicefrac{1}{2}({F^T F}-\id_3))]^2$ lead to the difference between the first fundamental form of the unknown midsurface parametrized by $m$ and the first fundamental form of the referential midsurface configuration parametrized by $y_0$, i.e., $\mathcal{G}_{\rm{Koiter}}:=\frac{1}{2}({{\rm I}_m}-{{\rm I}_{y_0}})$.  It is clear that the linearisation of these two strain measures for the change of metric are equal\footnote{In the linearisation process we  express the total midsurface deformation  as
$
	m(x_1,x_2)=y_0(x_1,x_2)+v(x_1,x_2),
$
	with $v:\omega\to \mathbb{R}^3$, the infinitesimal shell-midsurface displacement and we find that $\mathcal{G}_{\rm{Koiter}}^{\rm{lin}} \,\,:=\frac{1}{2}\big[{\rm I}_m - {\rm I}_{y_0}\big]^{\rm{lin}}
	= \sym\big[ (\nabla y_0)^{T}(\nabla v)\big]=\mathcal{G}_{\infty}^{\rm{lin}}\in {\rm Sym}(2)$} \cite{GhibaNeffPartV}, but in the nonlinear case they are different.\footnote{Note that the Biot-energy $W_{\rm Biot}$ is physically more realistic than the Saint-Venant-Kirchhoff energy $W_{\rm SVK}$.}

On the contrary, no agreement exists on the names and about which strain tensors should be used for $\mathcal{F}$, as well as  what $\mathcal{F}$ really measures. Sometimes the energy term of order $O(h^3)$, i.e., $W_2(\mathcal{F})$ is called
\begin{itemize}
	\item {\it bending energy} ($\mathcal{F}$ is called bending strain tensor) \cite{Koiter60,Eremeyev1,acharya2000nonlinear},
	\item {\it change of curvature energy} ($\mathcal{F}$ is called change of curvature strain tensor) \cite{Naghdi63,anicic1999formulation,anicic2002mesure,vsilhavycurvature},
	\item {\it flexural energy} ($\mathcal{F}$ is called change of curvature strain tensor) \cite{Ciarlet00}.
\end{itemize}
Therefore, the question arises: does the  strain tensor $\mathcal{F}$ in a shell model correspond to the physical significance of the given name, bending versus ~change of curvature? One source of the confusions is  the fact that for  $\mathcal{G}_{\rm{Koiter}}^{\rm{lin}}=0$ (infinitesimal pure flexure) or for  plates (flat shells),  no difference between bending and change of curvature measures can be observed.

In this respect, on one hand, Acharya's essential early invariance requirement from \cite{acharya2000nonlinear} should be recalled:  \textit{A vanishing bending strain at a point should be associated with any deformation  that leaves the orientation of the unit normal field locally unaltered around that point.} We consider relevant the conclusions given by Acharya on the iMechanica's blog\footnote{https://imechanica.org/node/1408} (the entire blog's text is given in the Appendix).

 Virga \cite[Section II]{virga2023pure} recently rediscovered the  problem posed by Acharya and he introduced a definition for a pure bending measure of   nonlinear shells: {\it a deformation measure is a pure bending measure if it undergoes bending-neutral deformations}, i.e. finite incremental changes of the plate’s shape bearing no further bending (in other words only rotations about the unit normal to the midsurface and stretches, that leave the tangent plane invariant). Acharya's and  Virga's minimal requirements on the bending strain measures are in fact  equivalent. In \cite{GhibaNeffPartIII} we already touched upon the proper invariance conditions for bending tensors in the framework of Cosserat shells. On the other hand, the change of curvature strain tensor should  measure the variation of the mathematical quantities defining the curvature of a surface \cite{vsilhavycurvature,anicic2002mesure}.

In our family of Cosserat shell models, we have the advantage to offer a greater level of generality compared to other frameworks. These models with rotational degree of freedom can then easily be specialized into more classical models by imposing specific constraints. This flexibility places us in an advantageous position to explore a wide range of model variants within a unified framework. See also \cite{sander2016numerical} and \cite{nebel2023geometrically} for numerical simulations in comparisons to  experiments.

 In our current study, we illustrate that the strain measures utilized in all our recent Cosserat shell models, as discussed in references such as \cite{GhibaNeffPartI,GhibaNeffPartII,GhibaNeffPartIII,GhibaNeffPartIV,GhibaNeffPartV,saem2023geometrically,saem2023explicit,gastel2022regularity}, have well-defined mathematical and physical interpretations that are consistent with the geometric explanations presented in references such as \cite{vsilhavycurvature,anicic2002mesure,anicic2003shell} and \cite{acharya2000nonlinear}. To provide a comprehensive overview of the models under consideration, the initial section of this paper introduces all the relevant strain tensors and establishes their connections to variational problems that define the shell models.

To the best of our knowledge, it appears that the models developed in \cite{GhibaNeffPartI,GhibaNeffPartIII,GhibaNeffPartIV}, and \cite{GhibaNeffPartV} (see also \cite{birsan2020derivation,birsan2021alternative}) are the first instances in the literature to possess the unique capability of explicitly specifying the influence of both bending and  (different) curvature measures. It    is worth noting that there has been some confusion between these two distinct measures, a confusion that will be  clarified in this present paper. A comprehensive contrast between our shell strain energy density and the one utilized in the 6-parameter shell theory, as described in \cite{Pietraszkiewicz-book04,Eremeyev08}, and \cite{Erem-Leb-Cloud10}, has been provided in \cite[Sect. 6]{GhibaNeffPartII}.

In the subsequent sections of this paper, we progressively demonstrate that our models encompass an entire family of models. Ultimately, the linearization of the deformation measures within our constrained Cosserat-shell model corresponds to the approaches advocated in later works by Sanders and Budiansky in \cite{budiansky1962best,budiansky1963best}, as well as by Koiter and Simmonds in \cite{koiter1973foundations}. They referred to the resulting theory as the "best first-order linear elastic shell theory." However, our change of the curvature measure aligns with those used by Anicic and Leg\'{e}r in \cite{anicic1999formulation}.

Given the varying terminology and interpretations of certain tensors in the existing literature, we start offering the comparative  Table \ref{tabelbc} for shells and for a simplified overview in Table \ref{tabelbcp} for plates, too.

\begin{landscape}
	\begin{table}[h!]
		\vspace*{-1cm}
		\centering 
		\footnotesize	{\bf Names and tensors used in the literature for linear and nonlinear shell models (factor 2 at places is correct)}
		\vspace*{2mm}\\
		\resizebox{25cm}{!}{	\begin{tabular}[h!]{|l|c|c|c|c|c|}
				\hline
				\footnotesize	\textbf{authors}	& \begin{minipage}{3cm}\footnotesize \textbf{change of metric}\end{minipage}& \begin{minipage}{4cm} \footnotesize\textbf{the candidate for ``the bending strain tensor''} \end{minipage}&\begin{minipage}{4cm} \footnotesize\textbf{the candidate for ``the change of curvature tensor''}  \end{minipage} & \begin{minipage}{2cm} \footnotesize\textbf{bending strain/ change of curvature  vanishing in infinitesimal pure stretch}  \end{minipage}&\begin{minipage}{1.5cm} \vspace*{1mm}
					\footnotesize	\textbf{bending strain/ change of curvature measures  variations in   $H$ and $K$} \\ \end{minipage} 
				\\
				\hline
				\footnotesize\begin{minipage}{2cm}\ \\ linear 	Koiter model \cite{Koiter60}\\\end{minipage}	& \begin{minipage}{4.5cm}\footnotesize$\mathcal{G}_{\rm{Koiter}}^{\rm{lin}} =\frac{1}{2}\big[{\rm I}_m - {\rm I}_{y_0}\big]^{\rm{lin}}\in{\rm Sym}(2)
					$\end{minipage}& \footnotesize$\times$ &  \footnotesize$\mathcal{R}_{\rm{Koiter}}^{\rm{lin}}=\big[{\rm II}_m - {\rm II}_{y_0}\big]^{\rm{lin}} \in{\rm Sym}(2)$ & $\times$ & $\times$\\
				\hline
				\footnotesize\begin{minipage}{2cm}\ \\	Sanders \cite{sanders1959improved}\\\end{minipage}	& \begin{minipage}{1cm}\footnotesize$\mathcal{G}_{\rm{Koiter}}^{\rm{lin}} 
					$\end{minipage}& \footnotesize$\mathcal{R}_{\rm{Koiter}}^{\rm{lin}} $   & \footnotesize$\times$ & $\times$ & $\times$\\
				\hline
				\footnotesize\begin{minipage}{2cm}\ \\	Budianski \\ \& Sanders \cite{budiansky1963best}\\\end{minipage} 	&  \begin{minipage}{1cm}\footnotesize$\mathcal{G}_{\rm{Koiter}}^{\rm{lin}} 
					$\end{minipage} &  \footnotesize $\mathcal{R}_{\rm KSB}= \mathcal{R}_{\rm{Koiter}}^{\rm{lin}} - {\bf 1}\, \sym[\,\mathcal{G}_{\rm{Koiter}}^{\rm{lin}} \,{\rm L}_{y_0}]$   $=\sym(\mathcal{R}_{\infty}^{\rm lin})$ &  \footnotesize$\times $ & $\surd$ & $\times$\\
				\hline
				\footnotesize\begin{minipage}{2.3cm}\ \\	Koiter \\ \& Simmonds	\cite{koiter1973foundations}\\\end{minipage}	&  \begin{minipage}{1cm}\footnotesize$\mathcal{G}_{\rm{Koiter}}^{\rm{lin}} 
					$\end{minipage}  & \footnotesize $\times$  &   \footnotesize $\mathcal{R}_{\rm KSB}= \mathcal{R}_{\rm{Koiter}}^{\rm{lin}} - {\bf 1}\, \sym[\,\mathcal{G}_{\rm{Koiter}}^{\rm{lin}} \,{\rm L}_{y_0}]$   $=\sym(\mathcal{R}_{\infty}^{\rm lin})$ & $\surd$ & $\times$\\
				\hline
				\footnotesize\begin{minipage}{2cm}\ \\	Anicic \\\& L\'eger	\cite{anicic1999formulation}\\ \end{minipage} &  \begin{minipage}{1cm}\footnotesize$\mathcal{G}_{\rm{Koiter}}^{\rm{lin}} 
					$\end{minipage}  & 
				\footnotesize $ \times $
				&  \footnotesize $\mathcal{R}_{\rm{AL}}^{\rm{lin}}= \mathcal{R}_{\rm{Koiter}}^{\rm{lin}} -{\bf 2}\, \sym[\,\mathcal{G}_{\rm{Koiter}}^{\rm{lin}} \,{\rm L}_{y_0}]$ & $\times$ & $\surd$ \\
				\hline
				\footnotesize\begin{minipage}{2cm}	\ \\Kirchoff-Love from {\v{S}}ilhav{\`y} \cite {vsilhavycurvature}	\\\end{minipage}&  \begin{minipage}{1cm}\footnotesize$\mathcal{G}_{\rm{Koiter}}^{\rm{lin}} 
					$\end{minipage} & $\times $  &  \footnotesize $\mathcal{R}_{\rm{AL}}^{\rm{lin}}$ &$\times$ & $\surd$ \\
				\hline
				\footnotesize	\begin{minipage}{2cm}\ \\Acharya \cite{acharya2000nonlinear}\\\end{minipage}	&  \begin{minipage}{1cm}\footnotesize$\mathcal{G}_{\rm{Koiter}}^{\rm{lin}} 
					$\end{minipage}  &  \footnotesize $\mathcal{R}_{\rm KSB}= \mathcal{R}_{\rm{Koiter}}^{\rm{lin}} - {\bf 1}\, \sym[\,\mathcal{G}_{\rm{Koiter}}^{\rm{lin}} \,{\rm L}_{y_0}]$   $=\sym(\mathcal{R}_{\infty}^{\rm lin})$ &  \footnotesize $ \times$ & $\surd$ & $\times$ \\
				\hline  
				\footnotesize	\begin{minipage}{2cm}our linear\\  modified \\	constrained\\ Cosserat model\\ \end{minipage}  	&  \centering \begin{minipage}{3.5cm}\footnotesize$\mathcal{G}^{\rm{lin}}_\infty =\mathcal{G}_{\rm{Koiter}}^{\rm{lin}}\in{\rm Sym}(2) 
					$\end{minipage} &  \footnotesize $\mathcal{R}_{\infty}^{\rm{lin}}=\mathcal{R}_{\rm{Koiter}}^{\rm{lin}} - {\bf 1}\, [\,\mathcal{G}_{\rm{Koiter}}^{\rm{lin}} \,{\rm L}_{y_0}]\not\in{\rm Sym}(2)$ & \footnotesize $\mathcal{R}^{\rm{lin}}_{\rm Koiter}-{\bf 2}\,\sym[\,\mathcal{G}^{\rm{lin}}_\infty \,{\rm L}_{y_0}]= \mathcal{R}_{\rm{AL}}^{\rm{lin}}$  & $\surd$ & $\surd$\\
				\hline
				\footnotesize\begin{minipage}{2cm}\ \\	our linear\\ Cosserat  model \end{minipage}	& \begin{minipage}{5.5cm}\ \\\footnotesize$\mathcal{G}^{\rm{lin}}=(\nabla y_0)^{T} ( \nabla v -\vartheta\times \nabla y_0 )\not\in{\rm Sym}(2),$  \\ \ \\$ \sym\, \mathcal{G}^{\rm{lin}}=\mathcal{G}_{\rm{Koiter}}^{\rm{lin}} 
					\in{\rm Sym}(2)$\\ \ \end{minipage} & \footnotesize   $\mathcal{R}^{\rm{lin}} =  - (\nabla y_0)^{T} ( \partial_{x_1}\vartheta\times n_0\;|\; \partial_{x_2}\vartheta\times n_0 )\not\in{\rm Sym}(2)$  &  \begin{minipage}{5.5cm}\footnotesize $\mathcal{R}^{\rm{lin}} - \mathcal{G}^{\rm{lin}} \,{\rm L}_{y_0}$ ,\\ \  additional vectors: $ \mathcal{T}^{\rm{lin}}, \mathcal{N}^{\rm{lin}}$\\ \end{minipage} & $\surd$ & $\surd$\\
				\hline
				\footnotesize \begin{minipage}{2cm}\ \\classical \\  nonlinear \\ Koiter \cite{Ciarlet00}\end{minipage} & \begin{minipage}{3cm}\footnotesize $\mathcal{G}_{\mathrm{Koiter}}=\frac{1}{2}\big[{\rm I}_m - {\rm I}_{y_0}\big]$\end{minipage}&  \begin{minipage}{5cm}\footnotesize $\mathcal{R}_{\mathrm{Koiter}}=\big[{\rm II}_m - {\rm II}_{y_0}\big]\in{\rm Sym}(2)$\end{minipage} & $\times$& $\times$ & $\times$\\
				\hline
				\footnotesize \begin{minipage}{2cm}\ \\our nonlinear \\ 
					modified\\ constrained \\ Cosserat model\\ \end{minipage} & \begin{minipage}{7cm}\footnotesize $\mathcal{G}_{\infty}^\flat=[\nabla\Theta ]^T\Big(\sqrt{[\nabla\Theta ]^{-T}\,\widehat{\rm I}_m\,\id_2^{\flat }\,[\nabla\Theta ]^{-1}}$\\
					\hspace*{1cm}	$-
					\sqrt{[\nabla\Theta ]^{-T}\,\widehat{\rm I}_{y_0}\,\id_2^{\flat }\,[\nabla\Theta ]^{-1}})\Big)[\nabla\Theta ]\in{\rm Sym}(3)$\end{minipage}&  \begin{minipage}{8.5cm}\footnotesize $\mathcal{R}_{\infty}^\flat=\,[\nabla\Theta \,]^{T}\Big(\sqrt{[\nabla\Theta ]\,\widehat{\rm I}_m^{-1}[\nabla\Theta ]^{T}}\,[\nabla\Theta ]^{-T} {\rm II}_m^\flat[\nabla\Theta ]^{-1}$\\\hspace*{1cm} $-\sqrt{[\nabla\Theta ]\,\widehat{\rm I}_{y_0}^{-1}[\nabla\Theta ]^{T}}[\nabla\Theta ]^{-T}{\rm II}_{y_0}^\flat [\nabla\Theta ]^{-1}\Big)\nabla\Theta\not\in {\rm Sym}(3)$\end{minipage} & $\sym (\mathcal{R}_{\infty}-\mathcal{G}_{\infty} \,{\rm L}_{y_0})$,   additional vector: $ \mathcal{N}_\infty$& $?$ & $?$\\
				\hline
				\footnotesize \begin{minipage}{2cm}\ \\generalized \\ non-linear Naghdi-type \cite{mardare2008derivation}\end{minipage} & \begin{minipage}{3cm}\footnotesize $\mathcal{G}_{\mathrm{Koiter}}=\frac{1}{2}\big[{\rm I}_m - {\rm I}_{y_0}\big]$\end{minipage}& $\times$&  \begin{minipage}{5cm}\footnotesize $\mathcal{R}_{\mathrm{Naghdi}}=-\sym\big[\nabla m)^T\nabla d - {\rm II}_{y_0}\big]$\\ additional vectors:  transverse shear\\
					$\Big(\langle d, \partial_{x_1}m\rangle,\langle d, \partial_{x_1}m\rangle\Big)$\end{minipage} & $\times$ & $\times$
				\\
				\hline
				\footnotesize \begin{minipage}{2cm}\ \\our nonlinear\\ Cosserat model\\\end{minipage} & \begin{minipage}{5cm}\footnotesize $
					\mathcal{G} =\, (\overline{Q}_{e,s} \nabla y_0)^{T} \nabla m- {\rm I}_{y_0}\not\in{\rm Sym}(2)$\end{minipage}&  \footnotesize  $\mathcal{R} = \, -(\overline{Q}_{e,s} \nabla y_0)^{T} \nabla (\overline{Q}_{e,s} n_0)- {\rm II}_{y_0}$ & \begin{minipage}{5.5cm}\footnotesize\ \\ $\mathcal{R}-\mathcal{G} \,{\rm L}_{y_0}$,  additional vectors: $ \mathcal{T}, \mathcal{N}$ \end{minipage}& ?& ?\\
				\hline
		\end{tabular}}
		\caption{\footnotesize  Here, $y_0$ denotes the initial surface and $m=y_0+v(x)$ is the midsurface deformation, with $v$ the midsurface   displacement, $\overline{Q}_{e,s}$ is the Cosserat rotation, ${\rm I}_{m}:=[{\nabla  m}]^T\,{\nabla  m}\in \mathbb{R}^{2\times 2}$ and  ${\rm II}_{m}:\,=\,-[{\nabla  m}]^T\,{\nabla  n}\in \mathbb{R}^{2\times 2}$ are  the matrix representations of the {\it first fundamental form (metric)} and the  {\it  second fundamental form}, respectively, $n\,=\,\dd\frac{\partial_{x_1}m\times \partial_{x_2}m}{\lVert \partial_{x_1}m\times \partial_{x_2}m\rVert} $, 	${\rm L}_{m}$ is  the {\it Weingarten map (or shape operator)}  defined by 
			$
			{\rm L}_{m}\,=\, {\rm I}_{m}^{-1} {\rm II}_{m}\in \mathbb{R}^{2\times 2}
			$, with similar definitions for ${\rm I}_{y_0}$, ${\rm II}_{y_0}$, ${n_0}$ and ${\rm L}_{y_0}$,  $\vartheta$ is the infinitesimal rotation and denotes the axial vector of $ \overline{A}_\vartheta $, such that 
			$ \overline{Q}_{e,s}=\exp(\overline{A}_\vartheta)\; = \;\id_3 + \overline{A}_\vartheta+\textrm{h.o.t.}$,  $ \mathcal{T}^{\rm{lin}}=  n_0^T ( \nabla v -\vartheta\times \nabla y_0) $ is the transverse shear deformation vector,   $\mathcal{N}^{\rm{lin}} = n_0^T (\nabla\vartheta)$ is the vector of drilling bendings.  We note that $\,\mathcal{G}_{\rm{Koiter}}^{\rm{lin}} \,{\rm L}_{y_0}\in {\rm Sym}(2)$ represents an additional constraint in the model. In the nonlinear models: 
			$\mathcal{T}:= \, (\overline{Q}_{e,s}  n_0)^{T} \nabla m\not\in {\rm Sym}(2)$ is  the transverse shear deformation vector and the \textit{vector of drilling bendings} in our nonlinear model is 
			$
			\mathcal{N} = n_0^T  \big(\mbox{axl}(\overline{Q}_{e,s}^T\partial_{x_1}\overline{Q}_{e,s})\,|\, \mbox{axl}(\overline{Q}_{e,s}^T\partial_{x_2}\overline{Q}_{e,s}) \big)$.}\label{tabelbc}
	\end{table}
\end{landscape}

\begin{landscape}
	\begin{table}[h!]
		\vspace*{-1cm}
		\centering 
		\footnotesize	{\bf Names and tensors used in the literature for linear and nonlinear plate (flat shell) models (factor 2 at places is correct)}
		\vspace*{2mm}\\
		\resizebox{19.5cm}{!}{	\begin{tabular}[h!]{|l|c|c|c|c|c|}
				\hline
				\footnotesize	\textbf{authors}	& \begin{minipage}{3cm}\footnotesize \textbf{change of metric}\end{minipage}& \begin{minipage}{4cm} \footnotesize\textbf{the candidate for ``the bending strain tensor''} \end{minipage}&\begin{minipage}{4cm} \footnotesize\textbf{the candidate for ``the change of curvature tensor''}  \end{minipage} & \begin{minipage}{2cm} \footnotesize\textbf{bending strain/ change of curvature  vanishing in infinitesimal pure stretch}  \end{minipage}&\begin{minipage}{1.5cm} \vspace*{1mm}
					\footnotesize	\textbf{bending strain/ change of curvature measures  variations in   $H$ and $K$} \\ \end{minipage} 
				\\
				\hline
				\footnotesize\begin{minipage}{2cm}\ \\ linear 	Koiter model \cite{Koiter60}\\\end{minipage}	& \begin{minipage}{4.5cm}\footnotesize$\mathcal{G}_{\rm{Koiter}}^{\rm{lin}} =\frac{1}{2}\big[{\rm I}_m - \id_2\big]^{\rm{lin}}\in{\rm Sym}(2)
					$\end{minipage}& \footnotesize$\times$ &  \footnotesize$\mathcal{R}_{\rm{Koiter}}^{\rm{lin}}=\big[{\rm II}_m\big]^{\rm{lin}} \in{\rm Sym}(2)$ & $\times$ & $\times$\\
				\hline
				\footnotesize\begin{minipage}{2cm}\ \\	Sanders \cite{sanders1959improved}\\\end{minipage}	& \begin{minipage}{1cm}\footnotesize$\mathcal{G}_{\rm{Koiter}}^{\rm{lin}} 
					$\end{minipage}& \footnotesize$\mathcal{R}_{\rm{Koiter}}^{\rm{lin}} $   & \footnotesize$\times$ & $\times$ & $\times$\\
				\hline
				\footnotesize\begin{minipage}{2cm}\ \\	Budianski \\ \& Sanders \cite{budiansky1963best}\\\end{minipage} 	&  \begin{minipage}{1cm}\footnotesize$\mathcal{G}_{\rm{Koiter}}^{\rm{lin}} 
					$\end{minipage} &  \footnotesize $\mathcal{R}_{\rm KSB}= \mathcal{R}_{\rm{Koiter}}=\sym(\mathcal{R}_{\infty}^{\rm lin})$ &  \footnotesize$\times $ & $\surd$ & $\times$\\
				\hline
				\footnotesize\begin{minipage}{2.3cm}\ \\	Koiter \\ \& Simmonds	\cite{koiter1973foundations}\\\end{minipage}	&  \begin{minipage}{1cm}\footnotesize$\mathcal{G}_{\rm{Koiter}}^{\rm{lin}} 
					$\end{minipage}  & \footnotesize $\times$  &   \footnotesize $\mathcal{R}_{\rm KSB}= \mathcal{R}_{\rm{Koiter}}^{\rm{lin}} =\sym(\mathcal{R}_{\infty}^{\rm lin})$ & $\surd$ & $\times$\\
				\hline
				\footnotesize\begin{minipage}{2cm}\ \\	Anicic \\\& L\'eger	\cite{anicic1999formulation}\\ \end{minipage} &  \begin{minipage}{1cm}\footnotesize$\mathcal{G}_{\rm{Koiter}}^{\rm{lin}} 
					$\end{minipage}  & 
				\footnotesize $ \times $
				&  \footnotesize $\mathcal{R}_{\rm{AL}}^{\rm{lin}}= \mathcal{R}_{\rm{Koiter}}^{\rm{lin}}$ & $\times$ & $\surd$ \\
				\hline
				\footnotesize\begin{minipage}{2cm}	
					Kirchoff-Love from \\{\v{S}}ilhav{\`y} \cite {vsilhavycurvature}	\\\end{minipage}&  \begin{minipage}{1cm}\footnotesize$\mathcal{G}_{\rm{Koiter}}^{\rm{lin}} 
					$\end{minipage} & $\times $  &  \footnotesize $\mathcal{R}_{\rm{AL}}^{\rm{lin}}$ &$\times$ & $\surd$ \\
				\hline
				\footnotesize	\begin{minipage}{2cm}\ \\Acharya \cite{acharya2000nonlinear}\\\end{minipage}	&  \begin{minipage}{1cm}\footnotesize$\mathcal{G}_{\rm{Koiter}}^{\rm{lin}} 
					$\end{minipage}  &  \footnotesize $\mathcal{R}_{\rm KSB}= \mathcal{R}_{\rm{Koiter}}^{\rm{lin}} $   $=\sym(\mathcal{R}_{\infty}^{\rm lin})$ &  \footnotesize $ \times$ & $\surd$ & $\times$ \\
				\hline  
				\footnotesize	\begin{minipage}{2cm}our linear\\  modified 	constrained model\\ \end{minipage}  	&  \centering \begin{minipage}{2.5cm}\footnotesize$\mathcal{G}^{\rm{lin}}_\infty =\mathcal{G}_{\rm{Koiter}}^{\rm{lin}} 
					$\end{minipage} &  \footnotesize $\mathcal{R}_{\infty}^{\rm{lin}}=\mathcal{R}_{\rm{Koiter}}^{\rm{lin}} \in{\rm Sym}(2)$ &\begin{minipage}{5.7cm}\vspace{2mm} \footnotesize $\mathcal{R}_{\infty}^{\rm{lin}}= \mathcal{R}_{\rm{AL}}^{\rm{lin}}$\\ additional vectors:\\ $ \mathcal{N}_\infty^{\rm lin}=\frac{1}{2} (\partial_{x_1}\curl (v_1,v_2)^T\,|\, \partial_{x_2}\curl (v_1,v_2)^T)$\vspace{2mm}\end{minipage}   & $\surd$ & $\surd$\\
				\hline
				\footnotesize\begin{minipage}{2cm}\ \\	our linear\\ Cosserat  model \end{minipage}	& \begin{minipage}{5cm}\ \\\footnotesize$\mathcal{G}^{\rm{lin}}=\left(\begin{array}{cc}
					\partial_{x_1}v_1&\partial_{x_2}v_1-\vartheta_3\\\partial_{x_1}v_2+\vartheta_3&\partial_{x_2}\vartheta_2
					\end{array}\right),$  \\ \ \\$ \sym\, \mathcal{G}^{\rm{lin}}=\mathcal{G}_{\rm{Koiter}}^{\rm{lin}} 
					$\\ \ \end{minipage} & \footnotesize   $\mathcal{R}^{\rm{lin}} =\left(\begin{array}{cc}
				\partial_{x_1}\vartheta_2&-\partial_{x_2}\vartheta_2\\\partial_{x_1}\vartheta_1&-\partial_{x_2}\vartheta_1
				\end{array}\right)$  &  \begin{minipage}{4.9cm}\footnotesize\vspace{1mm} $\mathcal{R}^{\rm{lin}} =\left(\begin{array}{cc}
					\partial_{x_1}\vartheta_2&-\partial_{x_2}\vartheta_2\\\partial_{x_1}\vartheta_1&-\partial_{x_2}\vartheta_1
					\end{array}\right)$ ,\\ \  additional vectors: \\$ \mathcal{T}^{\rm{lin}}=\left(\begin{array}{ccc}
					\partial_{x_1}v_3+\vartheta_2&|&\partial_{x_2}v_3-\vartheta_1
					\end{array}\right),$\\$ \mathcal{N}^{\rm{lin}}=\left(\begin{array}{ccc}
					\partial_{x_1}v_3&|&\partial_{x_2}v_3
					\end{array}\right)$\\ \end{minipage} & $\surd$ & $\surd$\\
				\hline
				\footnotesize \begin{minipage}{2cm}\ \\classical \\  nonlinear \\ Koiter \cite{Ciarlet00}\end{minipage} & \begin{minipage}{3cm}\footnotesize $\mathcal{G}_{\mathrm{Koiter}}=\frac{1}{2}\big[{\rm I}_m - \id_2\big]$\end{minipage}&  \begin{minipage}{3.5cm}\footnotesize $\mathcal{R}_{\mathrm{Koiter}}={\rm II}_m $\end{minipage}  & $\times$& $\times$ & $\times$\\
				\hline
				\footnotesize \begin{minipage}{2cm}\ \\our nonlinear \\ 
					modified\\ constrained \\ Cosserat model\\ \end{minipage} & \begin{minipage}{3cm}\footnotesize $\mathcal{G}_{\infty}=\sqrt{\,{\rm I}_m}-\id_2$\\\\
					$\mathcal{G}_{\infty}^{\rm lin}=\mathcal{G}_{\rm Koiter}^{\rm lin}$\end{minipage}&  \begin{minipage}{3.8cm}\footnotesize  $\mathcal{R}_{\infty}=\,\sqrt{\widehat{\rm I}_m^{-1}}\, {\rm II}_m$\end{minipage} &\begin{minipage}{5.9
						cm}\footnotesize $\sym (\mathcal{R}_{\infty})$,\\ additional vectors:\\ $ \mathcal{N}_\infty= e_3^T  \big(\mbox{axl}(\overline{Q}_\infty^T\partial_{x_1}\overline{Q}_\infty)\,|\, \mbox{axl}(\overline{Q}_\infty^T\partial_{x_2}\overline{Q}_\infty) \big)$\\\end{minipage}& $?$ & $?$\\
				\hline
				\footnotesize \begin{minipage}{2cm}\ \\generalized \\ non-linear Naghdi-type \cite{mardare2008derivation}\end{minipage} & \begin{minipage}{3cm}\footnotesize $\mathcal{G}_{\mathrm{Koiter}}=\frac{1}{2}\big[{\rm I}_m - \id_2\big]$\end{minipage}& $\times$&  \begin{minipage}{5.5cm}\footnotesize $\mathcal{R}_{\mathrm{Naghdi}}=-\sym\big[\nabla m)^T\nabla d\big]$\\ additional vectors:  transverse shear\\$\Big(\langle d, \partial_{x_1}m\rangle,\langle d, \partial_{x_1}m\rangle\Big)$\end{minipage} & $\times$ & $\times$\\
				\hline
				\footnotesize \begin{minipage}{2cm}\ \\our nonlinear\\ Cosserat model\\\end{minipage} & \begin{minipage}{5cm}\footnotesize $
					\mathcal{G} =\, (\overline{Q}_1|\overline{Q}_2)^{T} \nabla m- \id_2\not\in{\rm Sym}(2)$\end{minipage}&  \footnotesize  $\mathcal{R} = \, -(\overline{Q}_1|\overline{Q}_2)^{T} \nabla \overline{Q}_3$ & \begin{minipage}{5.5cm}\vspace*{5mm}\footnotesize\ $\mathcal{R} = \, -(\overline{Q}_1|\overline{Q}_2)^{T} \nabla \overline{Q}_3$,  additional vectors: $ \mathcal{T}=\overline{Q}_3^T\nabla m,$\\ 
					$
					\mathcal{N} = e_3^T  \big(\mbox{axl}(\overline{Q}^T\partial_{x_1}\overline{Q})\,|\, \mbox{axl}(\overline{Q}^T\partial_{x_2}\overline{Q}) \big)$\\ \end{minipage}& ?& ?\\
				\hline
		\end{tabular}}
		\caption{\footnotesize For plate models, $\nabla\Theta(x_3)\,=\,\id_3, $ $y_0(x_1,x_2)\,=\,(x_1,x_2,0),$  $n_0\,=\,e_3$, $ {\rm B}_{\rm id}\,=\,0_3,$  $ {\rm I}_{\rm id} \,=\,\id_2,$ $\
			{\rm II}_{\rm id} \,=\,0_2$ $
			{\rm L}_{\rm id} \,=\,0_2$. $m=\id+v$ is the midsurface deformation, $v$ is the midsurface displacemenet, $\vartheta$ is the infinitesimal rotation and denotes the axial vector of $ \overline{A}_\vartheta $, such that  the Cosserat microrotation
			$ \overline{Q}=\exp(\overline{A}_\vartheta)\; = \;\id_3 + \overline{A}_\vartheta+\textrm{h.o.t.}$. In the nonlinear models: 
			$\mathcal{T}$ is  the transverse shear deformation vector,
			and $
			\mathcal{N}$ is the \textit{vector of drilling bendings}, with their corresponding vector in the linearised models denoted by $\cdot^{\rm lin}$. {\bf We observe that the difference between ``bending tensors" and ``curvature tensors" disappears completely for classical flat shell models.}}\label{tabelbcp}
	\end{table}
\end{landscape}

The key findings of the paper highlight several advantages of our new Cosserat shell models:
	\begin{enumerate}
		\item The shell models are of order $O(h^5)$ concerning the thickness parameter, denoted as $h$. It    is worth noting that while the linear model developed by Anicic and L{\'{e}}ger \cite{anicic1999formulation} also includes higher-order terms in the thickness parameter, these terms are not explicitly stated. However, Anicic and L{\'{e}}ger's model does not  account for Cosserat effects, and it does not  stem directly from a parental 3D nonlinear theory. We assert that determining the internal energy for a shell model is challenging without first having a parent nonlinear model derived from a nonlinear 3D-model. It    is important to remember that a shell, despite being a 2D-approximation of a 3D model, is fundamentally a 3D object with inherent 3D properties.
		\item In addition to Anicic and L{\'{e}}ger's model, our model, even in its linearised  form, includes energy terms describing generalized transverse shear and drilling bendings. This is a result of the presence of Cosserat effects from the outset. Furthermore, the change of curvature measure utilized by Anicic and L{\'{e}}ger is not suitable as a measure for bending, whereas our model accounts for both these measures: one for bending and another dedicated one for the change of curvature.
		\item After linearization, our nonlinear constrained Cosserat shell theory reduces to the shell model with the Koiter-Sanders-Budiansky bending measure. This bending measure, when compared to other bending tensors proposed in the literature, has the property of vanishing in cases of infinitesimal pure stretch deformations of a quadrant of a cylinder. In contrast to Anicic and L{\'{e}}ger's model, it appears natural that the three-dimensional Biot-type energy leads, after dimensional reduction, to in-plane deformation-bending coupling terms rather than a quadratic form involving only the Koiter-Sanders-Budiansky bending measure or the Koiter bending measure.
		\item The inclusion of bending measures in our shell models stems from the incorporation of the Cosserat-curvature 3D energy within the parental 3D energy. It appears that without considering a Cosserat model or a couple stress model, a 3D energy derived solely from classical elasticity (e.g., $W_{\rm Biot}$ or $W_{\rm SVK}$) the appropriate strain measure for  pure bending measures will not appear. Instead,  a measure of curvature is present after dimension reduction.
		\item The internal energy within our nonlinear Cosserat shell theory contains quadratic expressions in strain tensors, which, after linearization, lead to a quadratic form akin to that proposed by Anicic and L{\'{e}}ger. The coercivity of the internal energy in the strain tensors indicates that vanishing energy implies no change     in the initial curvatures. These quadratic terms under discussion align with the recent study by {\v{S}}ilhav{\`y} \cite{vsilhavycurvature}.
		\item In our model we also include quadratic energies in terms of the Koiter-Sanders-Budiansky bending measure due to the presence of the $2D$ approximation of the $3D$ Cosserat curvature energy.

		\item In our linearised  constrained Cosserat-shell model, the constitutive coefficients are derived from the three-dimensional formulation. Additionally, the curved initial shell configuration explicitly factors into the expression of the coefficients for the energies in the reduced two-dimensional variational problem.
	\end{enumerate}

\section{Shell-kinematics and strain measures in  nonlinear  and linearised Cosserat-shell models}\setcounter{equation}{0}\label{seccomp}

\subsection{Shell-like thin domain and  reconstructed 3D deformation}\setcounter{equation}{0}
Let $\Omega_\xi\subset\mathbb{R}^3$ be a three-dimensional curved {\it shell-like thin domain}. Here, the domain  $ \Omega_\xi $ is referred to a  fixed right Cartesian coordinate frame with unit vectors $
 e_i$ along the axes $Ox_i$. A generic point of $\Omega_\xi$ will be denoted by $(\xi_1,\xi_2,\xi_3)$. The elastic material constituting the shell is assumed to be homogeneous and isotropic and the reference configuration $\Omega_\xi$ is assumed to be a natural state. 
The deformation of the body occupying the domain $\Omega_\xi$ is described by a vector map $\varphi_\xi:\Omega_\xi\subset\mathbb{R}^3\rightarrow\mathbb{R}^3$ (\textit{called deformation}) and by a \textit{microrotation}  tensor
$
\overline{R}_\xi:\Omega_\xi\subset\mathbb{R}^3\rightarrow {\rm SO}(3)\, 
$ attached at each point\footnote{ We let ${\rm Sym}(n)$ and ${\rm Sym}^+(n)$ denote the symmetric and positive definite symmetric tensors, respectively.  We adopt the usual abbreviations of Lie-group theory, i.e.,
	${\rm GL}(n)=\{X\in\mathbb{R}^{n\times n}\;|\det({X})\neq 0\}$ the general linear group ${\rm SO}(n)=\{X\in {\rm GL}(n)| X^TX=\id_n,\det({X})=1\}$ with
	corresponding Lie-algebras $\mathfrak{so}(n)=\{X\in\mathbb{R}^{n\times n}\;|X^T=-X\}$ of skew symmetric tensors
	and $\mathfrak{sl}(n)=\{X\in\mathbb{R}^{n\times n}\;| \,\tr({X})=0\}$ of traceless tensors. For all $X\in\mathbb{R}^{n\times n}$ we set ${\rm sym}\, X\,=\frac{1}{2}(X^T+X)\in{\rm Sym}(n)$, $\skw\,X\,=\frac{1}{2}(X-X^T)\in \mathfrak{so}(n)$ and the deviatoric part $\dev \,X\,=X-\frac{1}{n}\;\,\tr(X)\cdot\id_n\in \mathfrak{sl}(n)$  and we have
	the orthogonal Cartan-decomposition  of the Lie-algebra
	$
	\mathfrak{gl}(n)=\{\mathfrak{sl}(n)\cap {\rm Sym}(n)\}\oplus\mathfrak{so}(n) \oplus\mathbb{R}\!\cdot\! \id_n,$ $
	X=\dev\, \sym \,X\,+ \skw\,X\,+\frac{1}{n}\,\tr(X)\!\cdot\! \id_n\,.
	$}. 
We denote the current configuration (deformed configuration) by $\Omega_c:=\varphi_\xi(\Omega_\xi)\subset\mathbb{R}^3$. 

We also need to consider  the \textit{fictitious Cartesian (planar) configuration} of the body $\Omega_h $, see Figure \ref{Fig2}. This parameter domain $\Omega_h\subset\mathbb{R}^3$ is a right cylinder of the form
$$\Omega_h=\left\{ (x_1,x_2,x_3) \,\Big|\,\, (x_1,x_2)\in\omega, \,\,\,-\dfrac{h}{2}\,< x_3<\, \dfrac{h}{2}\, \right\} =\,\,\dd\,\omega\,\times\left(-\frac{h}{2},\,\frac{h}{2}\right),$$
where  $\omega\subset\mathbb{R}^2$ is a bounded domain with Lipschitz boundary
$\partial \omega$ and the constant length $h>0$ is the \textit{thickness of the shell}.
For shell--like bodies we consider   the  domain $\Omega_h $ to be {thin}, i.e., the thickness $h$ is {small}.

In the formulation of the minimization problem we  will consider the  {\it Weingarten map (or shape operator)}  defined by 
$
{\rm L}_{y_0}\,=\, {\rm I}_{y_0}^{-1} {\rm II}_{y_0}\in \mathbb{R}^{2\times 2},
$
where ${\rm I}_{y_0}:=[{\nabla  y_0}]^T\,{\nabla  y_0}\in \mathbb{R}^{2\times 2}$ and  ${\rm II}_{y_0}:\,=\,-[{\nabla  y_0}]^T\,{\nabla  n_0}\in \mathbb{R}^{2\times 2}$ are  the matrix representations of the {\it first fundamental form (metric)} and the  {\it  second fundamental form} of the surface, respectively\footnote{
	For  an open domain  $\Omega\subseteq\mathbb{R}^{3}$,
	the usual Lebesgue spaces of square integrable functions, vector or tensor fields on $\Omega$ with values in $\mathbb{R}$, $\mathbb{R}^3$, $\mathbb{R}^{3\times 3}$ or ${\rm SO}(3)$, respectively will be denoted by ${\rm L}^2(\Omega;\mathbb{R})$, ${\rm L}^2(\Omega;\mathbb{R}^3)$, ${\rm L}^2(\Omega; \mathbb{R}^{3\times 3})$ and ${\rm L}^2(\Omega; {\rm SO}(3))$, respectively. Moreover, we use the standard Sobolev spaces ${\rm H}^{1}(\Omega; \mathbb{R})$ \cite{Adams75,Raviart79,Leis86}
	of functions $u$.  For vector fields $u=\left(    u_1, u_2, u_3\right)^T$ with  $u_i\in {\rm H}^{1}(\Omega)$, $i=1,2,3$,
	we define
	$
	\nabla \,u:=\left(
	\nabla\,  u_1\,|\,
	\nabla\, u_2\,|\,
	\nabla\, u_3
	\right)^T.
	$
	The corresponding Sobolev-space will be denoted by
	$
	{\rm H}^1(\Omega; \mathbb{R}^{3})$. A tensor $Q:\Omega\to {\rm SO}(3)$ having the components in ${\rm H}^1(\Omega; \mathbb{R})$ belongs to ${\rm H}^1(\Omega; {\rm SO}(3))$.  }.  
Then, the {\it Gau{\ss} curvature} ${\rm K}$ of the surface  is determined by
$
{\rm K} :=\,{\rm det}\,{{\rm L}_{y_0}}\, 
$
and the {\it mean curvature} $\,{\rm H}\,$ through
$
2\,{\rm H}\, :={\rm tr}({{\rm L}_{y_0}}).
$  We will also use  the  tensors defined by
$
{\rm A}_{y_0}:=(\nabla y_0|0)\,\,[\nabla\Theta \,]^{-1}\in\mathbb{R}^{3\times 3}, \ 
{\rm B}_{y_0}:=-(\nabla n_0|0)\,\,[\nabla\Theta \,]^{-1}\in\mathbb{R}^{3\times 3},
$
and the so-called \textit{{alternator tensor}} \cite{Zhilin06} of the surface 
\begin{equation}
{\rm C}_{y_0}:=\det\nabla\Theta \,\, [	\nabla\Theta \,]^{-T}\,\begin{footnotesize}\left(\begin{array}{cc|c}
0&1&0 \\
-1&0&0 \\
\hline
0&0&0
\end{array}\right)\end{footnotesize}\,  [	\nabla\Theta \,]^{-1}.
\end{equation}

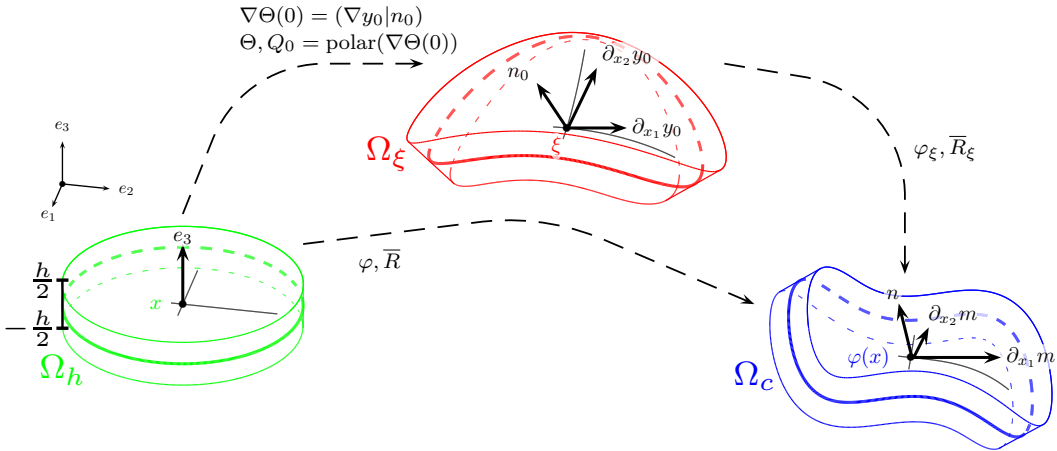
\begin{figure}[h!]
	\begin{center}
		\includegraphics[scale=1.6]{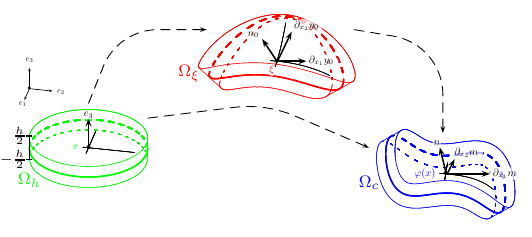}
		\put(-320,165){\footnotesize{$\Theta  ,  Q_0={\rm polar}(\nabla \Theta(0))$}} 
		\put(-320,175){\footnotesize{$\nabla\Theta(0)=(\nabla y_0|n_0)$}}
		\put(-275,82){\footnotesize{$\varphi  , \overline{ R}$}}
		\put(-65,125){\footnotesize{$\varphi_\xi  ,   \overline{ R}_\xi$}}
		\caption{\footnotesize Kinematics of the 3D-Cosserat model. In each point $\xi\in \Omega_\xi$ of the curvy reference  configuration, there is the deformation $\varphi_\xi:\Omega_\xi\to \mathbb{R}^3$ and the microrotation $\overline{ R}_\xi: \Omega_\xi\to {\rm SO}(3)$. We introduce a fictitious  flat configuration $\Omega_h$ and refer all fields to that configuration. This introduces a multiplicative split of the total deformation $\varphi:\Omega_h\to \mathbb{R}^3$ and total rotation $\overline{ R}: \Omega_h\to {\rm SO}(3)$ into ``elastic" parts ($\varphi_\xi: \Omega_\xi\to \mathbb{R}^3$ and $\overline{ R}_\xi: \Omega_\xi\to {\rm SO}(3)$) and compatible ``plastic" parts (given by $\Theta:\Omega_h\to \Omega_\xi$ and $Q_0:\Omega_h\to {\rm SO}(3)$). The "intermediate" configuration $\Omega_\xi$ is compatible by construction.}
		\label{Fig2}       
	\end{center}
\end{figure}
Now, let us  define the map
$
\varphi:\Omega_h\rightarrow \Omega_c,\  \varphi(x_1,x_2,x_3)=\varphi_\xi( \Theta(x_1,x_2,x_3)).
$
We view $\varphi$ as a function which maps the fictitious  planar reference configuration $\Omega_h$ into the deformed configuration $\Omega_c$.
We also consider the \textit{elastic microrotation}
$
\overline{Q}_{e,s}:\Omega_h\rightarrow{\rm SO}(3),\  \overline{Q}_{e,s}(x_1,x_2,x_3):=\overline{R}_\xi(\Theta(x_1,x_2,x_3))\,,
$ see Figure \ref{Fig2}.

The dimensional descent in \cite{GhibaNeffPartI} is done  by   assuming that  the elastic microrotation is constant through the thickness, i.e.
$
\overline{Q}_{e,s}(x_1,x_2,x_3)=\overline{Q}_{e,s}(x_1,x_2), 
$
and  by considering an \textit{8-parameter quadratic ansatz} in the thickness direction for the reconstructed total deformation $\varphi_s:\Omega_h\subset \mathbb{R}^3\rightarrow \mathbb{R}^3$ of the shell-like body, i.e.,
\begin{align}\label{ansatz}
\varphi_s(x_1,x_2,x_3)\,=\,&m(x_1,x_2)+\bigg(x_3\varrho_m(x_1,x_2)+\dd\frac{x_3^2}{2}\varrho_b(x_1,x_2)\bigg)\overline{Q}_{e,s}(x_1,x_2)\nabla\Theta.e_3\, .
\end{align}
Here $m:\omega\subset\mathbb{R}^2\to\mathbb{R}^3$ represents the total 
deformation of the midsurface,  $\varrho_m,\,\varrho_b:\omega\subset\mathbb{R}^2\to \mathbb{R}$ allow in principal for symmetric thickness stretch  ($\varrho_m\neq1$) and asymmetric thickness stretch ($\varrho_b\neq 0$) about the midsurface and they are given analytically by
\begin{align}\label{final_rho}
\varrho_m\,=\,&1-\frac{\lambda}{\lambda+2\,\mu}[\bigl\langle  \overline{Q}_{e,s}^T(\nabla m|0)[\nabla\Theta \,]^{-1},\id_3 \bigr\rangle -2]\;, \\
\dd\varrho_b\,=\,&-\frac{\lambda}{\lambda+2\,\mu}\bigl\langle  \overline{Q}_{e,s}^T(\nabla (\,\overline{Q}_{e,s}\nabla\Theta \,.e_3)|0)[\nabla\Theta \,]^{-1},\id_3 \bigr\rangle   +\frac{\lambda}{\lambda+2\,\mu}\bigl\langle  \overline{Q}_{e,s}^T(\nabla m|0)[\nabla\Theta \,]^{-1}(\nabla n_0|0)[\nabla\Theta \,]^{-1},\id_3 \bigr\rangle .\notag
\end{align}

\subsection{Shell strain tensors  in  Koiter type models}
The classical {\it change of metric}  tensor in the Koiter model \cite{Steigmann12,Steigmann13,Ciarlet00} is given by the difference between the first fundamental form of the unknown midsurface parametrized by $m$ and the first fundamental form of the referential midsurface configuration parametrized by $y_0$
\begin{align} \mathcal{G}_{\mathrm{Koiter}} := \dfrac12 \big[ ( \nabla m)^{T} \nabla m- {\rm I}_{y_0}\big]=\dfrac12\,({\rm I}_m-{\rm I}_{y_0})\in {\rm Sym}(2),\end{align}
with
$
{\rm I}_{m}\coloneqq ({\nabla  m})^T({\nabla  m})\in {\rm Sym}^+(2),
$
while the classical {\it bending strain tensor}  in the Koiter model is defined by
the difference between the second fundamental form of the unknown midsurface parametrized by $m$ and the second fundamental form of the referential midsurface configuration parametrized by $y_0$ \begin{align} \mathcal{R}_{\rm{Koiter}} :=  -(\nabla m)^{T} \nabla n-(-(\nabla y_0)^{T} \nabla n_0)={\rm II}_{m}- {\rm II}_{y_0}\in {\rm Sym}(2).\end{align}

In the linearised model, the total midsurface deformation is written as
\begin{align}
m(x_1,x_2)=y_0(x_1,x_2)+v(x_1,x_2),
\end{align}
with $v:\omega\to \mathbb{R}^3$ the infinitesimal midsurface displacement, and the strain measures \cite{Ciarlet00} of the linearised  Koiter model are given by
\begin{equation}
\label{equ11}
\mathcal{G}_{\rm{Koiter}}^{\rm{lin}} \,\,:=\frac{1}{2}\big[{\rm I}_m - {\rm I}_{y_0}\big]^{\rm{lin}}= \,\,\frac12\;\big[  (\nabla y_0)^{T}(\nabla v) +  (\nabla v)^T(\nabla y_0)\big]
= \sym\big[ (\nabla y_0)^{T}(\nabla v)\big]\in {\rm Sym}(2)
\;
\end{equation}
and
\begin{align}
\label{equ12}
\mathcal{R}_{\rm{Koiter}}^{\rm{lin}} \,\,:&= \,\, \big[{\rm II}_m - {\rm II}_{y_0}\big]^{\rm{lin}} \,= 	 \Big( \bigl\langle n_0 ,  \partial_{x_\alpha  x_\beta}\,v- \dd\sum_{\gamma=1,2}\Gamma^\gamma_{\alpha \beta}\,\partial_{x_\gamma}\,v\bigr\rangle a^\alpha\,\Big)_{\alpha\beta}\in {\rm Sym}(2).
\end{align}
The expression of $\mathcal{R}_{\rm{Koiter}}^{\rm{lin}}$ involves the Christoffel  symbols
$
\Gamma^\gamma_{\alpha\beta}
$  on the surface parametrized by $y_0$ given by
\begin{align}
\Gamma^\gamma_{\alpha\beta}=\bigl\langle a^\gamma, \partial_{x_\alpha} a_\beta\bigr\rangle=-\bigl\langle \partial_{x_\alpha} a^\gamma,  a_\beta\bigr\rangle=\Gamma^\gamma_{\beta\alpha}.
\end{align}
	Here, and in the rest of the paper, $a_1, a_2, a_3\,$ denote the columns of $\nabla\Theta \,$, while  $a^1, a^2, a^3\,$ denote the rows of $[\nabla\Theta \,]^{-1}$, i.e. 
$
\nabla\Theta =(\nabla y_0|\,n_0)\,=\,(a_1|\,a_2|\,a_3),\  [\nabla\Theta \,]^{-1}=(a^1|\,a^2|\,a^3)^T.
$
In fact,   $a_1, a_2\,$ are the covariant base vectors and $ a^1, a^2\,$ are the contravariant base vectors in the tangent plane given by
$
a_\alpha:=\,\partial_{x_\alpha}y_{0},\  \langle a^\beta, a_\alpha\rangle \,=\,\delta_\alpha^\beta,\ \alpha,\beta=1,2,
$
and $a_3\,= a^3=n_0\,$. The following relations hold \cite[page 95]{Ciarlet00}: 
$
\lVert a_1\times a_2\rVert=\sqrt{\det {\rm I}_{y_0}}, $ $  a_3\times a_1=\sqrt{\det {\rm I}_{y_0}}\, a^2, $ $ a_2\times a_3=\sqrt{\det {\rm I}_{y_0}}\, a^1.
$

	Other alternative (equivalent) forms of the change of metric tensor and  the change of curvature tensor   \cite[Page 181]{Ciarlet00} are
\begin{equation}\label{formK}
\mathcal{G}_{\rm{Koiter}}^{\rm{lin}} =  \Big( \frac{1}{2}(\partial_\beta v_\alpha+\partial_\alpha v_\beta)-\sum_{\gamma=1,2}\Gamma_{\alpha\beta}^\gamma v_\gamma-b_{\alpha\beta}v_3 \Big)_{\alpha\beta}\in {\rm Sym}(2),
\end{equation}
and \begin{align}\label{formR}
\mathcal{R}_{\rm{Koiter}}^{\rm{lin}} = & \Big( \partial_{x_\alpha x_\beta}v_3-\sum_{\gamma=1,2}\Gamma_{\alpha\beta}^\gamma \partial_{x_\gamma}v_3-\sum_{\gamma=1,2}b_\alpha^\gamma b_{\gamma\beta}v_3+\sum_{\gamma=1,2}b_{\alpha}^\gamma(\partial_{x_\beta}v_\gamma-\sum_{\tau=1,2}\Gamma_{\beta\gamma}^\tau v_\tau)\\&+\sum_{\gamma=1,2}b_{\beta}^\gamma(\partial_{x_\alpha}v_\gamma-\sum_{\tau=1,2}\Gamma_{\alpha\tau}^\gamma v_\gamma)
+\sum_{\tau=1,2}(\partial _{x_\alpha}b_\beta^\tau+\sum_{\gamma=1,2}\Gamma_{\alpha\gamma}^\tau b_\beta^\gamma-\sum_{\gamma=1,2}\Gamma_{\alpha\beta}^\gamma b_\gamma^\tau)v_\tau\Big)_{\alpha\beta}\in {\rm Sym}(2),\notag
\end{align}  respectively,
	where  $b_{\alpha\beta}(m)$ are the components of the second fundamental form corresponding to the map $m$, $b_{\alpha}^\beta(m)$ are the components of the matrix associated to the Weingarten map (shape operator).
	\subsection{Kinematics in the  Cosserat shell model}
\subsubsection{Strain tensors in the nonlinear Cosserat shell model}
In the resulting fully two-dimensional minimization problem, the reduced energy density is formulated using the following tensor fields which are also discussed in \cite{Libai98,Pietraszkiewicz-book04,Eremeyev06,NeffBirsan13}, and \cite{Birsan-Neff-L54-2014}, although with different contexts and reasons for their significance, all defined on the surface denoted as $\omega$. The tensor fields are\footnote{A matrix having the  three  column vectors $A_1,A_2, A_3$ will be written as 
$
(A_1\,|\, A_2\,|\,A_3).
$ 
We make use of the operator $\mathrm{axl}: \mathfrak{so}(3)\to\mathbb{R}^3$ associating with a matrix $A\in \mathfrak{so}(3)$ the vector $\mathrm{axl}({A})\coloneqq(-A_{23},A_{13},-A_{12})^T$. The inverse operator will be denoted by ${\rm Anti}: \mathbb{R}^3\to \mathfrak{so}(3)$.}
\begin{align}\label{e55}
\mathcal{E}_{m,s} & :\,=\,    \overline{Q}_{e,s}^T  (\nabla  m|\overline{Q}_{e,s}\nabla\Theta \,.e_3)[\nabla\Theta \,]^{-1}-\id_3\not\in {\rm Sym}(3),\qquad \quad \qquad\ \ \ \ \,  \text{{\it elastic shell strain tensor}} ,  \\
\mathcal{K}_{e,s} & :\,=\,  \Big(\mathrm{axl}(\overline{Q}_{e,s}^T\,\partial_{x_1} \overline{Q}_{e,s})\,|\, \mathrm{axl}(\overline{Q}_{e,s}^T\,\partial_{x_2} \overline{Q}_{e,s})\,|0\Big)[\nabla\Theta \,]^{-1}\not\in {\rm Sym}(3) \quad \text{\it\  elastic shell bending--curvature tensor}.\notag
\end{align}
Beside these two strain tensors, in the expression of the internal energy two other tensors are present (with different physical meaning, in comparison with the elastic shell strain tensor and the elastic shell bending-curvature tensor), namely
$
\mathrm{C}_{y_0} \mathcal{K}_{e,s}
$
and
$
\mathcal{E}_{\infty} {\rm B}_{y_0}  + \mathrm{C}_{y_0} \mathcal{K}_{e,s}.
$
In order to see what each of these four tensors measures, we observe that we can express the strain tensors as
\begin{align}\label{eq5}
\mathcal{E}_{m,s}=&\quad\ \, [\nabla\Theta \,]^{-T}
\begin{footnotesize}\left( \begin{array}{c|c}
(\overline{Q}_{e,s} \nabla y_0)^{T} \nabla m- {\rm I}_{y_0} & \mathbf{0}\\
\hline\vspace*{2mm}
\raisebox{-2pt}{$(\overline{Q}_{e,s}  n_0)^{T} \nabla m$} & \raisebox{-2pt}{$0$}
\end{array} \right)\end{footnotesize} [\nabla\Theta \,]^{-1}=
[\nabla\Theta \,]^{-T}
\begin{footnotesize}\left( \begin{array}{c|c}
\mathcal{G} & \mathbf{0}\\
\hline\vspace*{2mm}
\raisebox{-2pt}{$\mathcal{T}$}  & \raisebox{-2pt}{$0$}
\end{array} \right)\end{footnotesize} [\nabla\Theta \,]^{-1},
\\
\mathrm{C}_{y_0} \mathcal{K}_{e,s} = &\quad\ \, [\nabla\Theta \,]^{-T}
\begin{footnotesize}\left( \begin{array}{c|c}
(\overline{Q}_{e,s} \nabla y_0)^{T} \nabla (\overline{Q}_{e,s} n_0)+ {\rm II}_{y_0} & \mathbf{0}\\
\hline\vspace*{2mm}
\raisebox{-2pt}{$0$} & \raisebox{-2pt}{$0$}
\end{array} \right)\end{footnotesize} [\nabla\Theta \,]^{-1}= -[\nabla\Theta \,]^{-T}
\begin{footnotesize}\left( \begin{array}{c|c}
\mathcal{R} & \mathbf{0}\\
\hline\vspace*{2mm}
\raisebox{-2pt}{$0$} & \raisebox{-2pt}{$0$}
\end{array} \right)\end{footnotesize} [\nabla\Theta \,]^{-1},\notag\\
\mathcal{E}_{m,s} {\rm B}_{y_0}  + \mathrm{C}_{y_0} \mathcal{K}_{e,s} 
= &-\, [\nabla_x \Theta]^{-T}
\begin{footnotesize}\left( \begin{array}{c|c}
 \mathcal{R}-\mathcal{G} \,{\rm L}_{y_0} & \mathbf{0}\\
 \hline\vspace*{2mm}
\raisebox{-2pt}{$\mathcal{T} \,{\rm L}_{y_0}$} & \raisebox{-2pt}{$0$}
\end{array} \right)\end{footnotesize} [\nabla_x \Theta]^{-1}\notag
,
\end{align}
with
\begin{align}\label{eq41}
\mathcal{G} :=&\, (\overline{Q}_{e,s} \nabla y_0)^{T} \nabla m- {\rm I}_{y_0}\not\in {\rm Sym}(2)\qquad\qquad\qquad\qquad\quad\ \ \ \,\  \,\textrm{\it the change of metric tensor},
\\
\mathcal{T}:=& \, (\overline{Q}_{e,s}  n_0)^{T} \nabla m= \, \left(\bigl\langle\overline{Q}_{e,s}  n_0, \partial_{x_1} m\bigr\rangle,\bigl\langle\overline{Q}_{e,s}  n_0, \partial_{x_2} m\bigr\rangle\right)\qquad  \textrm{\it the transverse shear deformation (row) vector},\notag
\\
\mathcal{R} :=& \, -(\overline{Q}_{e,s} \nabla y_0)^{T} \nabla (\overline{Q}_{e,s} n_0)- {\rm II}_{y_0}\not\in {\rm Sym}(2)
\quad  \qquad\qquad \ \ \, \,\,\textrm{\it the bending  strain tensor}.
\notag\end{align}

The non-symmetric  quantity $\mathcal{R} -\mathbf{1}\, \mathcal{G} \, {\rm L}_{y_0}\not\in {\rm Sym}(2)$ serves as a representation of what we refer to as the {\it change of curvature} tensor. The rationale behind this nomenclature will become evident as this paper progresses. For now, it is  important to note that the definition of $\mathcal{G}$ is associated with the classical {\it change of metric} tensor in the Koiter model \cite{Steigmann12,Steigmann13,Ciarlet00}, denoted as $\mathcal{G}_{\mathrm{Koiter}}$. 

Our bending strain tensor $\mathcal{R}$ generalises the bending strain tensor in the Naghdi-type shell model  \cite[p.~11]{mardare2008derivation} with one independent director field  $d:\omega\subset\mathbb{R}^2\to\mathbb{R}^3$  given by
\begin{align}
\mathcal{R}_{\rm{Naghdi}} &\coloneqq   -[\sym((\nabla m)^{T} \nabla d)-(\nabla y_0)^{T} \nabla n_0]= -(\sym((\nabla m)^{T} \nabla d)-{\rm II}_{y_0})\in {\rm Sym}(2).\notag
\end{align}

Identifying $d=\overline{Q}_{e,s} n_0$, our transverse shear deformation row vector $\mathcal{T}$ may be seen in relation to the transverse shear deformation row vector in the Naghdi-type shell models, see  \cite[Section 6]{mardare2008derivation} and \cite{gastel2022regularity}, which is \begin{align}\mathcal{T}_{\rm Naghdi}:= \, d^{T} \nabla m=\bigl\langle d, \partial_{x_1} m\bigr\rangle,\bigl\langle d, \partial_{x_2} m\bigr\rangle.
	\end{align}
	
 The presence of  our additional strain tensor
\begin{align}\mathcal{C}:=\mathcal{R}-\mathcal{G} \,{\rm L}_{y_0}&=-(\overline{Q}_{e,s} \nabla y_0)^{T} \nabla (\overline{Q}_{e,s} n_0)- {\rm II}_{y_0} -(\overline{Q}_{e,s} \nabla y_0)^{T} \nabla m\,{\rm L}_{y_0}+ {\rm I}_{y_0}\,{\rm L}_{y_0},\notag\\
&=-(\overline{Q}_{e,s} \nabla y_0)^{T} \nabla (\overline{Q}_{e,s} n_0) -(\overline{Q}_{e,s} \nabla y_0)^{T} \nabla m\,{\rm L}_{y_0}
\end{align} named by us {\it change of curvature} is surprising at first. Our change of curvature tensor $\mathcal{C}$ replaces  the strain tensor 
\begin{align}
\mathcal{P}_{\rm Naghdi}=(\nabla d)^T(\nabla d)-{\rm III}_{y_0}
\end{align}
considered in some Naghdi-type models, see  \cite[Section 6]{mardare2008derivation},   but its expression is fundamentally different, primarily because it involves the deformation gradient $\nabla m$ and not only the gradient of the director $d$.

It is possible to express the tensor $ \mathcal{K}_{e,s}$  in terms of the tensor $ \mathrm{C}_{y_0}\, \mathcal{K}_{e,s}$ and the vector $ \mathcal{K}_{e,s}^T n_0$ as
\begin{align}  \label{descK0}
\mathcal{K}_{e,s} = {\rm A}_{y_0} \, \mathcal{K}_{e,s} +(0|0|n_0) \,(0|0|n_0)^T \, \mathcal{K}_{e,s}= \mathrm{C}_{y_0}( - \mathrm{C}_{y_0} \mathcal{K}_{e,s}) +(0|0|n_0) \,(0|0|\mathcal{K}_{e,s}^T\,n_0)^T\, \, .
\end{align} 
We have already seen that   ${\rm C}_{y_0} \mathcal{K}_{e,s}$ from the above decomposition can be expressed in terms of the {\it bending strain} tensor $ \mathcal{R} $,
while
the remaining vector $\mathcal{K}_{e,s}^T\,n_0$ from \eqref{descK0} is completely characterized by the row vector
\begin{equation}
\label{e5d0}
\mathcal{N} :=  n_0^T \big(\mbox{axl}(\overline{Q}_{e,s}^T\partial_{x_1}\overline{Q}_{e,s})\,|\, \mbox{axl}(\overline{Q}_{e,s}^T\partial_{x_2}\overline{Q}_{e,s}) \big) ,
\end{equation}
which is   called the row vector of {\it drilling bendings}. 

One aim of the present paper is to argue the following new names of the involved strain tensors
\begin{align}\label{e55}
\mathcal{E}_{m,s} & \hspace{1.5cm}  \text{{\it the change of metric-transverse shear deformation strain tensor}}, \notag \\
\mathcal{K}_{e,s} &  \hspace{1.5cm} \text{\it the  bending-drilling strain  tensor},\\
\mathrm{C}_{y_0} \mathcal{K}_{e,s}& \hspace{1.5cm}  \text{\it the  bending strain  tensor}, \notag\\
\mathcal{E}_{m,s} {\rm B}_{y_0}  + \mathrm{C}_{y_0} \mathcal{K}_{e,s}& \hspace{1.5cm}  \text{\it the change of curvature-transverse shear deformation strain tensor}.\notag
\end{align}
\subsubsection{Strain tensors in the linearised  Cosserat shell model}

In the linearised model, as usual,  the total midsurface deformation is written
$
m(x_1,x_2)=y_0(x_1,x_2)+v(x_1,x_2),
$
with $v:\omega\to \mathbb{R}^3$ the infinitesimal shell-midsurface displacement, while the elastic rotation tensor $ \overline{Q}_{e,s}\in\rm{SO}(3) $ 
is approximated by
$ \overline{Q}_{e,s}= \;\id_3 + \overline{A}_\vartheta+{\rm h.o.t},$
where
the skew-symmetric matrix $\overline{A}_\vartheta\in \mathfrak{so}(3)$ is   the infinitesimal elastic  microrotation $
\overline{A}_\vartheta:={\rm Anti}(\vartheta_1,\vartheta_2,\vartheta_3):=\begin{footnotesize}
\left(\begin{array}{ccc}
0&-\vartheta_3&\vartheta_2\\
\vartheta_3&0&-\vartheta_1\\
-\vartheta_2&\vartheta_1&0
\end{array}\right)\end{footnotesize}\in \mathfrak{so}(3)
$
and $ \vartheta={\rm axl}( \overline{A}_\vartheta) $ denotes the corresponding axial vector of $ \overline{A}_\vartheta $.
  Then, we linearise all the previous strain tensors and  obtain 
  \begin{align}\label{eq5}
  \mathcal{E}_{m,s}^{\rm lin}=&=
  [\nabla\Theta \,]^{-T}
  \begin{footnotesize}\left( \begin{array}{c|c}
  \mathcal{G}^{\rm lin} & \mathbf{0}\\
  \hline
  \raisebox{-2pt}{$\mathcal{T}^{\rm lin}$}  & \raisebox{-2pt}{$0$}
  \end{array} \right)\end{footnotesize} [\nabla\Theta \,]^{-1},
  \vspace{6pt}\\
  \mathrm{C}_{y_0} \mathcal{K}_{e,s}^{\rm lin} = & -[\nabla\Theta \,]^{-T}
  \begin{footnotesize}\left( \begin{array}{c|c}
  \mathcal{R}^{\rm lin} & \mathbf{0}\\\hline
  \raisebox{-2pt}{$0$} & \raisebox{-2pt}{$0$}
  \end{array} \right)\end{footnotesize} [\nabla\Theta \,]^{-1},\notag\\
  \mathcal{E}_{m,s}^{\rm lin} {\rm B}_{y_0}  + \mathrm{C}_{y_0} \mathcal{K}_{e,s} ^{\rm lin}
  = &\,- [\nabla_x \Theta]^{-T}
  \begin{footnotesize}\left( \begin{array}{c|c}
   \mathcal{R}^{\rm lin}-\mathcal{G}^{\rm lin} \,{\rm L}_{y_0} & \mathbf{0}\\\hline
 \raisebox{-2pt}{$ \mathcal{T}^{\rm lin} \,{\rm L}_{y_0}$} & \raisebox{-2pt}{$0$}
  \end{array} \right)\end{footnotesize} [\nabla_x \Theta]^{-1}\notag
  \end{align}
  and 
  \begin{align}
 \mathcal{K}_{e,s}^{\rm{lin}} =  (\nabla\vartheta\, |\, 0) \; [\nabla\Theta \,]^{-1},
  \end{align}
  where\footnote{
  		For any column vector $ q\in \mathbb{R}^3$ and any  matrix $ M=(M_1|M_2|M_3)\in \mathbb{R}^{3\times 3}  $ we define the cross-product
  $
  	q\times M :=\, (q\times M_1\,|\,q\times M_2\,|\,q\times M_3) \ (\mbox{operates on columns})\ \textrm{and}\ 
  	M^T \times q^T:=\, - (q\times M)^T\ (\mbox{operates on rows}).\notag
  $
  	Note that $ M $ can also be a $ 3\times 2 $ matrix, the definition remains the same.
  }
  \begin{align}
  \mathcal{G}^{\rm{lin}} &=    (\nabla y_0)^{T}  \nabla v + (\vartheta\times\nabla y_0 )^T\nabla y_0\,,\notag
 \qquad 
  \mathcal{T}^{\rm{lin}}  = 
  n_0^{T}  \nabla v + (\vartheta\times n_0)^T\nabla y_0 \;,
  \qquad 
  \mathcal{R}^{\rm{lin}} =    - ( n_0\times\nabla y_0)^{T}  \nabla \vartheta \;\notag
  \end{align}
  are the linearisation of $ \mathcal{G},  \mathcal{T}$ and $ \mathcal{R}$, respectively.

 Since the strain tensors from the Cosserat shell model are correlated to strain tensors from the Naghdi-type models, we give the linearised form of the tensors considered in these models \cite[Section 7]{mardare2008derivation} 
 \begin{align}
 \mathcal{R}_{\rm{Naghdi}}^{\rm lin} &\coloneqq   -[\sym((\nabla y_0)^{T} \nabla d)+\sym((\nabla v)^{T} \nabla n_0)]\in {\rm Sym}(2)
 \end{align}
 and  \
 \begin{align}\mathcal{T}_{\rm Naghdi}^{\rm lin}:= \, n_0^{T} \nabla v+\zeta^{T} \nabla y_0= \, \left(\bigl\langle n_0, \partial_{x_1} v\bigr\rangle,\bigl\langle n_0, \partial_{x_2} v\bigr\rangle\right)+\left(\bigl\langle \zeta, \partial_{x_1} y_0\bigr\rangle,\bigl\langle \zeta, \partial_{x_2} y_0\bigr\rangle\right),
 \end{align}
where $d=n_0+\zeta$. The linearisation of the Naghdi-type candidate for the change of curvature $\mathcal{C}_{\rm Naghdi}$ is \begin{align}
\mathcal{C}_{\rm Naghdi}^{\rm lin}=\frac{1}{2}[(\nabla n_0)^T(\nabla d)+(\nabla d)^T(\nabla n_0)]=\sym[(\nabla n_0)^T(\nabla d)] .
\end{align}

To  obtain a comparison with the classical linear Koiter-shell model, let us first present an alternative form of $\mathcal{G}^{\rm lin}$, i.e., we can express $\mathcal{G}^{\rm lin}$ also as
\begin{equation}\label{equ3}
\begin{array}{rcl}
\mathcal{G}^{\rm lin} &=&    (\nabla y_0)^{T}(\nabla v) + \bigl\langle\vartheta, n_0\bigr\rangle\begin{footnotesize}
\sqrt{\det {\rm I}_{y_0}}\left(\begin{array}{cc}
0 & 1
\\
-1 & 0
\end{array}\right)\end{footnotesize}.
\end{array}
\end{equation}  From \eqref{equ3}  we note the relation
\begin{equation}
\label{equ12,5}
\sym \,\mathcal{G}^{\rm{lin}} \; =\; \sym\big[ (\nabla y_0)^{T}(\nabla v)\big] \;=\; \mathcal{G}_{\rm{Koiter}}^{\rm{lin}} \;,
\end{equation}
therefore  $ \;\mathcal{G}_{\rm{Koiter}}^{\rm{lin}} $ corresponds to the symmetric part of our $ \mathcal{G}^{\rm{lin}} $ but it does not coincide with $ \mathcal{G}^{\rm{lin}}$.

It is not possible to establish an equivalence between the linear bending strain tensor in the Koiter model and the bending strain tensor addressed in our linear Cosserat-shell model. This difference arises because, whereas $\mathcal{R}_{\rm{Koiter}}^{\rm{lin}}$ relies solely on infinitesimal displacements, $\mathcal{R}^{\rm{lin}}$ is contingent on both infinitesimal displacements and infinitesimal elastic microrotations.

\subsection{Kinematics in the constrained Cosserat shell model}
\subsubsection{Strain tensors in the constrained modified nonlinear Cosserat shell model}

In the constrained Cosserat model, the microrotation  is not any more an independent unknown \cite{GhibaNeffPartIII,neff2004geometrically} and \cite{neff2013grioli} of the model, it now depends on the  deformation of the total midsurface through
\begin{align}
\overline{Q}_{ e,s}\equiv\overline{Q}_{ \infty }:={\rm polar}\big((\nabla  m|n) [\nabla\Theta ]^{-1}\big)=(\nabla m|n)[\nabla\Theta ]^{-1}\,\sqrt{[\nabla\Theta ]\,\widehat {\rm I}_{m}^{-1}\,[\nabla\Theta ]^{T}},
\end{align}
with the lifted quantity $\widehat{\rm I}_{m} \in \mathbb{R}^{3\times 3}$   given by
$
\widehat{\rm I}_{m}\coloneqq ({\nabla  m}|n)^T({\nabla  m}|n),\ \   n\,=\,\dd\frac{\partial_{x_1}m\times \partial_{x_2}m}{\lVert \partial_{x_1}m\times \partial_{x_2}m\rVert}.
$

Thus, the strain tensors of the unconstrained  nonlinear Cosserat shell model become the following strain tensors of the constrained  nonlinear Cosserat shell model
\begin{itemize}
	\item  the symmetric elastic shell strain tensor 	$\mathcal{E}_{ \infty }\in {\rm Sym}(3)$: \begin{align}\label{e551}
	\mathcal{E}_{ \infty }&:=   \overline{Q}_{\infty}^T  (\nabla  m|\overline{Q}_{\infty}\nabla\Theta \,.e_3)[\nabla\Theta \,]^{-1}-\id_3=[{\rm polar}\big((\nabla  m|n) [\nabla\Theta ]^{-1}\big)]^T(\nabla m|n)[\nabla\Theta ]^{-1}-\id_3\\	&\ = \sqrt{[\nabla\Theta ]^{-T}\,\widehat{\rm I}_m\,\id_2^{\flat }\,[\nabla\Theta ]^{-1}}-
	\sqrt{[\nabla\Theta ]^{-T}\,\widehat{\rm I}_{y_0}\,\id_2^{\flat }\,[\nabla\Theta ]^{-1}};\hspace{8cm}\notag
	\end{align}
	\item the (still non-symmetric) elastic shell bending--curvature tensor $\mathcal{K}_{\infty}\not\in {\rm Sym}(3)$: \begin{align}\label{e552}
	\mathcal{K}_{\infty} & :\,=\,  \Big(\mathrm{axl}(\overline{Q}_{\infty}^T\,\partial_{x_1} \overline{Q}_{\infty})\,|\, \mathrm{axl}(\overline{Q}_{\infty}^T\,\partial_{x_2} \overline{Q}_{\infty})\,|0\Big)[\nabla\Theta \,]^{-1}\notag\\&\ =\bigg(\mathrm{axl}(\, \sqrt{[\nabla\Theta ]\,\widehat{\rm I}_m^{-T}[\nabla\Theta ]^{T}}[\nabla\Theta ]^{-T}(\nabla m|n)^T\,\partial_{x_1} \Big((\nabla m|n)[\nabla\Theta ]^{-1} \sqrt{[\nabla\Theta ]\,\widehat{\rm I}_m^{-1}[\nabla\Theta ]^{T}}\Big)\big)\\
	& \ \ \ \ \    \ \,|\, \mathrm{axl}(\, \sqrt{[\nabla\Theta ]\,\widehat{\rm I}_m^{-T}[\nabla\Theta ]^{T}}[\nabla\Theta ]^{-T}(\nabla m|n)^T\,\partial_{x_2} \Big((\nabla m|n)[\nabla\Theta ]^{-1} \sqrt{[\nabla\Theta ]\,\widehat{\rm I}_m^{-1}[\nabla\Theta ]^{T}}\Big)\big)  \,|0\bigg)[\nabla\Theta ]^{-1}\hspace{3cm}.\notag
	\end{align}
\end{itemize}

Let us recall that in the constrained parental 3D Cosserat model, the microrotation is chosen to be \begin{align}Q_{\rm 3D}={\rm polar} (\nabla \varphi_{\rm 3D}),
\end{align} where $\varphi_{\rm 3D}$ is the 3D-deformation, and the 3D strain measure becomes the symmetric Biot-strain $\mathcal{E}_{\rm 3D}=\sqrt{(\nabla \varphi_{\rm 3D})^T (\nabla \varphi_{\rm 3D})}-\id_3\in {\rm Sym}(3)$. Therefore, in the constrained Cosserat shell  model, the reconstructed strain measure given by  
\begin{align}\label{extE}
{\mathcal{E}}_{\rm 3D}  \; =\; &\,\quad \,\,
1\,\Big[  \underbrace{\mathcal{E}_{\infty}}_{\in {\rm Sym}(3)} - \frac{\lambda}{\lambda+2\mu}\,\tr( \mathcal{E}_{\infty} )\; \underbrace{(0|0|n_0)\, (0|0|n_0)^T}_{\in {\rm Sym}(3)}  \Big]
\notag\vspace{2.5mm}\\
& 
+x_3\Big[ (\mathcal{E}_{\infty} \, {\rm B}_{y_0} +  {\rm C}_{y_0} \mathcal{K}_{\infty}) -
\frac{\lambda}{(\lambda+2\mu)}\, {\rm tr}  (\mathcal{E}_{\infty} {\rm B}_{y_0} + {\rm C}_{y_0}\mathcal{K}_{\infty} )\; \underbrace{(0|0|n_0)\,  (0|0|n_0)^T}_{\in {\rm Sym}(3)}  \Big]
\vspace{2.5mm}\\
& 
+x_3^2\Big[\,(\mathcal{E}_{\infty} \, {\rm B}_{y_0} +  {\rm C}_{y_0} \mathcal{K}_{\infty}) {\rm B}_{y_0} \Big]\;+\; O(x_3^3)\notag
\end{align}
{\bf should  be symmetric! }
Since $\{1,x_3,x_3^2\}$ are linear independent, from the last relation we see that the symmetry of the reconstructed 3D strain measures  ${\mathcal{E}}_{\rm 3D}$ in \eqref{extE} leads to the idea to use  only the symmetric parts of the tensors in the reconstruction, i.e., 
\begin{align}\label{eq5}
\mathcal{E}_{\infty}=&\,
[\nabla\Theta \,]^{-T}
\begin{footnotesize}\left( \begin{array}{c|c}
\mathcal{G}_{\infty} & \mathbf{0} \\\hline
\raisebox{-2pt}{$0$}  & \raisebox{-2pt}{$0$}
\end{array} \right)\end{footnotesize} [\nabla\Theta \,]^{-1},
\vspace{6pt}\notag\\
\\
\mathcal{E}_{\infty} {\rm B}_{y_0}  + \mathrm{C}_{y_0} \mathcal{K}_{\infty} 
= &\, -[\nabla\Theta \,]^{-T}
\begin{footnotesize}\left( \begin{array}{c|c}
\mathcal{R}_{\infty}-\mathcal{G}_{\infty} \,{\rm L}_{y_0} & \mathbf{0} \\\hline \raisebox{-2pt}{$0$} & \raisebox{-2pt}{$0$}
\end{array} \right)\end{footnotesize} [\nabla\Theta \,]^{-1}\notag
,
\end{align}
where
\begin{align}\label{eq41}
\mathcal{G}_{\infty} :=&\, (\overline{Q}_{\infty} \nabla y_0)^{T} \nabla m- {\rm I}_{y_0}\in {\rm Sym}(2), \qquad \qquad 	\mathcal{R}_{\infty} := \, -(\overline{Q}_{\infty} \nabla y_0)^{T} \nabla (\overline{Q}_{\infty} n_0)- {\rm II}_{y_0}\not\in {\rm Sym}(2).
\end{align}

Following this idea, in the modified constrained Cosserat-shell model we consider the symmetrized strain measures
\begin{align}\label{eq5m}
\sym\,\mathcal{E}_{\infty}=\mathcal{E}_{\infty}=&\,
[\nabla\Theta \,]^{-T}
\begin{footnotesize}\left( \begin{array}{c|c}
\mathcal{G}_{\infty} & \mathbf{0} \\\hline
\raisebox{-2pt}{$0$}  & \raisebox{-2pt}{$0$}
\end{array} \right)\end{footnotesize} [\nabla\Theta \,]^{-1},
\vspace{6pt}\notag\\
\sym(\mathcal{E}_{\infty} {\rm B}_{y_0}  + \mathrm{C}_{y_0} \mathcal{K}_\infty )
= &\, -[\nabla\Theta \,]^{-T}
\begin{footnotesize}\left( \begin{array}{c|c}
\sym(\mathcal{R}_{\infty}-\mathcal{G}_{\infty} \,{\rm L}_{y_0}) & \mathbf{0} \\\hline \raisebox{-2pt}{$0$} & \raisebox{-2pt}{$0$}
\end{array} \right)\end{footnotesize} [\nabla\Theta \,]^{-1}
,\\
\sym(\mathcal{E}_{\infty} {\rm B}_{y_0}^2  + \mathrm{C}_{y_0} \mathcal{K}_{\infty} {\rm B}_{y_0})
= &\, -[\nabla\Theta ]^{-T}
\begin{footnotesize}\left( \begin{array}{c|c}
\sym[(\mathcal{R}_{\infty} -\mathcal{G}_{\infty} \,{\rm L}_{y_0})\,{\rm L}_{y_0}]& \mathbf{0} \\\hline
\raisebox{-2pt}{$0$} & \raisebox{-2pt}{$0$}
\end{array} \right)\end{footnotesize} [\nabla\Theta ]^{-1},\notag
\end{align}
where we used that $\sym (X^T Y\, X )=X^T \sym (Y)\, X$, and the reconstructed symmetric 3D strain measure is then
\begin{align}\label{extE}
{\mathcal{E}}_{\rm 3D}  \; =\; &\,\quad \,\,
1\,\Big[  \mathcal{E}_{\infty} - \frac{\lambda}{\lambda+2\mu}\,\tr( \mathcal{E}_{\infty} )\; (0|0|n_0)\, (0|0|n_0)^T  \Big]
\notag\vspace{2.5mm}\\
& 
+x_3\Big[ \sym(\mathcal{E}_{\infty} \, {\rm B}_{y_0} +  {\rm C}_{y_0} \mathcal{K}_{\infty}) -
\frac{\lambda}{(\lambda+2\mu)}\, {\rm tr}  (\mathcal{E}_{\infty} {\rm B}_{y_0} + {\rm C}_{y_0}\mathcal{K}_{\infty} )\; (0|0|n_0)\,  (0|0|n_0)^T  \Big]
\vspace{2.5mm}\\
& 
+x_3^2\Big[\,\sym[(\mathcal{E}_{m,s} \, {\rm B}_{y_0} +  {\rm C}_{y_0} \mathcal{K}_{e,s}) {\rm B}_{y_0}] \Big]\;+\; O(x_3^3).\notag
\end{align}
Note that the symmetry of $ \mathcal{G}_\infty $ follows from the symmetry of $ \mathcal{E}_\infty$. The constrained Cosserat-shell model is not able to reflect the effect of the transverse shear vector $\mathcal{T}_\infty:= \, (\overline{Q}_{\infty}  n_0)^T (\nabla m) $, since from the constraint for the expression of $\overline{Q}_{\infty}$ it follows that the transverse shear vector vanishes.

We also notice that,  in the constrained model, too, we use the decomposition 
of  $ \mathcal{K}_{\infty}$  in terms of the tensor $ \mathrm{C}_{y_0}\, \mathcal{K}_{\infty}$ and the vector $ \mathcal{K}_{\infty}^T n_0$ as
\begin{align}  \label{descK0c}
\mathcal{K}_{\infty} = {\rm A}_{y_0} \, \mathcal{K}_{\infty} +(0|0|n_0) \,(0|0|n_0)^T \, \mathcal{K}_{\infty}= \mathrm{C}_{y_0}( - \mathrm{C}_{y_0} \mathcal{K}_{\infty}) +(0|0|n_0) \,(0|0|\mathcal{K}_{\infty}^T\,n_0)^T\, \, ,
\end{align} 
and that   ${\rm C}_{y_0} \mathcal{K}_{\infty}$ from the above decomposition can be expressed in terms of the {\it bending strain} tensor $ \mathcal{R}_{\infty} $ by \begin{align}
 \mathrm{C}_{y_0} \mathcal{K}_{\infty} = &\, -[\nabla\Theta \,]^{-T}
 \begin{footnotesize}\left( \begin{array}{c|c}
 \mathcal{R}_{\infty} & \mathbf{0} \\\hline
 \raisebox{-2pt}{$0$} & \raisebox{-2pt}{$0$}
 \end{array} \right)\end{footnotesize} [\nabla\Theta \,]^{-1},
 \end{align}
while
the remaining vector $\mathcal{K}_{\infty}^T\,n_0$ from \eqref{descK0c} is completely characterized by the row vector ({\it drilling bendings})
\begin{equation}
\label{e5d0c}
\mathcal{N}_{\infty} :=  n_0^T \big(\mbox{axl}(\overline{Q}_{\infty}^T\partial_{x_1}\overline{Q}_{\infty})\,|\, \mbox{axl}(\overline{Q}_{\infty}^T\partial_{x_2}\overline{Q}_{\infty}) \big) .
\end{equation}

	As we will explain in Subsection \ref{vcm}, we do not identify a modelling reason for considering only the symmetric part of $ \mathcal{R}_\infty$ as kinematic strain measure.
In the constrained Cosserat shell model, the still non-symmetric bending strain tensor $ \mathcal{R}_\infty$ has the following mixed expression in terms of the first and  second fundamental form
\begin{align}\,
\mathcal{R}_{\infty}^\flat=\,[\nabla\Theta \,]^{T}\Big(&\sqrt{[\nabla\Theta ]\,\widehat{\rm I}_m^{-1}[\nabla\Theta ]^{T}}\,[\nabla\Theta ]^{-T} {\rm II}_m^\flat[\nabla\Theta ]^{-1} -\sqrt{[\nabla\Theta ]\,\widehat{\rm I}_{y_0}^{-1}[\nabla\Theta ]^{T}}[\nabla\Theta ]^{-T}{\rm II}_{y_0}^\flat [\nabla\Theta ]^{-1}\Big)\nabla\Theta.
\end{align}
Moreover, for in-extensional deformations
${\rm I}_m={\rm I}_{y_0}$ (pure flexure), the bending strain tensor turns into
\begin{align}\,
\mathcal{R}_{\infty}^\flat=\,[\nabla\Theta \,]^{T}\sqrt{[\nabla\Theta ]\,\widehat{\rm I}_{y_0}^{-1}[\nabla\Theta ]^{T}}\,[\nabla\Theta ]^{-T}\big( {\rm II}_m^\flat -{\rm II}_{y_0}^\flat \big)[\nabla\Theta ]^{-1}\nabla\Theta={\rm II}_m^\flat -{\rm II}_{y_0}^\flat =\mathcal{R}_{\rm Koiter}^\flat\in{\rm Sym}(3)\notag.
\end{align}
Hence, in the pure flexure case $\mathcal{R}_{\infty}^\flat$ coincides with the classical Koiter bending tensor $\mathcal{R}_{\rm Koiter}^\flat$.

However, when it comes to coupled membrane bending or a change in membrane curvature (flexure), there is no immediate indication as to why the classical Koiter bending tensor $\mathcal{R}_{\rm Koiter}^\flat$ should serve as a suitable measure for bending or  curvature change. In fact, as we will explore, alternate variations of the classical Koiter tensor, some of which Koiter himself has acknowledged, or others supported by clear arguments presented by {\v{S}}ilhav{\`y} \cite{vsilhavycurvature}, can be contemplated within a coupled membrane bending or curvature alteration (flexural) model.

\subsubsection{Strain tensors in the linearised  constrained Cosserat shell model}

In the linearised constrained Cosserat shell model, as  reminiscent of the conditions imposed by  the constrained Cosserat shell model, the infinitesimal microrotation and the infinitesimal displacement are not independent any more and
\begin{align}
\overline{A}_{\vartheta_\infty}&\equiv{\rm Anti}\vartheta_\infty=-\skw( (\nabla v\,|\dd\sum_{\alpha=1,2}\bigl\langle n_0, \partial_{x_\alpha}v\bigr\rangle\, a^\alpha)[\nabla \Theta]^{-1})\in \mathfrak{so}(3),\notag\\
\vartheta_\infty &=  -\frac{1}{2}\,\tr(\mbox{skew}\,\big[ (\nabla y_0)^{T}(\nabla v)\big]\,{\rm C}^{-1})\,n_0-\sum_{\alpha=1,2}\bigl\langle n_0, \partial_{x_\alpha}v\bigr\rangle\, a^\alpha\in\mathbb{R}^3,
\end{align}
	where 
\begin{align}\label{defC2}
{\rm C}=\sqrt{\det {\rm I}_{y_0}}\begin{footnotesize}\left(\begin{array}{cc}
0 & 1 \\
-1 & 0 \\
\end{array}\right)\end{footnotesize}. 
\end{align}
Using these dependences, the linearisation of the strain tensors are
\begin{align}\label{eq12l2}
 \mathcal{E}_{ \infty }^{\rm{lin}}  = &\quad \  [\nabla\Theta \,]^{-T}
(\mathcal{G}_{\rm{Koiter}}^{\rm{lin}})^\flat [\nabla\Theta \,]^{-1},\notag\\
\mathrm{C}_{y_0} \mathcal{K}_{ \infty }^{\rm{lin}} = &\,- [\nabla\Theta \,]^{-T}(\mathcal{R}_{\rm{Koiter}}^{\rm{lin}} -\mathbf{1} \,\mathcal{G}_{\rm{Koiter}}^{\rm{lin}} \,{\rm L}_{y_0})^\flat [\nabla\Theta \,]^{-1}\\
\sym (\mathcal{E}_{ \infty }^{\rm{lin}} {\rm B}_{y_0}  + \mathrm{C}_{y_0} \mathcal{K}_{ \infty }^{\rm{lin}}) 
= & -\,[\nabla\Theta \,]^{-T} [\mathcal{R}_{\rm{Koiter}}^{\rm{lin}}-\mathbf{2}\,\sym (\mathcal{G}_{\rm{Koiter}}^{\rm{lin}} \,{\rm L}_{y_0}) ]^\flat
[\nabla\Theta \,]^{-1}\notag
,\\
 \mathcal{K}_\infty^{\rm{lin}} \,=& \quad \  (\nabla\vartheta_\infty \, |\, 0) \; [\nabla\Theta \,]^{-1}.\notag
\end{align}
In the above relations we have already used that the linearisation of $ \mathcal{G}_\infty$ and $ \mathcal{R}_\infty$ are given by
\begin{align}
\mathcal{G}_{\infty}^{\rm{lin}}=\mathcal{G}_{\rm{Koiter}}^{\rm{lin}},\qquad \qquad 
\mathcal{R}_{\infty}^{\rm{lin}}&=\mathcal{R}_{\rm{Koiter}}^{\rm{lin}}-{\bf 1}\,\mathcal{G}_{\rm{Koiter}}^{\rm{lin}} \,{\rm L}_{y_0}.
\end{align}

At this point, it is  important to note a significant distinction between the interpretation of the bending tensor in the constrained Cosserat shell model and the bending tensor employed in the Koiter model. As we will see in the following $\sym\, \mathcal{R}_{\infty}^{\rm{lin}}=\mathcal{R}_{\rm{Koiter}}^{\rm{lin}}-\mathbf{1}\sym (\mathcal{G}_{\rm{Koiter}}^{\rm{lin}} \,{\rm L}_{y_0})$ measures the bending and $\sym\,\mathcal{R}_{\infty}^{\rm{lin}}- \sym \,(\mathcal{G}_{\rm{Koiter}}^{\rm{lin}} \,{\rm L}_{y_0})=\mathcal{R}_{\rm{Koiter}}^{\rm{lin}}-\mathbf{2}\,\sym (\mathcal{G}_{\rm{Koiter}}^{\rm{lin}} \,{\rm L}_{y_0})$
 measures the change of curvature, while the  "bending tensor" employed in the Koiter model does not measure none of them. If $\mathcal{G}_{\rm{Koiter}}^{\rm{lin}}=0$ (infinitesimal pure flexure) or for a plate (flat shell), $\mathcal{R}_{\infty}^{\rm{lin}}=\mathcal{R}_{\rm{Koiter}}^{\rm{lin}}$ and no difference between bending and change of curvature occurs. 
\section{Concise description of the family of Cosserat-shell models}\setcounter{equation}{0}
\subsection{Variational problem for  nonlinear and linear Koiter models}
The  variational problem for the Koiter energy   is to find 
\begin{align}
\begin{cases}
\text{a deformation of the midsurface}\qquad \ \ \ \,\,
	m:\omega\subset\mathbb{R}^2\to\mathbb{R}^3, & \text{in the nonlinear Koiter model}\vspace{2mm}\\
\text{a midsurface displacement vector field} \ \ \ 
	v:\omega\subset\mathbb{R}^2\to\mathbb{R}^3,& \text{in the linear Koiter model}
\end{cases}
\end{align}  minimizing on the planar domain $\omega\subset \mathbb{R}^2$
\begin{equation}\label{Ap7matrixcon}
\begin{array}{l}
\dd\int_\omega \bigg\{h\,\bigg(
\mu\rVert   \mathcal{E} \rVert^2  +\dfrac{\,\lambda\,\mu}{\lambda+2\,\mu} \, [\mathrm{tr} (\mathcal{E})]^2\bigg) +\dd\frac{h^3}{12}\bigg(
\mu\rVert    \mathcal{F}\rVert^2 +\dfrac{\,\lambda\,\mu}{\lambda+2\,\mu} \, [\mathrm{tr} ( \mathcal{F} )]^2\bigg)\bigg\}\,{\rm det}\nabla\Theta \,\, {\rm d}a,
\end{array}
\end{equation}
where 
\begin{align}
\mathcal{E}&=\begin{cases} [\nabla\Theta]^{-T} (\mathcal{G}_{\rm Koiter})^\flat[\nabla\Theta]^{-1} , & \text{in the nonlinear Koiter model},\vspace{2mm}\\
[\nabla\Theta]^{-T}(\mathcal{G}_{\rm Koiter}^{\rm lin})^\flat[\nabla\Theta]^{-1},& \text{in the linear Koiter model},
\end{cases}
\end{align}
and
\begin{align}
\mathcal{F}&=\begin{cases} [\nabla\Theta]^{-T}(\mathcal{R}_{\rm Koiter})^\flat[\nabla\Theta]^{-1}, & \text{in the nonlinear Koiter model}\vspace{2mm}\\
[\nabla\Theta]^{-T}(\mathcal{R}_{\rm Koiter}^{\rm lin})^\flat[\nabla\Theta]^{-1},& \text{in the linear Koiter model}.
\end{cases}
\end{align}

The main feature of the classical Koiter model is that it is just the sum of the correctly identified membrane term and flexural terms (but only under inextensional deformation).\footnote{We are not delving here into a discussion whether the membrane energy needs an additional quasiconvexification step \cite{Raoult95b}.}

\subsection{Variational problem for geometrically nonlinear  and  linear\\ Cosserat-shell models}

The total internal energy of the models given in  \cite{GhibaNeffPartI,GhibaNeffPartIV},  is written with the help of the following quadratic/bilinear  forms in terms of some second order tensors $\mathcal{E}, \mathcal{K}\in \mathbb{R}^{3\times 3}$:
\begin{align}\label{quadraticforms}
W_{\mathrm{shell}}( X) & =   \mu\,\lVert  \mathrm{sym}\,X\rVert ^2 +  \mu_{\rm c}\lVert \mathrm{skew}\,X\rVert ^2 +\dfrac{\lambda\,\mu}{\lambda+2\,\mu}\,\big[ \mathrm{tr}   (X)\big]^2,\notag\\
\mathcal{W}_{\mathrm{shell}}(  X,  Y)& =   \mu\,\bigl\langle  \mathrm{sym}\,X,\,\mathrm{sym}\,   \,Y \bigr\rangle   +  \mu_{\rm c}\bigl\langle \mathrm{skew}\,X,\,\mathrm{skew}\,   \,Y \bigr\rangle   +\,\dfrac{\lambda\,\mu}{\lambda+2\,\mu}\,\mathrm{tr}   (X)\,\mathrm{tr}   (Y),  \vspace{2.5mm}\\
W_{\mathrm{mp}}(  X)&= \mu\,\lVert \mathrm{sym}\,X\rVert ^2+  \mu_{\rm c}\lVert \mathrm{skew}\,X\rVert ^2 +\,\dfrac{\lambda}{2}\,\big[  \tr(X)\,\big]^2=
\mathcal{W}_{\mathrm{shell}}(  X)+ \,\dfrac{\lambda^2}{2\,(\lambda+2\,\mu)}\,[\mathrm{tr} (X)]^2,\notag\vspace{2.5mm}\\
W_{\mathrm{curv}}(  X )&=\mu\, L_{\rm c}^2 \left( b_1\,\lVert  \dev\,\text{sym} \,X\rVert ^2+b_2\,\lVert \text{skew}\,X\rVert ^2+b_3\,
[\tr (X)]^2\right), \quad \forall\, X,Y\in \mathbb{R}^{3\times 3}.\notag
\end{align}
The parameters $\mu$ and $\lambda$ are the \textit{Lam\'e constants}
of classical isotropic elasticity, $\kappa=\frac{2\,\mu+3\,\lambda}{3}$ is the \textit{infinitesimal bulk modulus}, $b_1, b_2, b_3$ are \textit{non-dimensional constitutive curvature coefficients (weights)}, $\mu_{\rm c}\geq 0$ is called the \textit{{Cosserat couple modulus}} and ${L}_{\rm c}>0$ introduces an \textit{{internal length} } which is {characteristic} for the material, e.g., related to the grain size in a polycrystal. The
internal length ${L}_{\rm c}>0$ is responsible for \textit{size effects} in the sense that smaller samples are relatively stiffer than
larger samples. If not stated otherwise, we assume that $\mu>0$, $\kappa>0$, $\mu_{\rm c}>0$, $b_1>0$, $b_2>0$, $b_3> 0$. All the constitutive coefficients  are coming from the three-dimensional Cosserat formulation, without using any a posteriori fitting of some two-dimensional constitutive coefficients.

It is important to note that there is no counterpart of $W_{\mathrm{curv}}$ in classical shell theories since $W_{\mathrm{curv}}$ is coming exclusively from the 3D independent Cosserat curvature. 

The two-dimensional minimization problem  in the nonlinear and linear Cosserat-shell model is to find  
\begin{align}
\hspace*{-0.5cm}\begin{cases}\begin{cases}
\text{a deformation of the midsurface}\qquad \ \ \ \ \, \:\: 
m:\omega\subset\mathbb{R}^2\to\mathbb{R}^3 \ \ \text{and}\qquad\qquad  \text{in the nonlinear}\\
\text{an elastic microrotation}\qquad \qquad \quad \ \ \ \,\ 
\overline{Q}_{e,s}:\omega\subset\mathbb{R}^2\to{\rm SO}(3)\qquad \qquad\ \,\text{ Cosserat-shell model}\\\end{cases}  
\vspace{2mm}\\\begin{cases}
\text{a midsurface displacement vector field}  \ \quad \,\,
v:\omega\subset\mathbb{R}^2\to\mathbb{R}^3\ \ \text{and}\qquad \qquad \text{in the linear}\\
\text{an elastic microrotation vector}\qquad \quad  \ \ \ \ \ \,
\vartheta:\omega\subset\mathbb{R}^2\to\mathbb{R}^3 \qquad\qquad \quad  \ \ \ \, \text{Cosserat-shell model}\end{cases} 
\end{cases}
\end{align}  minimizing on $\omega\subset \mathbb{R}^2$
 the  functional
\begin{align}\label{e89}
I\!=\!\! \int_{\omega}   \!\!\Big[ & \Big(h+{\rm K}\,\dfrac{h^3}{12}\Big)\,
W_{\mathrm{shell}}\big(    \mathcal{E} \big)+   \Big(\dfrac{h^3}{12}\,-{\rm K}\,\dfrac{h^5}{80}\Big)\,
W_{\mathrm{shell}}  \big(   \mathcal{E} \, {\rm B}_{y_0} +   {\rm C}_{y_0} \mathcal{K}\big)\notag  \\&
-\dfrac{h^3}{3} \mathrm{ H}\,\mathcal{W}_{\mathrm{shell}}  \big(  \mathcal{E} ,
\mathcal{E}{\rm B}_{y_0}+{\rm C}_{y_0}\, \mathcal{K} \big)+
\dfrac{h^3}{6}\, \mathcal{W}_{\mathrm{shell}}  \big(  \mathcal{E} ,
( \mathcal{E}{\rm B}_{y_0}+{\rm C}_{y_0}\, \mathcal{K}){\rm B}_{y_0} \big)\vspace{2.5mm}\\&+ \,\dfrac{h^5}{80}\,\,
W_{\mathrm{mp}} \big((  \mathcal{E} \, {\rm B}_{y_0} +  {\rm C}_{y_0} \mathcal{K} )   {\rm B}_{y_0} \,\big),  \vspace{2.5mm}\notag\\
 &+\Big(h-{\rm K}\,\dfrac{h^3}{12}\Big)\,
W_{\mathrm{curv}}\big(  \mathcal{K} \big)    +  \Big(\dfrac{h^3}{12}\,-{\rm K}\,\dfrac{h^5}{80}\Big)\,
W_{\mathrm{curv}}\big(  \mathcal{K}  {\rm B}_{y_0} \,  \big)  + \,\dfrac{h^5}{80}\,\,
W_{\mathrm{curv}}\big(  \mathcal{K}   {\rm B}_{y_0}^2  \big)
\Big] \,{\rm det}\nabla\Theta        \, d a ,\notag
\end{align}
where 
\begin{align}
\mathcal{E}&=\begin{cases} \mathcal{E}_{m,s} , & \text{in the nonlinear Cosserat-shell model},\vspace{2mm}\\
\mathcal{E}_{m,s}^{\rm lin},& \text{in the linear  Cosserat-shell model},
\end{cases}
\end{align}
and
\begin{align}
\mathcal{K}&=\begin{cases} \mathcal{K}_{e,s}, & \text{in the nonlinear  Cosserat-shell model}\vspace{2mm}\\
\mathcal{K}_{e,s}^{\rm lin},& \text{in the linear  Cosserat-shell model}.
\end{cases}
\end{align}

\subsection{Variational problem for the modified nonlinear and modified linear constrained  Cosserat shell model}\label{vcm}
Instead of requiring directly the symmetry of the tensors $\mathcal{E}_{\rm m,s}$, $ \mathcal{E}_{\rm m,s}{\rm B}_{y_0}+{\rm C}_{y_0}\, \mathcal{K}_{\rm e,s}$ and $( \mathcal{E}_{\rm m,s}{\rm B}_{y_0}+{\rm C}_{y_0}\, \mathcal{K}_{\rm e,s}){\rm B}_{y_0}$ (the  coefficients of  $\{1,x_3,x_3^2\}$ of the decomposition of the symmetric 3D strain tensor) in the variational problem of the constrained Cosserat shell model, in the modified constrained Cosserat shell model the admissible space remains untouched, but instead only the symmetric parts of the above tensors are considered in the energy. 
	
Therefore,  
the  variational problem for the modified constrained Cosserat $O(h^5)$-shell model \cite{GhibaNeffPartIII}  is  to find 
\begin{align}
\hspace*{-0.25cm}\begin{cases}\begin{cases}
\text{a deformation of the midsurface} \ \ \ \,\,\ 
m:\omega\subset\mathbb{R}^2\to\mathbb{R}^3 \ \ \text{and}\qquad \quad \quad \text{in the modified constrained nonlinear}\\
\text{an elastic microrotation} \qquad \quad \ \ \ \,\ 
\overline{Q}_{\infty}:\omega\subset\mathbb{R}^2\to{\rm SO}(3)\quad \qquad \quad\ \,\,\, \text{Cosserat-shell model}\\\end{cases} 
\vspace{2mm}\\\begin{cases}
\text{a midsurface displacement vector field} \ \  \: \:
v:\omega\subset\mathbb{R}^2\to\mathbb{R}^3\ \ \text{and}\quad \quad \ \,  \text{in the modified constrained linear}\\
\text{an elastic microrotation vector} \qquad  \ \ \ \ \,
\vartheta_\infty:\omega\subset\mathbb{R}^2\to\mathbb{R}^3\quad \quad \qquad  \,\,\text{ Cosserat-shell model}\end{cases}
\end{cases}
\end{align}  minimizing on $\omega\subset\mathbb{R}^2$
the  functional
\begin{align}\label{e89}
I\!=\!\! \int_{\omega}   \!\!\Big[ & \Big(h+{\rm K}\,\dfrac{h^3}{12}\Big)\,
W_{\mathrm{shell}}^\infty\big(    \mathcal{E} \big)+   \Big(\dfrac{h^3}{12}\,-{\rm K}\,\dfrac{h^5}{80}\Big)\,
W_{\mathrm{shell}}^\infty  \big(   \mathcal{E} \, {\rm B}_{y_0} +   {\rm C}_{y_0} \mathcal{K}\big)\notag  \\&
-\dfrac{h^3}{3} \mathrm{ H}\,\mathcal{W}_{\mathrm{shell}}^\infty  \big(  \mathcal{E} ,
\mathcal{E}{\rm B}_{y_0}+{\rm C}_{y_0}\, \mathcal{K} \big)+
\dfrac{h^3}{6}\, \mathcal{W}_{\mathrm{shell}}^\infty  \big(  \mathcal{E} ,
( \mathcal{E}{\rm B}_{y_0}+{\rm C}_{y_0}\, \mathcal{K}){\rm B}_{y_0} \big)\vspace{2.5mm}\\&+ \,\dfrac{h^5}{80}\,\,
W_{\mathrm{mp}}^\infty \big((  \mathcal{E} \, {\rm B}_{y_0} +  {\rm C}_{y_0} \mathcal{K} )   {\rm B}_{y_0} \,\big),  \vspace{2.5mm}\notag\\
&+\Big(h-{\rm K}\,\dfrac{h^3}{12}\Big)\,
W_{\mathrm{curv}}\big(  \mathcal{K} \big)    +  \Big(\dfrac{h^3}{12}\,-{\rm K}\,\dfrac{h^5}{80}\Big)\,
W_{\mathrm{curv}}\big(  \mathcal{K}  {\rm B}_{y_0} \,  \big)  + \,\dfrac{h^5}{80}\,\,
W_{\mathrm{curv}}\big(  \mathcal{K}   {\rm B}_{y_0}^2  \big)
\Big] \,{\rm det}\nabla\Theta        \, d a ,\notag
\end{align}
where 
\begin{align}
\mathcal{E}&=\begin{cases} \mathcal{E}_\infty , & \text{in the nonlinear modified constrained Cosserat-shell model},\vspace{2mm}\\
\mathcal{E}_{\infty}^{\rm lin},& \text{in the linear modified constrained Cosserat-shell model},
\end{cases}
\end{align}
and
\begin{align}
\mathcal{K}&=\begin{cases} \mathcal{K}_{\infty}, & \text{in the nonlinear  modified constrained Cosserat-shell model},\vspace{2mm}\\
\mathcal{K}_{\infty}^{\rm lin},& \text{in the linear modified constrained  Cosserat-shell model},
\end{cases}
\end{align} 
with
\begin{align} 
W_{{\rm shell}}^{\infty}(  X)  &=   \mu\,\lVert\,   \sym X\rVert^2  +\,\dfrac{\lambda\,\mu}{\lambda+2\mu}\,\big[ \mathrm{tr}   \, (X)\big]^2,\qquad 
\mathcal{W}_{{\rm shell}}^{\infty}(  X ,   Y) =   \mu\,\bigl\langle  \sym X,   \sym Y\bigr\rangle+\,\dfrac{\lambda\,\mu}{\lambda+2\mu}\,\mathrm{tr}  (\sym X)\,\mathrm{tr}  (\sym Y), \vspace{2.5mm}\notag\\
W_{\mathrm{mp}}^{\infty}( X )&= \mu\,\lVert  \sym X\rVert^2+\,\dfrac{\lambda}{2}\,\big[ \mathrm{tr}\,   (\sym X)\big]^2 \qquad \quad\forall \ X,Y\in\mathbb{R}^{3\times 3}, \vspace{2.5mm}\\
W_{\mathrm{curv}}(  X )&=\mu\,L_c^2\left( b_1\,\lVert \dev\,\textrm{sym}\, X\rVert^2+b_2\,\lVert\text{skew} \,X\rVert^2+b_3\,
[\tr(X)]^2\right) \quad\qquad \forall\  X\in\mathbb{R}^{3\times 3}.\notag
\end{align}

In the dimensional reduction procedure, we do not identify a modelling reason for considering only the symmetric part of $\mathcal{K}$ in the reduced Cosserat-curvature energy\footnote{It is also not clear whether setting $b_2=0$ would still lead to a well-posed problem.} $W_{\rm curv}$. Therefore, since $\skw\, \mathcal{K}_\infty^{\rm lin}$ and $\skw \,\mathcal{K}_\infty$, respectively are present, 
	the tensors 
	\begin{align} \mathcal{R}_{\infty}^{\rm{lin}}&=\mathcal{R}_{\rm{Koiter}}^{\rm{lin}}-\mathcal{G}_{\rm{Koiter}}^{\rm{lin}} \,{\rm L}_{y_0} \quad \text{and}\quad 
\mathcal{R}_{\infty}=\mathcal{R}_{\rm{Koiter}}-\mathcal{G}_{\rm{Koiter}}^{} \,{\rm L}_{y_0}
\end{align}
 are the bending tensors candidates in our linear and nonlinear modified constrained models, respectively, instead of
\begin{align} \mathcal{R}_{\rm KSB}^{\rm{lin}}&:=\mathcal{R}_{\rm{Koiter}}^{\rm{lin}}-\mathbf{\bf sym}(\mathcal{G}_{\rm{Koiter}}^{\rm{lin}} \,{\rm L}_{y_0}  )
\end{align}
considered by Budiansky and Sanders \cite{budiansky1962best} and by Koiter \cite{koiter1973foundations}. It is easy to remark that
\begin{align} \mathcal{R}_{\rm KSB}^{\rm{lin}}&=\sym\,\mathcal{R}_{\infty}^{\rm{lin}},
\end{align}
but our bending tensor $\mathcal{R}_{\infty}^{\rm{lin}}$ does not coincide with the Koiter-Sander-Budiansky bending tensor $\mathcal{R}_{\rm KSB}^{\rm{lin}}$. Moreover, we also propose the bending tensor $\mathcal{R}_{\infty}$ in the nonlinear model, while $\mathcal{R}_{\rm KSB}^{\rm{lin}}$ is the linearisation of the second bending tensor ${\mathcal{R}_{\rm Acharya}}
\!=\sym\,\widetilde{\mathcal{R}}_{\rm Acharya}\in {\rm Sym}(3)$ introduced by Acharya \cite{acharya2000nonlinear}, where $\widetilde{\mathcal{R}}_{\rm Acharya}
=- {\sqrt{[\nabla\Theta ]^{-T}\,\widehat{\rm I}_m [\nabla\Theta ]^{-1}}}[\nabla\Theta \,]^{-T}\mathcal{R}_{\infty}^\flat[\nabla\Theta \,]^{-1}$. The linearisation of the first bending tensor $\widetilde{\mathcal{R}}_{\rm Acharya}$ considered by Acharya is our linear bending tensor $\mathcal{R}_{\infty}^{\rm{lin}}$.  Considering only its symmetric part was made according to  the Budiansky-Sanders-Koiter \cite{budiansky1962best,budiansky1963best,koiter1973foundations} demand to consider symmetrised measures ``to effect a reduction in the number of stress-couple resultant components that enter the theory", see \cite[Page 5521]{acharya2000nonlinear}. However, in a couple-stress theory the couple stress may be non-symmetric, at least in the constrained Cosserat 3D theory. Thus,  the remarks about the symmetry of the bending strain measure is  suitable only in a couple stress model with a symmetric couple stress. We continue this discussion and comparisons in the next section.

Contrary to what we have described in the above paragraph regarding the non-symmetry of our bending measure, the curvature measure considered by us, i.e.,
\begin{align}
\sym(\mathcal{E}_{\infty}^{\rm lin}{\rm B}_{y_0}+{\rm C}_{y_0}\mathcal{K}_\infty^{\rm lin})=\mathcal{R}_{\rm AL}^{\rm lin} &:=\mathcal{R}_{\rm{Koiter}}^{\rm lin}-\mathbf{2}\,\sym(\mathcal{G}_{\rm Koiter }^{\rm lin} \,{\rm L}_{y_0}  )
\end{align}
in the linear modified model and
\begin{align}
\sym(\mathcal{E}_{\infty}{\rm B}_{y_0}+{\rm C}_{y_0}\mathcal{K}_\infty)&:=\mathcal{R}_{\rm{Koiter}}-\mathbf{2}\,\sym(\mathcal{G}_{\rm{Koiter}} \,{\rm L}_{y_0}  )
\end{align}
in the nonlinear modified model, are both symmetric, through the construction of the model, and requested by the symmetry of the reconstructed 3D strain measure.

\section{The change of metric tensor}\setcounter{equation}{0}

In the linear elastic  models (which do not include Cosserat effects) the same measure was used for the change of metric tensor, i.e. the tensor $\mathcal{G}^{\rm lin}_{\rm Koiter}$ given by \eqref{equ11}. However,  in the nonlinear  models which do not include Cosserat effects there is a difference between our choice for the measure of the change of metric and other models, see Table \ref{tabelcm}.

There is a unanimous consensus regarding the choice of strain tensor for measuring metric changes in linearised  models. However, in nonlinear models, there are still diverging opinions. For example, when using the derivation approach\cite{Neff_plate04_cmt,GhibaNeffPartI}  or the expression of the Gamma-limit \cite{neff2007geometrically,saem2023geometrically} based on the Biot-type quadratic parental 3D energy, the strain tensor  $\sqrt{[\nabla\Theta ]^{-T}\,\widehat{\rm I}_m\,\id_2^{\flat }\,[\nabla\Theta ]^{-1}}-
\sqrt{[\nabla\Theta ]^{-T}\,\widehat{\rm I}_{y_0}\,\id_2^{\flat }\,[\nabla\Theta ]^{-1}}$ appears naturally in the formulation of the 2D variational problem.
On the other hand, models derived from the Saint-Venant-Kirchhoff energy yield a metric change represented as the difference between the first fundamental form of the current midsurface and that of the reference midsurface configuration. It is worth noting that sometimes this difference, ${\rm I}_m - {\rm I}_{y_0},$ is directly incorporated into the model construction without further elaboration. In essence, while the linearised  versions of these two strain measures for metric changes are equivalent, they differ in the nonlinear case.

Comparing to the general 6-parameter shell model \cite{Eremeyev06}, we have considered the same change of metric tensor in the Cosserat linear and in the nonlinear shell models. However, in our model there are  mixed energetic terms present  depending on the change of metric tensor which are not occuring in the general 6-parameter shell model.

\section{What does  bending mean? Scaling invariance of bending tensors}\label{sec:invariance}\setcounter{equation}{0}

\subsection{Idealized invariance requirements for a bending strain tensor}\label{subsec:Acharya}
What is the appropriate approach to modeling the physical concept of bending? A thorough grasp of bending measures must provide a precise definition of its complementary work counterpart when establishing equilibrium equations. Simultaneously, it is beneficial to develop a suitable formulation for traction boundary conditions when they are necessary. With this mindset, Acharya, as described in \cite[page 5519]{acharya2000nonlinear}, has presented a series of modeling criteria for a bending strain tensor in any first-order nonlinear shell theory:

\begin{description}[style=multiline,leftmargin=3em]
	\item[AR1] \textit{``Being a strain measure, it should be a tensor that vanishes in rigid deformations".}
	\item[AR2] \textit{``It should be based on a \textit{proper} tensorial comparison of the deformed and underformed curvature fields }[${\rm II}_{m}$ and ${\rm II}_{y_0}$]".
	\item[AR3] \textit{``A vanishing bending strain at a point should be associated with any deformation that leaves the orientation of the unit normal field locally unaltered around that point."}
\end{description}
\begin{figure}[h!]
	\hspace*{-2cm}
	\begin{center}
		\includegraphics[scale=1.6]{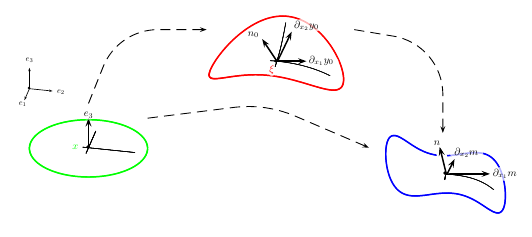}
				\put(-305,59){\footnotesize{${e}_2$}}
		 		\put(-330,85){\footnotesize{${e}_3$}}
				\put(-128,126){\footnotesize{$\varphi({x})\!=\!\varphi_\xi({\xi})$}} 
		\put(-320,165){\footnotesize{$\Theta  ,  Q_0={\rm polar}(\nabla \Theta)$}} 
		\put(-320,140){\footnotesize{$y_0$}} 
		\put(-250,102){\footnotesize{$m$}}
		\put(-65,125){\footnotesize{$\overline{ Q}_e, \ F_e $}}
		\put(-390,50){\footnotesize{$\omega$}}  
		\put(-15,30){\footnotesize{$\omega_c$}}  
		\put(-180,175){\footnotesize{$\omega_\xi=y_0(\omega)$}}  
		\caption{\footnotesize Kinematics of the 2D-constrained Cosserat shell model.  Here,  ${Q}_{ e }  $ is the elastic rotation field, ${Q}_{0}$ is the  initial rotation from the fictitious planar Cartesian reference $\omega$ configuration to the initial  configuration $\omega_\xi$.}
		\label{Fig2}       
	\end{center}
\end{figure}
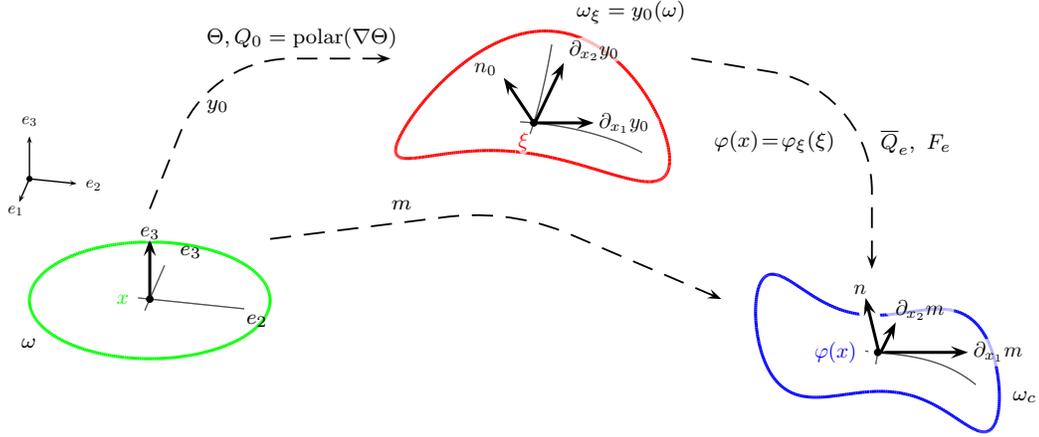
The first two criteria, \textbf{AR1} and \textbf{AR2}, are met by the nonlinear bending tensors discussed in this paper, namely, $\mathcal{R}$ and $\mathcal{R}_{\rm Koiter}$. These two requirements are intuitively sound from a physical perspective. However, the third requirement, \textbf{AR3}, implies that a non-zero bending tensor should only be associated with a change in the orientation of tangent planes. For example, in the case of a radial expansion of a cylinder, this should result in a zero bending strain measure, as it {\bf  does not  induce additional bending deformation of the shell (though it alters the curvature)}.
In \cite{GhibaNeffPartIII} we have shown that    $\mathcal{R}_{\rm{Koiter}}={\rm II}_{m}-{\rm II}_{y_0}$ satisfies \textbf{AR1} and \textbf{AR2}, since rigid deformations keep the second fundamental form invariant,  but  $\mathcal{R}_{\rm{Koiter}}$ does not satisfy \textbf{AR3}. 

Let us consider  $\omega_\xi=y_0(\omega)$  the deformation of a planar surface $\omega\subset \mathbb{R}^2$ through the mapping $y_0$, then \textbf{AR3} asserts that a deformation measure qualifies as a pure measure of bending if the bending tensor remains invariant under a $C^2$-mapping ${m}:\mathbb{R}^2\to \mathbb{R}^3$ that changes  the  surface  $\omega_\xi=y_0(\omega)\subset \mathbb{R}^3$, see Figure \ref{Fig2}, into another surface by having the relative (elastic) reconstructed ``gradient''
	\begin{align}\label{FRU}
	F_e:=(\nabla {m}(\xi)\,|\, {n}(\xi))(\nabla y_0\,|\,n_0)^{-1}={R}_e(\xi)\, {U}_e(\xi),
	\end{align}
	where ${U}_e(\xi)\in {\rm Sym}^+(3)$ is a linear mapping of the tangent plane to the initial surface at $\xi=y_0(x)$ into itself  and 
	\begin{align}
	{R}_e(\xi)=R_0\,R_{n_0}(\xi),
	\end{align}
	with $R_0\in {\rm SO}(3)$ a uniform rotation, i.e., independent of position, and $R_{n_0}(\xi)\in{\rm SO}(3)$ belongs to the group of rotations about the unit normal $n_0(\xi)$ to the surface at $\xi=y_0(x)$ (pure drill). In other words, \textbf{AR3} considers  mappings ${m}$ which produce a local stretching, described by ${U}_e$, a twist about $n_0$ (a drill about $n_0$), described by $R_{n_0}$, and an overall rotation given by $R_0$.

While condition \textbf{AR3} holds strong physical appeal, it may, in practice, prove to be too stringent for general application. Therefore, already in \cite{GhibaNeffPartIII}, we introduced and explored a less restrictive suitable invariance requirement. To facilitate this discussion, let us revisit the following definition from \cite{GhibaNeffPartIII}:
	\begin{definition}\label{defpure}
		Let $m$ induce  a deformation of the midsurface $y_0$. Denoting by $n$ and $n_0$ normal fields on the surface $m$ and $y_0$, respectively, we say that the midsurface deformation $m$ is obtained from a \textbf{pure elastic stretch} provided that ~$\U=F_e\coloneqq(\nabla m \,|\, n)\,(\nabla y_0\,|\,n_0)^{-1} = (\nabla m \,|\, n)\,[\nabla \Theta]^{-1}$ is already symmetric and positive-definite, i.e., belongs to $\operatorname{Sym}^+(3)$.
\end{definition}

In particular, the relation \eqref{FRU} is satisfied for a pure elastic stretch of the surface $\omega_\xi=m(\omega)$ as in Definition \ref{defpure} that in addition leave the unit normal field unaltered in a point\footnote{We have already shown in \cite{GhibaNeffPartIII} that for {pure elastic stretch} that in addition leave the unit normal field unaltered it holds $\U n_0=n_0$ and  ${\rm U}_e^{-1}n_0=n_0$.}, i.e., ${n}=n_0$ and $\U=(\nabla {m} \,|\, {n})\,(\nabla y_0\,|\,n_0)^{-1})$ is symmetric and positive definite. In other words, when 
	$
	F_e:=(\nabla {m}(\xi)\,|\, {n}(\xi))(\nabla y_0\,|\,n_0)^{-1}$ $={U}_e(\xi),
	$ a case already considered by Swabowicz \cite{szwabowicz2008pure}.\footnote{In fact, Swabowicz has shown more, namely \begin{itemize}
			\item 
			pure strain maps of surfaces preserve the third fundamental form of a surface.
			\item pure strain maps of non-umbilical surfaces preserve the lines of principal
			curvature.
			\item under a pure strain map the principal directions of stretch  and second fundamental forms of the original surface and of its image   all coincide.
			\item a map between two non-umbilical surfaces  is a pure strain map if and only if it preserves the third fundamental form and the lines of principal curvature.
	\end{itemize}} Another particular case in  \eqref{FRU} is a pure drill deformation, i.e., $
	F_e:=(\nabla {m}(\xi)\,|\, {n}(\xi))(\nabla y_0\,|\,n_0)^{-1}= R_e(\xi)
	$, considered by  Saem, Lewintan and Neff\footnote{ Regarding pure drill, Saem, Lewintan and Neff \cite[Proposition 5.2.]{mohammadi2021plane} have shown the following rigidity  result:
			Let $\omega\subset \mathbb{R}^2$ be a bounded Lipschitz domain. Assume that $m,y_0\in C^2(\overline{\omega},\mathbb{R}^3)$ are regular surfaces, $Q\in C^1(\overline{\omega},{\rm SO}(3))$ and 
			\begin{align*}
			\nonumber	\nabla m(x)&=Q(x)\nabla y_0(x), \qquad x\in \overline{\omega}\hspace{0.75cm}\Leftrightarrow\hspace{0.75cm}(\nabla m(x)\,|n(x))(\nabla y_0(x)\,|n_0(x))^{-1}=Q(x), \qquad x\in \overline{\omega}\,\\ Q(x)\,n_0(x)&=n_0(x)\,,\qquad \qquad \ \ x\in \overline{\omega}\,\hspace{4.77cm}Q(x)\,n_0(x)=n_0(x)\,,\ \ \ \ \,  x\in \overline{\omega}
			\\
			m|_{\gamma_{d}}&=y_0|_{\gamma_{d}}\hspace{7.9cm} m|_{\gamma_{d}}=y_0|_{\gamma_{d}}\,,
			\end{align*}
			where $n_0=\frac{\partial_1 y_0\times \partial_2 y_0}{\norm{\partial_1 y_0\times \partial_2 y_0}}$ denotes the normal field on $y_0(\omega)$ and $\gamma_d$ is a relatively open, non-empty subset of the boundary $\partial\omega$.
			Then $m\equiv y_0$.
	} in \cite{mohammadi2021plane}.

In  \cite[Eq.~(8) and (10)]{acharya2000nonlinear}, Acharya introduced two nonlinear bending strain tensors for a first-order nonlinear elastic shell theory, expressed in our notation as follows:
\begin{align}\label{nAch}
\textrm{the first proposal:} \ \  
&\widetilde{\mathcal{R}}_{\rm Acharya}
\! =\!\!- \left([\nabla\Theta \,]^{-T} {\rm II}_{m}^\flat  [\nabla\Theta \,]^{-1}\!\! -\!\!
\sqrt{[\nabla\Theta \,]^{-T}\; {\rm I}_{m}^\flat \; [\nabla\Theta \,]^{-1}}
[\nabla\Theta \,]^{-T} {\rm II}_{y_0}^\flat \, [\nabla\Theta \,]^{-1}\right) \!\!\not\in {\rm Sym}(3),\notag \\\textrm{the second proposal:}\ \  & {\mathcal{R}_{\rm Acharya}}
\!=\sym(\widetilde{\mathcal{R}}_{\rm Acharya})\in {\rm Sym}(3).
\end{align}

According to Acharya \cite{acharya2000nonlinear}, the nonlinear bending strain tensor $\mathcal{R}_{\rm Acharya}$ would satisfy all three requirements \textbf{AR1 - AR3} ``if locally pure stretch deformations are the only ones that leave the orientation of tangent planes unaltered locally under deformation."\footnote{However, this is not true as the case of pure drill deformation shows \cite{mohammadi2021plane}.}

Incidentally, the second tensor \eqref{nAch}$_2$ introduced by Acharya reduces after linearization as well to the Koiter-Sanders-Budiansky ``best'' bending measure \cite{budiansky1962best,budiansky1963best,koiter1973foundations}
\begin{equation}
{\mathcal{R}}^{\rm{lin}}_{\rm Acharya}\, =\mathcal{R}^{\rm{lin}}_{\rm KSB}\,  = \,\, \mathcal{R}_{\rm{Koiter}}^{\rm{lin}} -{\bf 1}\, \sym[\,\mathcal{G}_{\rm{Koiter}}^{\rm{lin}} \,{\rm L}_{y_0}]\in{\rm Sym}(2),
\end{equation}
while the first tensor introduced by Acharya reduces after linearisation to our   bending measure 
\begin{equation}
\widetilde{\mathcal{R}}^{\rm{lin}}_{\rm Acharya}=\mathcal{R}^{\rm{lin}}_{\infty}\,  = \,\, \mathcal{R}_{\rm{Koiter}}^{\rm{lin}} - [\,\mathcal{G}_{\rm{Koiter}}^{\rm{lin}} \,{\rm L}_{y_0}]\not\in{\rm Sym}(2).
\end{equation} It is then interesting to note that our tensor  $\mathcal{R}_{\infty}^\flat$ has the same properties as $\widetilde{\mathcal{R}}_{\rm Acharya}$, since 
Acharya's bending tensors are similar to but do not coincide with the bending tensor appearing  naturally in our  nonlinear constrained Cosserat-shell model, i.e., we can express
	\begin{align}\label{newrA}\widetilde{\mathcal{R}}_{\rm Acharya}
	=- \underbrace{\sqrt{[\nabla\Theta ]^{-T}\,\widehat{\rm I}_m [\nabla\Theta ]^{-1}}}_{\text{invertible}}[\nabla\Theta \,]^{-T}\mathcal{R}_{\infty}^\flat[\nabla\Theta \,]^{-1}.
	\end{align}
	Therefore, 
	\begin{align}
	\widetilde{\mathcal{R}}_{\rm Acharya}=0\qquad\qquad  \Longleftrightarrow \qquad \qquad  \mathcal{R}_{\infty}^\flat=0.
	\end{align}

 Acharya demonstrated that his nonlinear bending strain measure is zero in cases of pure stretch deformations that maintain the normal direction unaltered. In contrast, classical bending strain measures do not exhibit this behavior. However, it is  important to note again that $\mathcal{R}_{\rm Acharya}$ may not necessarily vanish for deformations characterized by a ``pure drill" rotation (a rotation about the normal leaving the tangent plane fixed) in their rotation tensor. 
 According to Acharya, his nonlinear bending measure  should only be seen as a mathematical ``better alternative"   for modelling the physical bending process since ``the set of deformations that leave the orientation of tangent planes unaltered locally can be divided into two classes\,-\,deformations that have a pure stretch deformation gradient locally, and those that have their local rotation tensor field consisting of either [in-plane] \textit{drill} rotations or the identity tensor."  Acharya has shown that   his nonlinear  bending strain measure vanishes in pure stretch deformations that leave the normal unaltered, while the other classical bending strain measures fail to do so, but $\mathcal{R}_{\rm Acharya}$  does not necessarily vanish for deformations whose rotation tensor is a ``drill" rotation.  Consequently, $\mathcal{R}_{\rm Acharya}$ does not consistently conform to \textbf{AR3}. 

Taking into account that pure in-plane drill does not leave the  $\mathcal{R}$ tensor invariant, but that pure in-plane drill is absent once Dirichlet boundary conditions are prescribed \cite{mohammadi2021plane} (see also  footnote 14), 
motivates that Acharya's essential invariance requirement may be modified  to \begin{description}[style=multiline,leftmargin=3em]
	\item[AR3$^*$] \textit{A vanishing bending strain at a point should be associated with any deformation obtained from a pure elastic stretch that leaves the orientation of the unit normal field locally unaltered around that point, i.e., $\U=F_e\coloneqq(\nabla m \,|\, \boldsymbol{n})\,(\nabla y_0\,|\,n_0)^{-1}  \stackrel{!}{=} (\nabla m \,|\, \boldsymbol{n}_0)\,(\nabla y_0\,|\,n_0)^{-1}$ is  symmetric and positive-definite.}
\end{description}

We have shown in  \cite{GhibaNeffPartIII} that both tensors $\mathcal{R}_{\rm Acharya}$ and $\mathcal{R}_{\infty}^\flat$ satisfy the three requirements \textbf{AR1}, \textbf{AR2} and \textbf{AR3}$^*$, while  $\mathcal{R}_{\rm Koiter}= {\rm II}_{m}-{\rm II}_{y_0}$ satisfies  \textbf{AR1}, \textbf{AR2}  but it does not satisfy \textbf{AR3}$^*$ either.

Let us notice that the minimal requirement for a bending tensor (for shells and plates) is that it should  be invariant under the simple scaling of the midsurface  $m\to \alpha\, m$, $\alpha>0$. Since the normal is preserved under such a  scaling of the midsurface   the first fundamental form, the second fundamental and the Weingarten maps are
\begin{align}
 {\rm I}_{\alpha\, m}:=&\alpha^2[{\nabla   m}]^T\,{\nabla   m}=\alpha^2\,{\rm I}_{ m}, \qquad {\rm II}_{\alpha\, m}:\,=\,-\alpha [{\nabla  m}]^T\,{\nabla  n}=\alpha\, {\rm II}_{ m},\\  {\rm L}_{\alpha\, m}:=&\frac{1}{\alpha}\, {\rm I}_{m}^{-1} {\rm II}_{m}=\frac{1}{\alpha}\, {\rm L}_{ m}, \qquad  \quad \, \  {\rm III}_{\alpha\, m}:=[{\nabla  n}]^T\,{\nabla  n}= {\rm III}_{ m}.\notag
\end{align}
Hence, the Koiter bending tensor
$
\mathcal{R}_{\rm Koiter}(\alpha\, m)=\alpha{\rm II}_{m}-{\rm II}_{y_0}\neq \mathcal{R}_{\rm Koiter}(m)
$
is {\bf not preserved under scaling}.
The situation is different when we look at our bending tensor, since
\begin{align}
\mathcal{R}_{\infty}^\flat(\alpha\, m)&=\,[\nabla\Theta \,]^{T}\Big(\sqrt{[\nabla\Theta ]\,\frac{1}{\alpha^2}\widehat{\rm I}_m^{-1}[\nabla\Theta ]^{T}}\,[\nabla\Theta ]^{-T} \alpha{\rm II}_m^\flat[\nabla\Theta ]^{-1} -\sqrt{[\nabla\Theta ]\,\widehat{\rm I}_{y_0}^{-1}[\nabla\Theta ]^{T}}[\nabla\Theta ]^{-T}{\rm II}_{y_0}^\flat [\nabla\Theta ]^{-1}\Big)\nabla\Theta\notag\\&=\mathcal{R}_{\infty}^\flat(m).
\end{align} Thus, {\bf our bending tensor $\mathcal{R}_{\infty}^\flat$ is invariant under this scaling.}
{The invariance under scaling is not satisfied by Acharya's bending tensors},  since
\begin{align}\label{newrAni}\widetilde{\mathcal{R}}_{\rm Acharya}(\alpha\, m)
&=- \sqrt{[\nabla\Theta ]^{-T}\,{\alpha^2}\widehat{\rm I}_m [\nabla\Theta ]^{-1}}[\nabla\Theta \,]^{-T}\mathcal{R}_{\infty}^\flat(\alpha\, m)[\nabla\Theta \,]^{-1}\notag\\&=-{\alpha} \sqrt{[\nabla\Theta ]^{-T}\,\widehat{\rm I}_m [\nabla\Theta ]^{-1}}[\nabla\Theta \,]^{-T}\mathcal{R}_{\infty}^\flat(
 m)[\nabla\Theta \,]^{-1}={\alpha}\,\widetilde{\mathcal{R}}_{\rm Acharya}(m)\neq \widetilde{\mathcal{R}}_{\rm Acharya}( m).
\end{align}
In the context of flat shell models, Virga \cite{virga2023pure} has proposed a measure of bending 
\begin{align}\widetilde{\mathcal{R}}_{\rm Virga}^{\rm plate}=(\nabla n)^T \nabla n.
\end{align}
The tensor $\widetilde{\mathcal{R}}_{\rm Virga}^{\rm plate}$ satisfies the condition \textbf{AR2}, since for a rigid deformation $y_0\to m =\widehat{Q}y_0$, $\widehat{Q}\in\operatorname{SO}(3)$ the fundamental forms coincide, 
$
{\rm I}_{m} = {\rm I}_{y_0} \quad \text{and} \quad {\rm II}_{m}={\rm II}_{y_0},
$
due to the identities $\nabla {m}=\widehat{Q} \nabla y_0, \ \nabla n_0=\nabla(\widehat{Q}\, n)=\widehat{Q}\,\nabla\, n_0$. From here, it is also clear that the definition of $\widetilde{\mathcal{R}}_{\rm Virga}^{\rm plate}$ satisfies  \textbf{AR1}, too.
We observe that the measure of bending considered by Virga is the	 {\it third fundamental form} (denoted by us ${\rm III}_{m}$) of the surface parametrized by {$m$}. Hence, since
$\nabla m\, {\rm L}_{m}\,=\,-\nabla n,
$
the measure of bending $\widetilde{\mathcal{R}}_{\rm Virga}^{\rm plate}$ reads
\begin{align}
\widetilde{\mathcal{R}}_{\rm Virga}^{\rm plate}&=(\nabla m\, {\rm L}_{m})^T (\nabla m\, {\rm L}_{m})={\rm L}_{m}^T (\nabla m)^T\, (\nabla m)\, {\rm L}_{m})={\rm L}_{m}^T {\rm I}_m\, {\rm L}_{m}=({\rm I}_{m}^{-1} {\rm II}_{m})^T{\rm I}_m\, {\rm L}_{m}({\rm I}_{m}^{-1} {\rm II}_{m})\notag\\&=
{\rm II}_{m}^T{\rm I}_{m}^{-1}{\rm I}_m\, ({\rm I}_{m}^{-1} {\rm II}_{m})={\rm II}_{m}^T{\rm I}_{m}^{-1}\, {\rm II}_{m}={\rm I}_{m}\, {\rm L}_{m}^2.
\end{align}

Let us strengthen further the additional invariance requirements on the bending tensor by postulating 
\begin{description}[style=multiline,leftmargin=5em]
	\item[AR3$^*_{\rm plate}$] \textit{For a planar reference geometry $({\rm II}_{y_0}\equiv\mathbf{0}_2)$ the bending tensor should be invariant under the scaling of the midsurface  $m\to \alpha\, m$, $\alpha>0$.}
\end{description}
It is   evident that a straightforward scaling transformation, such as $m \to \alpha\, m$ with $\alpha > 0$, represents an additional in-plane stretching without introducing any additional bending. Therefore, \textbf{AR3$^*_{\rm plate}$} should be considered as the appropriate requirement for a genuine expression of a bending tensor. Moreover, this new condition, \textbf{AR3$^*_{\rm plate}$}, allows to distinguish between Acharya's ad hoc bending tensors, denoted as $\widetilde{\mathcal{R}}_{\rm Acharya}$ and ${\mathcal{R}}_{\rm Acharya}$, and our derived bending tensor $\mathcal{R}_{\infty}^\flat$. In fact, in \cite{GhibaNeffPartIII}, it has been demonstrated that our  bending tensor $\mathcal{R}_{\infty}^\flat$ satisfies \textbf{AR3$^*_{\rm plate}$} for pure elastic stretch which do not change the normal, while $\widetilde{\mathcal{R}}_{\rm Acharya}$ and ${\mathcal{R}}_{\rm Acharya}$ lack this particular invariance property.
For plates (flat shells), $y_0={\rm id}$, Definition \ref{defpure} becomes \begin{definition}
	Let $m$ be a deformation of the flat reference domain. Denoting by $n$ and $e_3$ normal fields on the surface $m$ and on the referential configuration, respectively, we say that the midsurface deformation $m$ is obtained from a \textbf{pure elastic stretch} provided that ~$\U=F_e\coloneqq(\nabla m \,|\, n)$ is symmetric and positive-definite, i.e., ~belongs to $\operatorname{Sym}^+(3)$.
\end{definition}
Since the normal is preserved under scaling,  {\bf Virga's bending tensor ${\mathcal{R}}_{\rm Virga}^{\rm plate}$ is also preserved} under the scaling $m\mapsto \alpha\, m$.
 As proven in \cite{virga2023pure},   {\bf Virga's bending tensor $\widetilde{\mathcal{R}}_{\rm Virga}^{\rm plate}$ also satisfies \textbf{AR3}}.

For  flat shells ($y_0={\rm id}$) we have
\begin{align}
n=\frac{\partial_{x_1} m\times \partial_{x_2} m}{\sqrt{\det ((\nabla m)^T\nabla m)}}=&\,e_3+\left(\partial_{x_1} y_0\times \partial_{x_2} v+\partial_{x_1} v\times \partial_{x_2} y_0+\text{h.o.t}\right)
 -\tr((\, \sym ((\nabla y_0)^T\nabla v) )\,e_3.
\end{align}
Therefore the linearisation of $\widetilde{\mathcal{R}}_{\rm Virga}^{\rm plate}$ is
$
{\mathcal{R}}_{\rm Virga}^{\rm plate, lin}=\mathbf{O}_2\neq {\mathcal{R}}_{\rm Koiter}^{\rm plate, lin}.
$
Here, we have used that 	 for  flat shells ($y_0={\rm id}$) the Christoffel symbols $\Gamma^\gamma_{\alpha \beta}$ vanishes and \begin{align}
\label{equ12}
\mathcal{R}_{\rm{Koiter}}^{\rm{plate,lin}} \,\,:&=  \Big( \bigl\langle e_3 ,  \partial_{x_\alpha  x_\beta}\,v- \dd\sum_{\gamma=1,2}\Gamma^\gamma_{\alpha \beta}\,\partial_{x_\gamma}\,v\bigr\rangle \,\Big)_{\alpha\beta}=\Big( \bigl\langle e_3 ,  \partial_{x_\alpha  x_\beta}\,v\bigr\rangle \,\Big)_{\alpha\beta}\in {\rm Sym}(2),
\end{align}
which does not vanish for every $v$.

The bending tensor considered by Virga is consistent with the tensor 
\begin{align}
\mathcal{P}_{\rm Naghdi}=(\nabla d)^T(\nabla d)-{\rm III}_{y_0},
\end{align}
considered in the Naghdi-type models, see \cite[Section  6]{mardare2008derivation}, since upon constraining  $d\to n$  we have 
\begin{align}
\mathcal{P}_{\rm Naghdi}^\infty=(\nabla n)^T(\nabla n)-{\rm III}_{y_0}={\rm III}_{m}-{\rm III}_{y_0}.
\end{align}After linearisation, this tensor becomes
\begin{align}
\mathcal{P}_{\rm Naghdi}^{\rm lin}=\left((\nabla n_0)^T(\nabla \zeta)+(\nabla \zeta)^T(\nabla n_0)\right),
\end{align} 
where $m=y_0+u$ and $d=n_0+\zeta$.
\subsection{The bending measures  of the first-order linear shell theory}

While there appears to be a general consensus, particularly in linearised  models, regarding the choice of a measure for the change in the metric tensor, the same cannot be said for the measurement of bending, as evidenced in Table \ref{tabelbcb}. It is crucial to distinguish between the measures of bending and those of the change in curvature. The distinctions, as well as the tensors discussed in the existing literature, will be explored in the following two subsections.

It is   worth noting that the bending tensor utilized in our model is the same as that in the comprehensive 6-parameter theory. However, our Cosserat shell model incorporates additional energetic terms and couplings.

In the ensuing discussion, our attention is directed towards both the constrained and linearised  constrained models, where we compare the measures used for bending with those employed in classical models that do not incorporate Cosserat effects. It    is important to emphasize that even in the modified constrained Cosserat shell model, not only the symmetric part of the bending strain tensor $\mathcal{R}_\infty$ impacts the overall internal energy but the entire $\mathcal{R}_\infty$ (due to the additional independent Cosserat curvature term).

The second proposal of Acharya, the tensor $\widetilde{\mathcal{R}}_{\rm Acharya}$, reduces after linearization to the Koiter-Sanders-Budiansky ``best'' bending measure  \cite{Koiter60,sanders1959improved,budiansky1962best,budiansky1963best}
\begin{equation}
\widetilde{\mathcal{R}}^{\rm{lin}}_{\rm Acharya}\,=\mathcal{R}^{\rm{lin}}_{\rm KSB}\,  = \,\, \mathcal{R}_{\rm{Koiter}}^{\rm{lin}} - \mathbf{1}\,\sym[\,\mathcal{G}_{\rm{Koiter}}^{\rm{lin}} \,{\rm L}_{y_0}]\in{\rm Sym}(2) 
\end{equation}
of the first-order linear shell theory, while the first proposal of Acharya ${\mathcal{R}}_{\rm Acharya}$ and our bending tensor $\mathcal{R}_\infty$ reduce after linearization to 
\begin{equation}
\mathcal{R}^{\rm{lin}}_\infty\,= \, \mathcal{R}_{\rm{Koiter}}^{\rm{lin}} - \mathbf{1}\,\mathcal{G}_{\rm{Koiter}}^{\rm{lin}} \,{\rm L}_{y_0}\not\in{\rm Sym}(2) .
\end{equation}
The tensor $\mathcal{R}^{\rm{lin}}_\infty$ is not symmetric as long as no additional a priori constraint, e.g., $ \,\mathcal{G}_{\rm{Koiter}}^{\rm{lin}} \,{\rm L}_{y_0}\in {\rm Sym}(2)$, is imposed. However, it is clear that the symmetric part of $\mathcal{R}^{\rm{lin}}_\infty$ is present in our linearised  constrained Cosserat shell model and that 
\begin{equation}
\sym\,\mathcal{R}^{\rm{lin}}_\infty\,= \, \mathcal{R}_{\rm{Koiter}}^{\rm{lin}} - \sym[\,\mathcal{G}_{\rm{Koiter}}^{\rm{lin}} \,{\rm L}_{y_0}]=\mathcal{R}^{\rm{lin}}_{\rm KSB}\in{\rm Sym}(2). 
\end{equation}

There are several compelling reasons for favoring the use of the Koiter-Sanders-Budiansky bending measure $\mathcal{R}_{\rm{KSB}}=\sym\,\mathcal{R}^{\rm{lin}}_\infty$ over the simpler tensor $\mathcal{R}_{\rm{Koiter}}^{\rm{lin}}$. Our argument begins by noting that Naghdi and Green \cite{naghdi1963nonlinear} regarded the straightforward application of differences in the first and second fundamental forms between two states of shells, derived from the works of Sanders \cite{sanders1961nonlinear} and Leonard \cite{leonard1961nonlinear}, as merely based on a "heuristic argument." Simultaneously, Koiter \cite{Koiter60} independently arrived at nearly the same conclusion as Sanders and Leonard.

\begin{landscape}
	
	\begin{table}[h!]
		\centering 
		\vspace*{1.5cm}
		\resizebox{25cm}{!}{	\begin{tabular}[h!]{|l c|c|c|c|c|c|c|c|}
				\hline
				\begin{minipage}{2cm}\end{minipage}& \begin{minipage}{4cm}\footnotesize \vspace*{2mm}\textbf{our models}\\\end{minipage}& \begin{minipage}{4cm} \footnotesize\vspace*{2mm} \textbf{Koiter-type models}\\ \end{minipage}& \begin{minipage}{2cm} \footnotesize \vspace*{2mm} \textbf{Anicic}\textbf{-L\'{e}ger}\\ \end{minipage}&\begin{minipage}{3cm} \footnotesize\vspace*{2mm} \textbf{Budiansky}\textbf{-Sanders}\\ \end{minipage}& \begin{minipage}{1.2cm} \footnotesize\vspace*{2mm} \textbf{Acharya}\\\end{minipage}& \begin{minipage}{2cm} \footnotesize\vspace*{2mm} \textbf{6-parameter}\\ \end{minipage}&\begin{minipage}{2.5cm}\footnotesize\ \\\textbf{generalized} \\ \textbf{Naghdi-type}\end{minipage}\\
				\hline
				\begin{minipage}{2cm} \footnotesize \vspace*{2mm} nonlinear\\ unconstrained\\ Cosserat \\\end{minipage}&  $\mathcal{R}  = \, -(\overline{Q}_{e,s} \nabla y_0)^{T} \nabla (\overline{Q}_{e,s} n_0)- {\rm II}_{y_0}\not\in {\rm Sym}(2)$ & $\times$ & $\times$ & $\times$ &  $\times$&$\mathcal{R} $& \\
				\hline
				\begin{minipage}{2cm} \footnotesize \vspace*{2mm}linear\\ unconstrained\\ Cosserat \\\end{minipage}& $ \mathcal{R}^{\rm{lin}} =    (\nabla y_0)^{T} ( n_0\times\nabla \vartheta )$ & $\times$ & $\times$ & $\times$ &  $\times$ & $ \mathcal{R}^{\rm{lin}} $&\\
				\hline
				\begin{minipage}{2cm}\vspace*{2mm} \footnotesize nonlinear\\modified \\constrained\\ Cosserat \\\end{minipage}&\begin{minipage}{9.5cm}$\mathcal{R}_{\infty}^\flat=\,[\nabla\Theta \,]^{T}\Big(\sqrt{[\nabla\Theta ]\,\widehat{\rm I}_m^{-1}[\nabla\Theta ]^{T}}\,[\nabla\Theta ]^{-T} {\rm II}_m^\flat[\nabla\Theta ]^{-1}$\\\hspace*{3cm}$ -\sqrt{[\nabla\Theta ]\,\widehat{\rm I}_{y_0}^{-1}[\nabla\Theta ]^{T}}[\nabla\Theta ]^{-T}{\rm II}_{y_0}^\flat [\nabla\Theta ]^{-1}\Big)\nabla\Theta$\\\hspace*{1.2cm}$\not\in {\rm Sym}(3)$\end{minipage} &\begin{minipage}{6cm} $\mathcal{R}_{\rm Koiter}^\flat  =[\nabla\Theta ]^{T}[{\rm II}_m^{\flat }-
					{\rm II}_{y_0}^{\flat }]\,[\nabla\Theta ]^{-1}$\\\hspace*{3cm}$\in{\rm Sym}(3)$\end{minipage} & $\times$ & $\times$& $\widetilde{\mathcal{R}}_{\rm Acharya} $ &$\mathcal{R}_{\infty}^\flat$&\\
				\hline
				\begin{minipage}{2cm} \footnotesize\vspace*{2mm}linear\\
					modified
					\\ constrained \\ Cosserat \\\end{minipage}& \begin{minipage}{7cm} \vspace*{2mm}$\mathcal{R}^{\rm{lin}}_{\rm KSB}\,  = \,\, \mathcal{R}_{\rm{Koiter}}^{\rm{lin}} - [\,\mathcal{G}_{\rm{Koiter}}^{\rm{lin}} \,{\rm L}_{y_0}]$\\\hspace*{3cm}$\not\in{\rm Sym}(2) $\end{minipage}& \begin{minipage}{6.5cm} \vspace*{2mm}$(\mathcal{R}^{\rm lin}_{\rm Koiter})^\flat=[[\nabla\Theta ]^{T}[{\rm II}_m^{\flat }-
					{\rm II}_{y_0}^{\flat }]\,[\nabla\Theta ]^{-1}]^{\rm lin}$\\\hspace*{3cm}$\in{\rm Sym}(3) $
				\end{minipage} & \begin{minipage}{5.5cm} $\mathcal{R}^{\rm{lin}}_{\rm AL}\,  = \,\, \mathcal{R}_{\rm{Koiter}}^{\rm{lin}} - \mathbf{2}\,\sym[\,\mathcal{G}_{\rm{Koiter}}^{\rm{lin}} \,{\rm L}_{y_0}]$\\\hspace*{3cm}$\in{\rm Sym}(2) $\end{minipage}& \begin{minipage}{6cm} $\mathcal{R}^{\rm{lin}}_{\infty}\,  = \,\, \mathcal{R}_{\rm{Koiter}}^{\rm{lin}} - {\bf 1}\sym[\,\mathcal{G}_{\rm{Koiter}}^{\rm{lin}} \,{\rm L}_{y_0}]$\\\hspace*{3cm}$\in{\rm Sym}(2) $	\end{minipage}& $\mathcal{R}^{\rm{lin}}_{\rm KSB}$ &$\mathcal{R}^{\rm lin}_{\rm KSB}$ &\\
				\hline
		\end{tabular}}
		\caption{\footnotesize The bending measures in various models. }\label{tabelbcb}
	\end{table}
	
	\vspace*{1cm}
	\begin{table}[h!]
		\centering 
		\resizebox{25cm}{!}
		{	\begin{tabular}[h!]{|l c|c|c|c|c|c|c|}
				\hline
				\begin{minipage}{1.7cm}\end{minipage}& \begin{minipage}{4cm}\footnotesize \vspace*{2mm}\textbf{our models}\\\end{minipage}& \begin{minipage}{4cm} \footnotesize\vspace*{2mm} \textbf{Koiter-type models}\\ \end{minipage}& \begin{minipage}{2cm} \footnotesize \vspace*{2mm} \textbf{Anicic}\textbf{-L\'{e}ger}\\ \end{minipage}&\begin{minipage}{3cm} \footnotesize\vspace*{2mm} \textbf{Budiansky}\textbf{-Sanders}\\ \end{minipage}& \begin{minipage}{1.2cm} \footnotesize\vspace*{2mm} \textbf{Acharya}\\\end{minipage}& \begin{minipage}{2cm} \footnotesize\vspace*{2mm} \textbf{6-parameter}\\ \end{minipage}&\begin{minipage}{2.5cm}\footnotesize\ \\\textbf{generalized} \\ \textbf{Naghdi-type}\end{minipage}\\
				\hline
				\begin{minipage}{1.7cm} \footnotesize \vspace*{2mm} nonlinear \\unconstrained \\Cosserat \\\end{minipage}&  $\mathcal{G}=(\overline{Q}_{e,s} \nabla y_0)^{T} \nabla m- {\rm I}_{y_0}\not\in {\rm Sym}(2)$ & $\times$ & $\times$ & $\times$ &  $\times$ &$\mathcal{G}$&$\times$\\
				\hline
				\begin{minipage}{1.7cm} \footnotesize \vspace*{2mm}linear\\ unconstrained \\Cosserat \\\end{minipage}& $ \mathcal{G}^{\rm{lin}} =  (\nabla y_0)^{T} ( \nabla v - \overline{A}_\vartheta\nabla y_0 ) \not\in {\rm Sym}(2)$ & $\times$ & $\times$ & $\times$ &  $\times$ &$\mathcal{G}^{\rm{lin}}$&$\times$\\
				\hline
				\begin{minipage}{1.7cm}\footnotesize \vspace*{2mm} nonlinear\\
					modified\\ constrained \\Cosserat \\\end{minipage}& 	\begin{minipage}{8.5cm}$\mathcal{G}_{ \infty }^\flat =[\nabla\Theta ]^{T}\Big(\sqrt{[\nabla\Theta ]^{-T}\,{\rm I}_m^{\flat }\,[\nabla\Theta ]^{-1}}$\\ \hspace*{2cm}$-
					\sqrt{[\nabla\Theta ]^{-T}\,{\rm I}_{y_0}^{\flat }\,[\nabla\Theta ]^{-1}}\,\Big)[\nabla\Theta ]^{-1}\in{\rm Sym}(3)$\end{minipage} & $\mathcal{G}_{\rm Koiter}^\flat  =\frac{1}{2}[\nabla\Theta ]^{T}[{\rm I}_m^{\flat }-
				{\rm I}_{y_0}^{\flat }]\,[\nabla\Theta ]^{-1}\in{\rm Sym}(3)$ & $\times$ & $\times$& $\mathcal{G}_{\rm Koiter} $ &$\mathcal{G}_{ \infty }$&
				$\mathcal{G}_{\mathrm{Koiter}}$\\
				\hline
				\begin{minipage}{1.7cm} \footnotesize \vspace*{2mm}linear\\
					modified\\ constrained\\ Cosserat \\\end{minipage}& \begin{minipage}{8cm} \vspace*{2mm}$\mathcal{G}^{\rm lin}_{\rm Koiter}=[[\nabla\Theta ]^{T}\Big(\sqrt{[\nabla\Theta ]^{-T}\,{\rm I}_m^{\flat }\,[\nabla\Theta ]^{-1}}$\\\hspace*{3cm}$-
					\sqrt{[\nabla\Theta ]^{-T}\,{\rm I}_{y_0}^{\flat }\,[\nabla\Theta ]^{-1}}\,\Big)[\nabla\Theta ]^{-1}]^{\rm lin}$
					\\
					\hspace*{0.3cm}	$\qquad =\sym(\mathcal{G}^{\rm{lin}} )$\\\end{minipage}& \begin{minipage}{6cm} \vspace*{2mm}$\mathcal{G}^{\rm lin}_{\rm Koiter}=\frac{1}{2}[[\nabla\Theta ]^{T}[{\rm I}_m^{\flat }-
					{\rm I}_{y_0}^{\flat }]\,[\nabla\Theta ]^{-1}]^{\rm lin}$
					\\
					\hspace*{0.3cm}	$\qquad =\sym(\mathcal{G}^{\rm{lin}} )$\end{minipage} & $\mathcal{G}^{\rm lin}_{\rm Koiter}$ & $\mathcal{G}^{\rm lin}_{\rm Koiter}$& $\mathcal{G}^{\rm lin}_{\rm Koiter}$ &$\mathcal{G}^{\rm lin}_{\rm Koiter}$&
				$\mathcal{G}_{\mathrm{Koiter}}^{\rm lin}$\\
				\hline
		\end{tabular}}
		\caption{\footnotesize  The change of metric measures in various models.}\label{tabelcm}
	\end{table}
\end{landscape}

The rationale behind why Koiter, Sanders, Leonard, and other subsequent authors initially favored the simpler strain measures, $\mathcal{R}_{\rm{Koiter}}$ and $\mathcal{G}_{\rm{Koiter}}$, in the finite strain regime is evident \cite{leonard1961nonlinear,sanders1959improved,koiter1973foundations}. This preference is rooted in the fact that, within this regime, knowledge of the first and second fundamental forms, subject to the Gau\ss \ and Codazzi integrability conditions, suffices to determine the deformed middle surface of the shell, barring any infinitesimal rigid body motion. Additionally, it is  worth noting that Naghdi himself considered alternative expressions for the bending strain measure, which differ by the difference between the second fundamental forms, in the nonlinear modeling of isotropic elastic shells. These alternative expressions coincide with $\mathcal{R}_{\rm{Koiter}}^{\rm{lin}}$ in the linearised  model, as shown in \cite{Naghdi72} and \cite[Eq. (7.22)]{vsilhavycurvature}.

It is noteworthy that Sanders \cite{sanders1959improved, sanders1961nonlinear, sanders1963nonlinear} was among the first to employ the same bending tensor found in our linear constrained Cosserat shell model, namely, our tensor $\mathcal{R}_\infty^{\rm{lin}}=\mathcal{R}_{\rm{Koiter}}^{\rm{lin}} - \mathcal{G}_{\rm{Koiter}}^{\rm{lin}} \,{\rm L}_{y_0}$. He was the first to anticipate the significance of using the strain measure $\mathcal{R}_{\rm{Koiter}}^{\rm{lin}} - \mathcal{G}_{\rm{Koiter}}^{\rm{lin}} \,{\rm L}_{y_0}$ instead of $\mathcal{R}_{\rm{Koiter}}^{\rm{lin}}$, as observed in \cite[Eq. 48 and Eqs. 23, 24]{sanders1961nonlinear}. This choice was made because $\mathcal{R}^{\rm{lin}}$ and $\mathcal{G}_{\rm{Koiter}}^{\rm{lin}}$ are present as a work conjugate pair of  some stress tensors, a result of the derivation of the equations for the displacement, which are used to write the equilibrium equations for the deformed middle surface, as described in \cite{green1954theoretical} and \cite{Ericksen58}. In the  context considered in \cite{sanders1959improved, sanders1961nonlinear, sanders1963nonlinear} $\mathcal{R}_{\rm{Koiter}}^{\rm{lin}} - \mathcal{G}_{\rm{Koiter}}^{\rm{lin}} \,{\rm L}_{y_0}$ is a work conjugate pair to a symmetric tensor, which means that it can be replaced by $\mathcal{R}^{\rm{lin}}_{\rm KSB} = \mathcal{R}_{\rm{Koiter}}^{\rm{lin}} - {\rm \bf sym}[\mathcal{G}_{\rm{Koiter}}^{\rm{lin}} \,{\rm L}_{y_0}]$, considering that $\mathcal{R}_{\rm{Koiter}}^{\rm{lin}}$ is already symmetric.
To accommodate this unexpected strain measure, Sanders made certain choices regarding the membrane stress tensor and the bending moment tensor, considering modified versions $\mathcal{R}^{\rm{lin}}_{\rm KSB}$ of these tensors that were assumed to be symmetric from the outset. These new membrane stress and bending moment tensors were designed to be the work conjugate pair for the difference between the first fundamental forms and the difference between the second fundamental forms between the two states of the shells, respectively (both symmetric). Sanders acknowledged that the introduction of these modified stresses might appear somewhat unusual, but he argued that these quantities could be adopted as the stress components in the theory instead of those typically used in the three-dimensional formulation of the deformation of the middle surface \cite{green1954theoretical}. Sanders noted that this indicated that the equilibrium equations containing the usual membrane stress tensor and the usual bending moment tensor were slightly more general than necessary for a theory constrained by Kirchhoff's normality hypotheses. Additionally, Sanders mentioned that there are an equal number of stress quantities as strain quantities, and for both a principle of minimum potential energy and a principle of minimum complementary energy to be applicable in the theory, the constitutive relations needed to be invertible. This was only feasible if there were an equal number of stress quantities and strain quantities, therefore the introduction of the symmetrical variants.

Building on insights from \cite{sanders1959improved, sanders1961nonlinear} and \cite{Koiter60}, later researchers, including Budiansky and Sanders \cite{budiansky1962best, budiansky1963best}, and Koiter and Simmonds \cite{koiter1973foundations}, reevaluated the use of the strain measures $\mathcal{R}_{\rm{Koiter}}^{\rm{lin}}$ and $\mathcal{G}_{\rm{Koiter}}^{\rm{lin}}$. They recognized that the choice of the membrane stress tensor and the new bending moment tensor was not unique, but had to be selected as a work conjugate term of a ``good" measure for bending.

Budiansky and Sanders, in their work, emphasized that ``Koiter himself \cite{Koiter60} prefers the expression\break [$\mathcal{R}_{\rm{Koiter}}^{\rm{lin}} - {\bf 1}\, \sym[\mathcal{G}_{\rm{Koiter}}^{\rm{lin}} \,{\rm L}_{y_0}]$]." They also pointed out that Koiter demonstrated that errors in Love's uncoupled strain energy expression (consistent with uncoupled stress-strain relations) were essentially the same, regardless of which alternative, such as a linear combination of $\mathcal{R}_{\rm{Koiter}}^{\rm{lin}}$ and $\mathcal{G}_{\rm{Koiter}}^{\rm{lin}}$ from a given list, was used, as detailed in \cite[Eqs. 18-20]{budiansky1962best}. In their concluding remarks, Budiansky and Sanders, based on the features presented, referred to their theory as the ``best" linear first-order theory of elastic shells. This nomenclature was subsequently adopted by Koiter and Simmonds \cite[page 152]{koiter1973foundations} when they expressed their equations in terms of the ``modified tensor of changes of curvature," which essentially equates to $\sym\,\mathcal{R}^{\rm{lin}}_\infty = \mathcal{R}^{\rm{lin}}_{\rm KSB} = \mathcal{R}_{\rm{Koiter}}^{\rm{lin}} - \sym[\mathcal{G}_{\rm{Koiter}}^{\rm{lin}} \,{\rm L}_{y_0}]\in{\rm Sym}(2)$.

The term known as the ``modified tensor of changes of curvature" $\sym\,\mathcal{R}^{\rm{lin}}= \mathcal{R}_{\rm{Koiter}}^{\rm{lin}} - \sym[\mathcal{G}_{\rm{Koiter}}^{\rm{lin}} \,{\rm L}_{y_0}]\in{\rm Sym}(2)$, employed by Koiter \cite{Koiter60}, Koiter and Simmonds \cite{koiter1973foundations} (while Sanders \cite{sanders1959improved, sanders1961nonlinear} and Budiansky and Sanders \cite{budiansky1962best, budiansky1963best} named it bending), possesses certain exceptional properties not shared by the classical bending strain tensor $\mathcal{R}_{\rm{Koiter}}^{\rm{lin}}$. However, it is worth clarifying that the term ``modified tensor of changes of curvature" is not appropriate for this particular concept of a bending strain tensor.

From a purely kinematical perspective, whether we use $\mathcal{R}_{\rm{Koiter}}^{\rm{lin}}$ and $\mathcal{G}_{\rm{Koiter}}^{\rm{lin}}$ or any other linear combinations of these two tensors may not be of paramount importance. This is because physically reasonable constitutive variables can always be reformulated in a manner suitable for different strain measures. Nevertheless, it is crucial to comprehend the relationship between the physical concept of bending and the mathematical measures of bending. As of now, a clear articulation of this aspect remains elusive. Referring again to Koiter and Simmonds \cite[page 173]{koiter1973foundations}, ``In the approximations, it is essential to bear in mind the physical interpretation of intermediate results at every stage in the analysis, and to apply appropriate corrections to the initial assumptions where this is required by the physics of the problem. It is indeed quite dangerous to derive a physical theory by a systematic and rigorous mathematical development of initial (approximate) assumptions unless due account is taken of the physical consequences at every step in the analysis. To physicists and engineers, these remarks will look like the forcing of an open door, but experience with quite a few papers on shell theory published in the last five years [1973] shows the need for a repetition of such cautionary remarks."\footnote{The situation has not changed much 50 years later.}

In conclusion, the bending measure utilized in our linearised  constrained Cosserat-shell model aligns perfectly with the Koiter-Sanders-Budiansky bending measure $\mathcal{R}^{\rm{lin}}_{\rm KSB}$, which is often considered the "best" possible choice. Meanwhile, $\mathcal{R}$ and $\mathcal{R}_\infty$ represent nonlinear generalizations of the Koiter-Sanders-Budiansky bending measure in the unconstrained and constrained Cosserat-shell models, respectively. The Koiter-Sanders-Budiansky bending measure is a component in the expression of our bending-curvature energy density through $\sym\,\mathcal{R}$, but the bending-curvature energy density also depends on $\skw\,\mathcal{R}$.
		
		In light of our discussion on the bending strain measure, all the above arguments substantiate that we have used appropriate and meaningful terminology for the bending measures that we employed in the Cosserat family.

\section{What does the change of curvature tensor describe?}\setcounter{equation}{0}
Let us remark that in the final variational problem of the modified linear constrained Cosserat shell model the energy terms, excepting the bending energy derived from the 3D curvature energy (called bending-curvature energy in \cite{GhibaNeffPartI}), are not written in terms of the bending strain measure  $\mathcal{R}^{\rm{lin}}_{ \infty }=\mathcal{R}_{\rm{Koiter}}^{\rm{lin}} - \,[\mathcal{G}_{\rm{Koiter}}^{\rm{lin}} \,{\rm L}_{y_0}]$, but rather in terms of the tensor
$\mathcal{R}^{\rm{lin}}_{ \infty } - \sym\,[\mathcal{G}^{\rm{lin}}_{ \infty }\,{\rm L}_{y_0}]=\mathcal{R}_{\rm{Koiter}}^{\rm{lin}} - 2\,\sym[\mathcal{G}_{\rm{Koiter}}^{\rm{lin}} \,{\rm L}_{y_0}]$. It is clear that we may always rewrite our internal energy as a quadratic form in terms of $\mathcal{R}^{\rm{lin}}_{ \infty }
$ and $\mathcal{G}_{\rm{Koiter}}^{\rm{lin}}$ or even in terms of $\mathcal{R}_{\rm{Koiter}}^{\rm{lin}}
$ and $\mathcal{G}_{\rm{Koiter}}^{\rm{lin}}$, without any difficulty, since, e.g.,
\begin{align}
\lVert\mathcal{R}^{\rm{lin}}_{ \infty }\rVert^2&=\lVert\mathcal{R}_{\rm{Koiter}}^{\rm{lin}} - \,[\mathcal{G}_{\rm{Koiter}}^{\rm{lin}} \,{\rm L}_{y_0}]\rVert^2\notag\\&=\lVert\mathcal{R}_{\rm{Koiter}}^{\rm{lin}}\rVert^2 -2\,\bigl\langle\mathcal{R}_{\rm{Koiter}}^{\rm{lin}},\mathcal{G}_{\rm{Koiter}}^{\rm{lin}} \,{\rm L}_{y_0} \bigr\rangle + \lVert\mathcal{G}_{\rm{Koiter}}^{\rm{lin}} \,{\rm L}_{y_0}\rVert^2.
\end{align} 
However, we believe that such a rewriting of the energy  is not necessary, since all the involved strain tensors have  clear meanings:
\begin{itemize}
	\item $\mathcal{G}_{\rm{Koiter}}^{\rm{lin}}$ measures the change of metric;
	\item $\mathcal{R}^{\rm{lin}}_{ \infty }=\mathcal{R}_{\rm{Koiter}}^{\rm{lin}} - {\bf 1}\,[\mathcal{G}_{\rm{Koiter}}^{\rm{lin}} \,{\rm L}_{y_0}]$ measures the bending;
	\item $\mathcal{R}_{\rm{Koiter}}^{\rm{lin}} - {\bf 2}\,\sym[\mathcal{G}_{\rm{Koiter}}^{\rm{lin}} \,{\rm L}_{y_0}]$ measures the change of curvature (this aspect is discussed in the rest of this subsection).
\end{itemize}

Acharya has shown that $ \mathcal{R}_{\rm{Koiter}}^{\rm{lin}}$ does not vanish in (infinitesimal) pure stretch deformation of a quadrant of a cylindrical surface, while the Koiter-Sanders-Budiansky bending measure $\mathcal{R}^{\rm{lin}}_{ \infty }$  does. Nothing is said by Acharya about the relation between  the Koiter-Sanders-Budiansky bending measure and the variations of the Gauss curvature  ${\rm K}$ and of the mean curvature ${\rm H}$.

Thus,  in the spirit of the definition by Acharya \cite{acharya2000nonlinear}, we may impose {\bf reasonable requirements for a change of curvature tensor}, i.e 
\begin{enumerate}
	\item[{\bf C1.}] it should be a tensor that vanishes in rigid deformations,
	\item[{\bf C2.}] it should be based on  ${\rm II}_{m}$ and ${\rm II}_{y_0}$, and its norm should be invariant when $m$ and $y_0$ interchange the roles (the inverse mapping produces the same energy),
	\item[{\bf C3.}] a vanishing  {\bf curvature tensor} should lead to zero variations of the {\bf Gauss curvature} ${\rm K}$ and of the {\bf mean curvature} ${\rm H}$ of the midsurface.
\end{enumerate}

\begin{landscape}
	\vspace*{0.5cm}
	\begin{table}[h!]
		\centering 
		\resizebox{25cm}{!}{	\begin{tabular}[h!]{|l c|c|c|c|c|c|c|c|}
				\hline
				\begin{minipage}{1.7cm} \footnotesize\end{minipage}& \begin{minipage}{3cm}\footnotesize \vspace*{2mm}\textbf{our models}\\\end{minipage}& \begin{minipage}{3.5cm} \footnotesize\vspace*{2mm} \textbf{Koiter-type models}\\ \end{minipage}& \begin{minipage}{2cm} \footnotesize \vspace*{2mm} \textbf{Anicic}\textbf{-L\'{e}ger}\\ \end{minipage}&\begin{minipage}{3cm} \footnotesize\vspace*{2mm} \textbf{Budiansky}\textbf{-Sanders}\\ \end{minipage}& \begin{minipage}{1.2cm} \footnotesize\vspace*{2mm} \textbf{Acharya}\\\end{minipage}& \begin{minipage}{2cm} \footnotesize\vspace*{2mm} \textbf{6-parameter}\\ \end{minipage}& \begin{minipage}{2.5cm}\footnotesize\ \\\textbf{generalized} \\ \textbf{Naghdi-type}\end{minipage}\\
				\hline
				\begin{minipage}{1.7cm} \footnotesize  \vspace*{2mm} nonlinear \\unconstrained\\ Cosserat \\\end{minipage}&  \begin{minipage}{6cm}\vspace*{2mm}$\mathcal{T}:=\, (\overline{Q}_{e,s}  n_0)^{T} \nabla m$\\\hspace*{0.5cm}$ = \, \left(\bigl\langle\overline{Q}_{e,s}  n_0, \partial_{x_1} m\bigr\rangle,\bigl\langle\overline{Q}_{e,s}  n_0, \partial_{x_2} m\bigr\rangle\right)$ \end{minipage}& $\times$ & $\times$ & $\times$ &  $\times$&$\mathcal{T} $& $\mathcal{T}_{\rm Naghdi}=\Big(\langle d, \partial_{x_1}m\rangle,\langle d, \partial_{x_1}m\rangle\Big)$\\
				\hline
				\begin{minipage}{1.7cm} \footnotesize \vspace*{2mm}linear\\ unconstrained\\ Cosserat \\\end{minipage}& \begin{minipage}{4.7cm}\vspace*{2mm}$\mathcal{T}^{\rm{lin}} =   n_0^{T} ( \nabla v - \vartheta\times\nabla y_0 )$ \end{minipage} & $\times$ & $\times$ & $\times$ &  $\times$ & $\mathcal{T}^{\rm{lin}} $&\begin{minipage}{5.2cm}$\mathcal{T}_{\rm Naghdi}^{\rm lin}=\Big(\langle n_0, \partial_{x_1}v\rangle,\langle n_0, \partial_{x_1}v\rangle\Big)$\\\hspace*{1.6cm}$+\Big(\langle \zeta, \partial_{x_1}y_0\rangle,\langle \zeta, \partial_{x_1}y_0\rangle\Big)$\end{minipage}\\
				\hline
				\begin{minipage}{1.7cm} \footnotesize\vspace*{2mm} nonlinear\\
					modified\\ constrained \\Cosserat \\\end{minipage}&\begin{minipage}{5cm}\vspace*{2mm}$0$ \end{minipage} & $\times$ & $\times$ & $\times$ &  $\times$ & $0 $&0\\
				\hline
				\begin{minipage}{1.7cm} \footnotesize \vspace*{2mm}linear\\ modified\\constrained\\ Cosserat \\\end{minipage}& \begin{minipage}{5cm}\vspace*{2mm}$0$ \end{minipage} & $\times$ & $\times$ & $\times$ &  $\times$ & $0  $&0\\
				\hline
		\end{tabular}}
		\caption{\footnotesize The transverse shear deformation  measures. }\label{tabeltrans}
	\end{table}
	\vspace*{0.5cm}
	\begin{table}[h!]
		\centering 
		\resizebox{25cm}{!}{	\begin{tabular}[h!]{|l c|c|c|c|c|c|c|c|}
				\hline
				\begin{minipage}{3cm}\end{minipage}& \begin{minipage}{2cm}\footnotesize \vspace*{2mm}\textbf{our models}\\\end{minipage}& \begin{minipage}{4cm} \footnotesize\vspace*{2mm} \textbf{Koiter-type models}\\ \end{minipage}& \begin{minipage}{2cm} \footnotesize \vspace*{2mm} \textbf{Anicic}\textbf{-L\'{e}ger}\\ \end{minipage}&\begin{minipage}{3.4cm} \footnotesize\vspace*{2mm} \textbf{Budiansky}\textbf{-Sanders}\\ \end{minipage}& \begin{minipage}{1.2cm} \footnotesize\vspace*{2mm} \textbf{Acharya}\\\end{minipage}& \begin{minipage}{2cm} \footnotesize\vspace*{2mm} \textbf{6-parameter}\\ \end{minipage}& \begin{minipage}{2.5cm}\footnotesize\ \\\textbf{generalized} \\ \textbf{Naghdi-type}\end{minipage}\\
				\hline
				\begin{minipage}{2cm} \footnotesize \vspace*{2mm} nonlinear\\ unconstrained\\ Cosserat \\\end{minipage}&  \begin{minipage}{7cm}\vspace*{2mm}$\mathcal{R}-\mathcal{G} \,{\rm L}_{y_0}$,\\ \hspace*{0.01mm} $ \mathcal{G}=(\overline{Q}_{e,s} \nabla y_0)^{T} \nabla m- {\rm I}_{y_0}\not\in {\rm Sym}(2)$,\\ \hspace*{0.5mm}$\mathcal{R}  = \, -(\overline{Q}_{e,s} \nabla y_0)^{T} \nabla (\overline{Q}_{e,s} n_0)- {\rm II}_{y_0}\not\in {\rm Sym}(2)$ \end{minipage}& $\times$ & $\times$ & $\times$ &  $\times$&$\times $ &\\
				\hline
				\begin{minipage}{2cm} \footnotesize \vspace*{2mm}linear \\unconstrained\\ Cosserat \\\end{minipage}& \begin{minipage}{7cm}\vspace*{2mm}$\mathcal{R}^{\rm{lin}}-\mathcal{G}^{\rm{lin}} \,{\rm L}_{y_0}$,\\ \hspace*{0.7mm}$ \mathcal{G}^{\rm{lin}} =  (\nabla y_0)^{T} ( \nabla v - \overline{A}_\vartheta\nabla y_0 ) \not\in {\rm Sym}(2)$,\\ $ \mathcal{R}^{\rm{lin}} =    (\nabla y_0)^{T} ( n_0\times\nabla \vartheta )$ \end{minipage} & $\times$ & $\times$ & $\times$ &  $\times$ & $\times $& \\
				\hline
				\begin{minipage}{2cm} \footnotesize\vspace*{2mm} nonlinear\\
					modified\\ constrained \\Cosserat \\\end{minipage}&\begin{minipage}{9cm}\vspace*{2mm}$\mathcal{R}_{\infty}^\flat-\sym(\mathcal{G}_{\infty}^\flat \,{\rm L}_{y_0})$,\\ $\mathcal{G}_{ \infty }^\flat =[\nabla\Theta ]^{T}\Big(\sqrt{[\nabla\Theta ]^{-T}\,{\rm I}_m^{\flat }\,[\nabla\Theta ]^{-1}}$\\\hspace*{2cm}$-
					\sqrt{[\nabla\Theta ]^{-T}\,{\rm I}_{y_0}^{\flat }\,[\nabla\Theta ]^{-1}}\,\Big)[\nabla\Theta ]^{-1}\in{\rm Sym}(3)$,\\ $\mathcal{R}_{\infty}^\flat=\,[\nabla\Theta \,]^{T}\Big(\sqrt{[\nabla\Theta ]\,\widehat{\rm I}_m^{-1}[\nabla\Theta ]^{T}}\,[\nabla\Theta ]^{-T} {\rm II}_m^\flat[\nabla\Theta ]^{-1} $\\\hspace*{2cm}$-\sqrt{[\nabla\Theta ]\,\widehat{\rm I}_{y_0}^{-1}[\nabla\Theta ]^{T}}[\nabla\Theta ]^{-T}{\rm II}_{y_0}^\flat [\nabla\Theta ]^{-1}\Big)\nabla\Theta\\\\\hspace*{2.5cm}\not\in{\rm Sym}(2)$\\ \end{minipage} & \begin{minipage}{6cm} $\mathcal{R}_{\rm Koiter}^\flat  =[\nabla\Theta ]^{T}[{\rm II}_m^{\flat }-
					{\rm II}_{y_0}^{\flat }]\,[\nabla\Theta ]^{-1}$\\\\\hspace*{2.3cm}$\in{\rm Sym}(3)$ \end{minipage}& $\times$ & $\times$& $\times $ &$\mathcal{R}_{\infty}^\flat$&\begin{minipage}{5cm}\footnotesize $\mathcal{R}_{\mathrm{Naghdi}}=-\sym\big[\nabla m)^T\nabla d - {\rm II}_{y_0}\big]$
				\end{minipage}\\
				\hline
				\begin{minipage}{2cm} \footnotesize \vspace*{2mm}linear\\
					modified \\constrained\\ Cosserat \\\end{minipage}& \begin{minipage}{8cm}\vspace*{2mm}$ \,\, \mathcal{R}_{\rm{Koiter}}^{\rm{lin}} -{\bf 2}\, \sym[\,\mathcal{G}_{\rm{Koiter}}^{\rm{lin}} \,{\rm L}_{y_0}]\\\\\hspace*{1.1cm}\in{\rm Sym}(2)$ \end{minipage}& \begin{minipage}{6cm} \vspace*{2mm}$\mathcal{R}^{\rm lin}_{\rm Koiter}=[[\nabla\Theta ]^{T}[{\rm II}_m^{\flat }-
					{\rm II}_{y_0}^{\flat }]\,[\nabla\Theta ]^{-1}]^{\rm lin}$\\\\\hspace*{2.5cm}$\in {\rm Sym}(3)$
				\end{minipage} & \begin{minipage}{6cm} $\mathcal{R}^{\rm{lin}}_{\rm AL}\,  = \,\, \mathcal{R}_{\rm{Koiter}}^{\rm{lin}} - {\bf 2}\,\sym[\,\mathcal{G}_{\rm{Koiter}}^{\rm{lin}} \,{\rm L}_{y_0}]$\\\\\hspace*{2.5cm}$\in{\rm Sym}(2) $\end{minipage}& \begin{minipage}{6cm} $\mathcal{R}^{\rm{lin}}_{\rm KSB}\,  = \,\, \mathcal{R}_{\rm{Koiter}}^{\rm{lin}} - {\bf 1}\sym[\,\mathcal{G}_{\rm{Koiter}}^{\rm{lin}} \,{\rm L}_{y_0}]$\\\\\hspace*{2.5cm}$\in{\rm Sym}(2) $\end{minipage}& $\times$ &\begin{minipage}{2cm}$\mathcal{R}_{\rm{Koiter}}^{\rm{lin}}$\\\\$\in{\rm Sym}(2)$\end{minipage}&\\
				\hline
		\end{tabular}}
		\caption{\footnotesize The change of curvature in various models. }\label{tabelcurv}
	\end{table}
	
\end{landscape}

Anicic and L{\'{e}}ger \cite{anicic1999formulation,anicic2002mesure,anicic2003shell},  have provided a linear change of curvature tensor, in the sense of the above definition. They  also derived a linear Kirchhoff-Love shell model which is in close connection to  our linear constraint Cosserat-shell model. They  proved that considering a  family of deformations $\{y_0+\eta\, v\,|\, \eta\in \mathbb{R}, \ v\in C^{2}(\omega) \ \ \text{such that} \ \ y_0+\eta \,v\ \ \text{defines a regular surface}\}$  of the middle surface  the following change of incremental curvature tensor
\begin{align}
\mathcal{R}_{\rm{AL}}^{\rm{lin}}(v)=\mathcal{R}_{\rm{Koiter}}^{\rm{lin}} - {\bf 2}\,\sym[\mathcal{G}_{\rm{Koiter}}^{\rm{lin}} \,{\rm L}_{y_0}]\in{\rm Sym}(2),
\end{align}
with $\mathcal{G}_{\rm{Koiter}}^{\rm{lin}} \,\,:= \,\,\frac12\;\big[  (\nabla y_0)^{T}(\nabla v) +  (\nabla v)^T(\nabla y_0)\big]
= \sym\big[ (\nabla y_0)^{T}(\nabla v)\big]\in {\rm Sym}(2)
\;$ and $
\mathcal{R}_{\rm{Koiter}}^{\rm{lin}} \,\,:= 	 \Big( \bigl\langle n_0 ,  \partial_{x_\alpha  x_\beta}\,v- \dd\sum_{\gamma=1,2}\Gamma^\gamma_{\alpha \beta}\,\partial_{x_\gamma}\,v\bigr\rangle a^\alpha\,\Big)_{\alpha\beta}\in {\rm Sym}(2),
$
has the alternative expression
\begin{align}\label{Ral}
\mathcal{R}_{\rm{AL}}^{\rm{lin}}(v)=\dfrac{1}{2}\left({\rm I}_{y_0}\dfrac{d\,{\rm L}_{y_0+\eta\, v}}{d\, \eta}\Big|_{\eta=0}+\dfrac{d\,{\rm L}_{y_0+\eta\, v}^T}{d\, \eta}\Big|_{\eta=0}{\rm I}_{y_0}\right)=\sym\left({\rm I}_{y_0}\dfrac{d\,{\rm L}_{y_0+\eta\, v}}{d\, \eta}\Big|_{\eta=0}\right) \in{\rm Sym}(2).
\end{align}

The proof is based on the formulae established by Blouza and Le Dret \cite{blouza1994lemme}, see also  \cite{anicic2002mesure},
\begin{align}\label{fromdret}
\mathcal{G}_{\rm{Koiter}}^{\rm{lin}} =\frac{1}{2}\dfrac{d\,{\rm I}_{y_0+\eta\, v}}{d\, \eta}\Big|_{\eta=0} \quad \text{and}\quad\mathcal{R}_{\rm{Koiter}}^{\rm{lin}} =\dfrac{d\,{\rm II}_{y_0+\eta\, v}}{d\, \eta}\Big|_{\eta=0}.
\end{align} Indeed, an interesting fact is that the local variations of the Weingarten map along the family of surfaces ${y_0+\eta\, v}$, with respect to $\eta$ is
\begin{align}
\dfrac{d\,{\rm L}_{y_0+\eta\, v}}{d\, \eta}\Big|_{\eta=0}={\rm I}_{y_0}^{-1}(\mathcal{R}_{\rm{Koiter}}^{\rm{lin}} - {\bf 2}\,[\mathcal{G}_{\rm{Koiter}}^{\rm{lin}} \,{\rm L}_{y_0}])={\rm I}_{y_0}^{-1}\,\mathcal{R}^{\rm{lin}}_{ \infty }.
\end{align}
  Indeed,  from ${\rm I}_{y_0+\eta\, v}^{-1} {\rm I}_{y_0+\eta\, v}=\id_2$ follows
\begin{align}
\dfrac{d\,{\rm I}_{y_0+\eta\, v}^{-1}}{d\, \eta}\, {\rm I}_{y_0+\eta\, v}=-{\rm I}_{y_0+\eta\, v}^{-1}\, \dfrac{d\,{\rm I}_{y_0+\eta\, v}}{d\, \eta}.
\end{align}
Thus, we have
\begin{align}\label{derivI-1}
\dfrac{d\,{\rm I}_{y_0+\eta\, v}^{-1}}{d\, \eta}=-{\rm I}_{y_0+\eta\, v}^{-1}\, \dfrac{d\,{\rm I}_{y_0+\eta\, v}}{d\, \eta}\, {\rm I}_{y_0+\eta\, v}^{-1},
\end{align}
and \begin{align}
\dfrac{d\,{\rm L}_{y_0+\eta\, v}}{d\, \eta}&=\dfrac{d}{d\eta}\,\left[{\rm I}_{y_0+\eta\, v}^{-1}{\rm II}_{y_0+\eta\, v}\right]=
\dfrac{d}{d\eta}\,{\rm I}_{y_0+\eta\, v}^{-1}{\rm II}_{y_0+\eta\, v}+
{\rm I}_{y_0+\eta\, v}^{-1}\dfrac{d}{d\eta}\,{\rm II}_{y_0+\eta\, v}\notag\\
&=-{\rm I}_{y_0+\eta\, v}^{-1}\, \dfrac{d\,{\rm I}_{y_0+\eta\, v}}{d\, \eta}\, {\rm I}_{y_0+\eta\, v}^{-1}{\rm II}_{y_0+\eta\, v}+
{\rm I}_{y_0+\eta\, v}^{-1}\dfrac{d}{d\eta}\,{\rm II}_{y_0+\eta\, v}
\\
&=-{\rm I}_{y_0+\eta\, v}^{-1}\, \dfrac{d\,{\rm I}_{y_0+\eta\, v}}{d\, \eta}\, {\rm L}_{y_0+\eta\, v}^{-1}+
{\rm I}_{y_0+\eta\, v}^{-1}\dfrac{d}{d\eta}\,{\rm II}_{y_0+\eta\, v}.\notag
\end{align}
Thus
\begin{align}
\dfrac{d\,{\rm L}_{y_0+\eta\, v}}{d\, \eta}\Big|_{\eta=0}
&=
{\rm I}_{y_0}^{-1}\,\mathcal{R}_{\rm{Koiter}}^{\rm{lin}}-{\rm I}_{y_0}^{-1}\, 2\,\mathcal{G}_{\rm{Koiter}}^{\rm{lin}}\, {\rm L}_{y_0}^{-1}
\end{align}
and
\begin{align}
{\rm I}_{y_0}\,\dfrac{d\,{\rm L}_{y_0+\eta\, v}}{d\, \eta}\Big|_{\eta=0}
&=
\mathcal{R}_{\rm{Koiter}}^{\rm{lin}}- 2\,\mathcal{G}_{\rm{Koiter}}^{\rm{lin}}\, {\rm L}_{y_0}^{-1},
\end{align}
which leads to
\begin{align}
\sym\left({\rm I}_{y_0}\,\dfrac{d\,{\rm L}_{y_0+\eta\, v}}{d\, \eta}\Big|_{\eta=0}\right)
&=
\mathcal{R}_{\rm{Koiter}}^{\rm{lin}}- 2\,\mathcal{G}_{\rm{Koiter}}^{\rm{lin}}\, {\rm L}_{y_0}^{-1}=\mathcal{R}_{\rm{AL}}^{\rm{lin}},
\end{align}
and \eqref{Ral} is proven.

Next, we show that
\begin{align}\label{ALrez}
\mathcal{R}_{\rm{AL}}^{\rm{lin}}(v)=0\qquad \Rightarrow\qquad \dfrac{d\,{\rm H}}{d\, \eta}(y_0+\eta\, v)\Big|_{\eta=0}=0\quad \text{and}\quad \dfrac{d\,{\rm K}}{d\, \eta}(y_0+\eta\, v)\Big|_{\eta=0}=0.
\end{align}

The components  of the first and second fundamental form of the surface are defined by $a_{\alpha\beta}(y_0+\eta\, v)=\bigl\langle a_\alpha(y_0+\eta\, v),a_\beta(y_0+\eta\, v)\bigr\rangle$ and $b_{\alpha\beta}(y_0+\eta\, v)=\bigl\langle\partial_\alpha a_\beta(y_0+\eta\, v),a_3(y_0+\eta\, v)\bigr\rangle$, respectively. The mixed components of the second fundamental form read $b^\beta_\alpha(y_0+\eta\, v)=b_{\alpha \rho}(y_0+\eta\, v)a^{\rho\beta}(y_0+\eta\, v)\,$, where $\big(a^{\alpha\beta}(y_0+\eta\, v)\big)$ is the matrix inverse of $\big(a_{\alpha\beta}(y_0+\eta\, v)\big)$. Thus, the mean curvature ${\rm H}$ and Gaussian curvature ${\rm K}$, read in terms of the coefficients of the fundamental forms,
\begin{align}
2\,{\rm H}(y_0+\eta\, v) = \tr\big(b^\beta_\alpha(y_0+\eta\, v)\big) = b^1_1(y_0+\eta\, v) + b^2_2(y_0+\eta\, v),
\end{align}
 \text{and}\begin{align} {\rm K} = \det \big(b^\beta_\alpha(y_0+\eta\, v)\big) = b^1_1(y_0+\eta\, v)\,b^2_2(y_0+\eta\, v)-b^1_2(y_0+\eta\, v)\,b^2_1(y_0+\eta\, v).
\end{align}

The condition  from the left hand side of \eqref{Ral}, i.e.,  $\mathcal{R}_{\rm{AL}}^{\rm{lin}}(v)=0$ \eqref{Ral}, see \cite[Eq. 9]{anicic1999formulation}, reads
\begin{align}
\frac{d b^\rho_\alpha(y_0+\eta\, v)}{d  \eta}\Big|_{\eta=0}\,a_{\rho\beta}(y_0) + \frac{d  b^\rho_\beta(y_0+\eta\, v)}{d  \eta}\Big|_{\eta=0}\,a_{\rho\alpha}(y_0)=0,
\end{align}
which explicits into
\begin{align}
\alpha=\beta=1\ &: &  0&=\frac{d  b^1_1(y_0+\eta\, v)}{d  \eta}\Big|_{\eta=0}\,a_{11}(y_0) + \frac{d  b^2_1}{d  \eta}\Big|_{\eta=0}\,a_{21}(y_0)\notag\\
\alpha=\beta=2\ &: &  0&=\frac{d  b^1_2(y_0+\eta\, v)}{d  \eta}\Big|_{\eta=0}\,a_{12}(y_0) + \frac{d  b^2_2(y_0+\eta\, v)}{d  \eta}\Big|_{\eta=0}\,a_{22}\notag
\\
\alpha=1, 2; \beta=2, 1\ &: &  0&=\frac{d  b^1_1(y_0+\eta\, v)}{d  \eta}\Big|_{\eta=0}\,a_{12}(y_0) + \frac{d  b^2_1(y_0+\eta\, v)}{d  \eta}\Big|_{\eta=0}\,a_{22}(y_0)\\&\  &\  &\qquad +\frac{d  b^1_2(y_0+\eta\, v)}{d  \eta}\Big|_{\eta=0}\,a_{11}(y_0) + \frac{d  b^2_2(y_0+\eta\, v)}{d  \eta}\Big|_{\eta=0}\,a_{21}(y_0).\notag
\end{align}
Using the symmetry $a_{\alpha \beta}= a_{\beta\alpha}$ we can rewrite the last system of equations using matrices in the following way:
\begin{align}
\left(\begin{array}{cc} a_{11}(y_0) & a_{12}(y_0) \\ a_{12}(y_0) & a_{22}(y_0) \end{array}\right)
\left[\frac{d }{d \eta} \left(\begin{array}{cc} b^1_1(y_0+\eta\, v) & b^1_2(y_0+\eta\, v) \\ b^2_1(y_0+\eta\, v) & b^2_2(y_0+\eta\, v) \end{array}\right) \right]\Big|_{\eta=0}
= \left(\begin{array}{cc}  0 & k \\ -k & 0 \end{array}\right),
\end{align}
where we have set $k=\frac{d  b^1_2(y_0+\eta\, v)}{d  \eta}\Big|_{\eta=0}\,a_{11}(y_0) + \frac{d  b^2_2(y_0+\eta\, v)}{d  \eta}\Big|_{\eta=0}\,a_{21}(y_0)$. Multiplying both sides with the inverse matrix $\big(a^{\alpha\beta}(y_0)\big)$ we arrive at
\begin{align}
\left[\frac{d }{d \eta} \left(\begin{array}{cc} b^1_1(y_0+\eta\, v) & b^1_2(y_0+\eta\, v) \\ b^2_1(y_0+\eta\, v) & b^2_2(y_0+\eta\, v) \end{array}\right) \right]\Big|_{\eta=0}
= \left(\begin{array}{cc}  -k\,a^{12}(y_0) & k\,a^{11}(y_0) \\ -k\,a^{22}(y_0) & k\,a^{12}(y_0) \end{array}\right).
\end{align}
Thus
\begin{align}
\frac{d  {\rm H}(y_0+\eta\, v)}{d \eta}(0)= \left. \frac{d }{d \eta}(b^1_1(y_0+\eta\, v)+b^2_2(y_0+\eta\, v))\right|_{\eta=0} = -k\,a^{12}(y_0)+k\,a^{12}(y_0) =0
\end{align}
and
\begin{align}
\frac{d  {\rm K}(y_0+\eta\, v)}{d \eta}\Big|_{\eta=0}&= \Big[\frac{d  b^1_1(y_0+\eta\, v)}{d \eta}\,b^2_2(y_0)+b^1_1(y_0)\,\frac{d  b^2_2(y_0+\eta\, v)}{d \eta}\notag\\&\qquad -\frac{d  b^1_2(y_0+\eta\, v)}{d \eta}\,b^2_1(y_0)-b^1_2(y_0)\,\frac{d  b^2_1(y_0+\eta\, v)}{d \eta } \Big]\Big|_{\eta=0}\\& = k\, [-a^{12}(y_0)\,b^2_2(y_0)+a^{12}(y_0)\,b^1_1(y_0)-a^{11}\,b^2_1(y_0)+a^{22}(y_0)\,b^1_2(y_0)] = 0,\notag
\end{align}
where the last step follows from the expression of the mixed components together with the symmetry $b_{\alpha \beta}= b_{\beta\alpha}$.

It is an interesting issue to understand a nonlinear version of the foregoing results.  The minimal requirement for a nonlinear change of curvature tensor is that its linearisation should   characterise the local variations of the mean curvature and the {Gau\ss}  curvature. This is automatically satisfied if the linearisation coincides with  $\mathcal{R}_{\rm{AL}}^{\rm{lin}}=\mathcal{R}_{\rm{Koiter}}^{\rm{lin}} - {\bf 2}\,\sym[\mathcal{G}_{\rm{Koiter}}^{\rm{lin}} \,{\rm L}_{y_0}]$.

We observe the surprising fact that in our constrained Cosserat shell model the membrane-change of curvature energy is actually written in terms of the Anicic-L\'eger's change of curvature tensor $\mathcal{R}_{\rm{AL}}^{\rm{lin}}$ and not in terms of the Koiter-Sanders-Budiansky bending measure $\mathcal{R}_{\infty}^{\rm{lin}}$, although everything can be rearranged to be expressed in the Koiter strain measure $\mathcal{R}_{\rm{AL}}^{\rm{lin}}$ and $\mathcal{G}_{\rm{Koiter}}^{\rm{lin}}$. Another coincidence is that quadratic coupling energies in terms of $\mathcal{R}_{\rm{AL}}^{\rm{lin}}$ and $\mathcal{G}_{\rm{Koiter}}^{\rm{lin}}$ are present both in Anicic-L\'eger's linear shell model and in our linear constrained Cosserat shell model. Anicic and L\'eger have also obtained explicit forms of the constitutive coefficients of the shell model in terms of the initial curvatures and the three-dimensional constitutive coefficients.

Just for  completeness of the discussion, we can do similar  calculations for the third fundamental form ${\rm III}_m={\rm II}_m\,{\rm I}_m^{-1}\, {\rm II}_m$ and we obtain
\begin{align}
\dfrac{d\,{\rm III}_{y_0+\eta\, v}}{d\, \eta}&=\dfrac{d}{d\eta}\,\left[{\rm II}_{y_0+\eta\, v}\,{\rm I}_{y_0+\eta\, v}^{-1}{\rm II}_{y_0+\eta\, v}\right]\\&=\dfrac{d}{d\eta}{\rm II}_{y_0+\eta\, v}\,{\rm I}_{y_0+\eta\, v}^{-1}{\rm II}_{y_0+\eta\, v} +{\rm II}_{y_0+\eta\, v}\,\dfrac{d}{d\eta}{\rm I}_{y_0+\eta\, v}^{-1}{\rm II}_{y_0+\eta\, v} +{\rm II}_{y_0+\eta\, v}\,{\rm I}_{y_0+\eta\, v}^{-1}\dfrac{d}{d\eta}{\rm II}_{y_0+\eta\, v}\notag
\end{align}
which, after using \eqref{fromdret}, and \eqref{derivI-1} leads to
\begin{align}\label{deriv3ff}
\dfrac{d\,{\rm III}_{y_0+\eta\, v}}{d\, \eta}\Big|_{\eta=0}&=\mathcal{R}_{\rm{Koiter}}^{\rm{lin}}\,{\rm I}_{y_0}^{-1}{\rm II}_{y_0} -{\rm II}_{y_0}\,{\rm I}_{y_0}^{-1}\, \dfrac{d\,{\rm I}_{y_0+\eta\, v}}{d\, \eta}\Big|_{\eta=0}\, {\rm I}_{y_0}^{-1}{\rm II}_{y_0} +{\rm II}_{y_0}\,{\rm I}_{y_0}^{-1}\mathcal{R}_{\rm{Koiter}}^{\rm{lin}}\notag\\
&=\mathcal{R}_{\rm{Koiter}}^{\rm{lin}}\,{\rm L}_{y_0} -{\rm L}_{y_0}^T 2\,\mathcal{G}_{\rm{Koiter}}^{\rm{lin}}\, {\rm L}_{y_0} +{\rm L}_{y_0}^T\mathcal{R}_{\rm{Koiter}}^{\rm{lin}}=2\,\sym(\mathcal{R}_{\rm{Koiter}}^{\rm{lin}}\,{\rm L}_{y_0} -{\rm L}_{y_0}^T \mathcal{G}_{\rm{Koiter}}^{\rm{lin}}\, {\rm L}_{y_0} )\\
&=2\,\sym({\rm L}_{y_0}^T\mathcal{R}_{\rm{Koiter}}^{\rm{lin}}\, -{\rm L}_{y_0}^T \mathcal{G}_{\rm{Koiter}}^{\rm{lin}}\, {\rm L}_{y_0} )=2\,\sym({\rm L}_{y_0}^T[\mathcal{R}_{\rm{Koiter}}^{\rm{lin}}\, - \mathcal{G}_{\rm{Koiter}}^{\rm{lin}}\, {\rm L}_{y_0}] )
.\notag
\end{align}
When we have constructed the constrained Cosserat shell model we have seen that we must impose the symmetry of $\mathcal{E}_{\infty}$, \ $\mathcal{E}_{\infty} \, {\rm B}_{y_0} +  {\rm C}_{y_0} \mathcal{K}_{\infty}$ and 
$(\mathcal{E}_{\infty} \, {\rm B}_{y_0} +  {\rm C}_{y_0} \mathcal{K}_{\infty}) {\rm B}_{y_0}$. In the modified constrained Cosserat shell model we have avoided this issue by considering only their symmetric parts in the variational problem. However, in the original constrained Cosserat shell model these symmetries remain as constraint in the variational problems, and   using \eqref{eq5m}, we know that the symmetry of $\mathcal{E}_{\infty}$,\  $\mathcal{E}_{\infty} \, {\rm B}_{y_0} +  {\rm C}_{y_0} \mathcal{K}_{\infty})$ and 
$(\mathcal{E}_{\infty} \, {\rm B}_{y_0} +  {\rm C}_{y_0} \mathcal{K}_{\infty}) {\rm B}_{y_0}$ are equivalent to the symmetry of $\mathcal{G}_{\infty}$ and $(\mathcal{R}_{\infty} -\mathcal{G}_{\infty} \,{\rm L}_{y_0})$ and $(\mathcal{R}_{\infty} -\mathcal{G}_{\infty} \,{\rm L}_{y_0})\,{\rm L}_{y_0}$. Then, we should have
\begin{align}\label{cond}
\mathcal{G}_{\infty}=\mathcal{G}_{\infty}^T, \qquad (\mathcal{R}_{\infty} -\mathcal{G}_{\infty} \,{\rm L}_{y_0})=(\mathcal{R}_{\infty} -\mathcal{G}_{\infty} \,{\rm L}_{y_0})^T,\qquad (\mathcal{R}_{\infty} -\mathcal{G}_{\infty} \,{\rm L}_{y_0})\,{\rm L}_{y_0}=[(\mathcal{R}_{\infty} -\mathcal{G}_{\infty} \,{\rm L}_{y_0})\,{\rm L}_{y_0}]^T.
\end{align}
Let us look at the last condition \eqref{cond}$_3$. Using \eqref{cond}$_{1,2}$ we see that \eqref{cond}$_3$ is equivalent to 
\begin{align}
(\mathcal{R}_{\infty} -\mathcal{G}_{\infty} \,{\rm L}_{y_0})\,{\rm L}_{y_0}=\,{\rm L}_{y_0}^T(\mathcal{R}_{\infty} -\mathcal{G}_{\infty} \,{\rm L}_{y_0}).
\end{align}
In the linearised constrained Cosserat shell model, the constraint \eqref{cond}$_1$ is automatically satisfied, since $\mathcal{G}_{\rm{Koiter}}^{\rm{lin}}$ is symmetric, while  \eqref{cond}$_{2,3}$ turn into
\begin{align}\label{ar}
(\mathcal{R}_{\rm{Koiter}}^{\rm{lin}} - {\bf 2}\,[\mathcal{G}_{\rm{Koiter}}^{\rm{lin}} \,{\rm L}_{y_0}])&=(\mathcal{R}_{\rm{Koiter}}^{\rm{lin}} - {\bf 2}\,[\mathcal{G}_{\rm{Koiter}}^{\rm{lin}} \,{\rm L}_{y_0}])^T\,, \notag\\ (\mathcal{R}_{\rm{Koiter}}^{\rm{lin}} - {\bf 2}\,[\mathcal{G}_{\rm{Koiter}}^{\rm{lin}} \,{\rm L}_{y_0}]){\rm L}_{y_0}&={\rm L}_{y_0}^T(\mathcal{R}_{\rm{Koiter}}^{\rm{lin}} - {\bf 2}\,[\mathcal{G}_{\rm{Koiter}}^{\rm{lin}} \,{\rm L}_{y_0}])^T.
\end{align}
Due to the symmetry of $\mathcal{R}_{\rm{Koiter}}^{\rm{lin}}$ the  relations \eqref{ar} are equivalent to
\begin{align}
\mathcal{G}_{\rm{Koiter}}^{\rm{lin}} \,{\rm L}_{y_0}&={\rm L}_{y_0}^T\mathcal{G}_{\rm{Koiter}}^{\rm{lin}} \,, \qquad \qquad  \mathcal{R}_{\rm{Koiter}}^{\rm{lin}} {\rm L}_{y_0}={\rm L}_{y_0}^T\mathcal{R}_{\rm{Koiter}}^{\rm{lin}}.
\end{align}
Hence, from \eqref{deriv3ff}, we have that in the constrained linear Cosserat shell model (not modified) we get\begin{align}\label{deriv3ffc}
\dfrac{d\,{\rm III}_{y_0+\eta\, v}}{d\, \eta}\Big|_{\eta=0}&=2\,\sym([\mathcal{R}_{\rm{Koiter}}^{\rm{lin}}\, - \mathcal{G}_{\rm{Koiter}}^{\rm{lin}}\, {\rm L}_{y_0}] {\rm L}_{y_0}),\notag
\end{align}
which attributes a clear   geometrical meaning of the energy terms containing the strain tensor $(  \mathcal{E} \, {\rm B}_{y_0} +  {\rm C}_{y_0} \mathcal{K} )   {\rm B}_{y_0} $, beside the geometrical meaning which we have given to $(  \mathcal{E} \, {\rm B}_{y_0} +  {\rm C}_{y_0} \mathcal{K} )$.

\medskip

Since the coercivity inequality
\begin{align}
W^{\infty}(\mathcal{E}_{ \infty }, \mathcal{K}_{ \infty })\geq\,&
\dfrac{h}{12} a_1^+ \lVert \mathcal{E}_{ \infty }\rVert ^2+\dfrac{h^3}{12}\, a_2^+ \lVert
\sym(\mathcal{E}_{ \infty }{\rm B}_{y_0}+{\rm C}_{y_0}\, \mathcal{K}_{ \infty })  \rVert ^2 + a_3^+\frac{ h^3}{6}\lVert  \mathcal{K}_{ \infty }\rVert ^2, \ \ a_i^+>0,
\end{align}
from the nonlinear case turns, upon linearization, into
\begin{align}\label{coecivityl}
W^{\infty}(\mathcal{E}_{ \infty }^{\rm lin}, \mathcal{K}_{ \infty }^{\rm lin})\geq\,&
\dfrac{h}{12} a_1^+ \lVert [\nabla\Theta \,]^{-T}
(\mathcal{G}_{\rm{Koiter}}^{\rm{lin}})^\flat [\nabla\Theta \,]^{-1} \rVert ^2\\&+\dfrac{h^3}{12}\, a_2^+ \lVert
[\nabla\Theta \,]^{-T} [\underbrace{\mathcal{R}_{\rm{Koiter}}^{\rm{lin}}-{\bf 2}\,\sym(\mathcal{G}_{\rm{Koiter}}^{\rm{lin}} \,{\rm L}_{y_0})}_{\mathcal{R}_{\rm{AL}}^{\rm{lin}}} ]^\flat [\nabla\Theta \,]^{-1} \rVert ^2 + a_3^+\frac{ h^3}{6}\lVert  \mathcal{K}_{ \infty }^{\rm lin}\rVert ^2, \ \ a_i^+>0,\notag
\end{align}
using the Anicic and L\'eger's  result \eqref{ALrez}, it is clear that vanishing elastic energy implies $\mathcal{G}_{\rm{Koiter}}^{\rm{lin}}=0$, $\mathcal{R}_{\rm{AL}}^{\rm{lin}}=0$ and  $\mathcal{K}_{ \infty }^{\rm lin}=0$, i.e. no changes of the metric, no variations of the Gauss curvature ${\rm K}$ and of the mean curvature ${\rm H}$ and zero bending, too.  

We note  that $\mathcal{K}_{ \infty }^{\rm lin}=0$ implies that  $\mathcal{R}^{\rm{lin}}_\infty=0$ (no bending) and $\mathcal{N}^{\rm{lin}}_\infty=0$ (no drill).

Therefore,  with regard  to Table \ref{tabelcurv}, we can claim that
\begin{enumerate}
	\item 
	the linearised modified constrained Cosserat shell model represents a generalization of Anicic and L\'eger's model by including the effect of the curvature energy and incorporating  also the bending effects, the transverse shear effect and drilling effect,
	\item   the  nonlinear modified constrained Cosserat shell model represents a nonlinear generalization of the Anicic and L\'eger's shell model, but starting from a 3D-Biot type energy,
	\item the  nonlinear unconstrained Cosserat shell model represents a nonlinear generalization of the Anicic and L\'eger's shell model by considering additional degrees of freedom and transverse shear deformations, as well as in-plane drill.
	\item the change of curvature tensors $ \mathcal{R}_{ \infty }-\mathcal{G}_{ \infty } \,{\rm L}_{y_0} $ and $ \mathcal{R}-\mathcal{G} \,{\rm L}_{y_0} $  represent generalizations of the  Anicic and L\'eger's  change of curvature tensor, in the nonlinear  constrained and unconstrained Cosserat shell models, respectively.
\end{enumerate}

It is worth noting that, based on different arguments, {\v{S}}ilhav{\`y} \cite{vsilhavycurvature} recently arrived at the conclusion that Anicic-L\'eger's change of curvature tensor $\mathcal{R}_{\rm{AL}}^{\rm{lin}}$ is more suitable for use as a curvature measure in a linear Kirchhoff-Love shell theory than the Koiter-Sanders-Budiansky bending measure. {\v{S}}ilhav{\'y}'s approach involves determining the three-dimensional strain tensor of a shear deformation of a shell-like body and then linearizing it with respect to the displacement and the distance of a point from the middle surface.

However, it is  important to note that Anicic-L\'eger's change of curvature tensor $\mathcal{R}_{\rm{AL}}^{\rm{lin}}$ does not vanish in pure stretch deformations, whereas the Koiter-Sanders-Budiansky bending measure $\mathcal{R}_{\infty}^{\rm{lin}}$ does not possess a similar property as $\mathcal{R}_{\rm{AL}}^{\rm{lin}}$ concerning the variations of the curvatures ${\rm H}$ and ${\rm K}$, as seen in \eqref{ALrez}.\footnote{Shell theory is not a wish concert! It seems one would think that $\mathcal{R}_{\rm{KSB}}^{\rm{lin}}$
and $\mathcal{R}_{\rm{AL}}^{\rm{lin}}$ should coincide in an ideal world but they simply do not, since $1\neq 2$.}

In summary, all the tensors in question, including the Koiter bending measure $\mathcal{R}_{\rm{Koiter}}^{\rm{lin}} = \big[{\rm II}_m - {\rm II}_{y_0}\big]^{\rm{lin}}$, the Koiter-Sanders-Budiansky bending measure  $\mathcal{R}^{\rm lin}_{\rm{KSB}}=\mathcal{R}_{\infty}^{\rm{Koiter}}-{\bf 1}\,\sym(\mathcal{G}_{\infty}^{\rm{Koiter}}\,{\rm L}_{y_0})$, and the change of curvature tensor  $\mathcal{R}^{\rm lin}_{\rm{AL}}=\mathcal{R}_{\infty}^{\rm{Koiter}}-{\bf 2}\,\sym(\mathcal{G}_{\infty}^{\rm{Koiter}}\,{\rm L}_{y_0})$, can each be accepted with their distinct and proper physical meanings. To the best of our knowledge, there is no existing statement in the literature asserting that the Koiter-Sanders-Budiansky bending measure has the same properties as Anicic-L\'eger's change of curvature tensor $\mathcal{R}^{\rm lin}_{\rm{AL}}$, or vice versa, i.e., if  a vanishing Koiter-Sanders-Budiansky tensor lead to a vanishing variations of the mean curvature and {Gau\ss} curvature and if the Anicic-L\'eger's satisfies the axioms put forward for a bending measure.
\section{Transverse shear in Cosserat, Ressner-Mindlin  and Naghdi shell models  }\setcounter{equation}{0}

By \eqref{eq5} we observe that the elastic shell strain tensor $\mathcal{E}_{m,s}$ is capable of measuring the change of metric but the transverse shear deformation, too.

To the contrary, in classical elastic theories, i.e., when  Cosserat effects are ignored, the  transverse shear deformation is missing, see Table \ref{tabeltrans}. The Cosserat shell theory, as well as  the 6-parameter shells, is able to show how the transverse shear deformation influences the energy density, through the transverse shear deformation vector (a row) 

\begin{equation}
\label{equ8t}
\mathcal{T}=\, (\overline{Q}_{e,s}  n_0)^{T} \nabla m= \, \left(\bigl\langle\overline{Q}_{e,s}  n_0, \partial_{x_1} m\bigr\rangle,\bigl\langle\overline{Q}_{e,s}  n_0, \partial_{x_2} m\bigr\rangle\right)\;
\end{equation}
in the nonlinear unconstrained Cosserat shell theory, while in the linear unconstrained Cosserat shell theory by
\begin{equation}
\label{equ8tl}
\mathcal{T}^{\rm{lin}} =   n_0^{T} ( \nabla v - \vartheta\times\nabla y_0 ).
\end{equation}
In the constrained nonlinear Cosserat shell model, hence also in the constrained linear Cosserat shell model, as a consequence of the constraint,  the transverse shear deformation vector vanishes
$
\mathcal{T}_\infty=0\  \text{and}\  \mathcal{T}^{\rm{lin}}_\infty=0.
$

The transverse shear deformation row vector is also present in the Naghdi-type models, see  \cite[Section 6]{mardare2008derivation} and \cite{gastel2022regularity,sky2023reissner} or Reissner-Mindlin 5-parameter model given by   \begin{align}\mathcal{T}_{\rm Naghdi}:= d^{T} \nabla m= \bigl\langle d, \partial_{x_1} m\bigr\rangle,\bigl\langle d, \partial_{x_2} m\bigr\rangle,
\end{align}
with its linearisation
\begin{align}\mathcal{T}_{\rm Naghdi}^{\rm lin}:= \, n_0^{T} \nabla v+\zeta^{T} \nabla y_0= \, \left(\bigl\langle n_0, \partial_{x_1} v\bigr\rangle,\bigl\langle n_0, \partial_{x_2} v\bigr\rangle\right)+\left(\bigl\langle \zeta, \partial_{x_1} y_0\bigr\rangle,\bigl\langle \zeta, \partial_{x_2} y_0\bigr\rangle\right),
\end{align}
where $d=n_0+\zeta$.
\section{Drilling appears only in Cosserat shell models}\setcounter{equation}{0}

The Cosserat shell theory, as well as  the 6-parameter shells, is able to describe the effect of drilling bending in
shells \cite{mohammadi2021plane}. These effects are absent in 5-parameter Reissner-Mindlin and Naghdi-type shell models. As already mentioned, the drilling bending effect is incorporated in the bending-curvature tensor $\mathcal{K}_{e,s}$, through the vector (row) of drilling bendings
\begin{align}
\mathcal{N} = n_0^T\, \big(\mbox{axl}(\overline{Q}_{e,s}^T\partial_{x_1}\overline{Q}_{e,s})\,|\, \mbox{axl}(\overline{Q}_{e,s}^T\partial_{x_2}\overline{Q}_{e,s}) \big)
\end{align}
in the nonlinear theory and by its linearisation 
\begin{equation}
\label{equ8d}
\mathcal{N}^{\rm{lin}} \,= \, n_0^T\,(\partial_{x_1}\vartheta \,|\, \partial_{x_2}\vartheta)= \,n_0^T\,  (\nabla\vartheta)
\end{equation}
in the linear model.
The capture of the drilling bending is absent in the other theories,  see Table \ref{tabeldrilling}, excepting 6-parameter shells model, presented in this comparison.
\begin{table}[h!]
	\centering 
	\hspace*{-0.3cm}
	\resizebox{17.3cm}{!}{	\begin{tabular}[h!]{|lc|c|c|c|c|c|c|c|}
			\hline
			\begin{minipage}{2cm}\end{minipage}& \begin{minipage}{4cm}\footnotesize \vspace*{2mm}\textbf{our models}\\\end{minipage}& \begin{minipage}{4cm} \footnotesize\vspace*{2mm} \textbf{Koiter-type models}\\ \end{minipage}& \begin{minipage}{2cm} \footnotesize \vspace*{2mm} \textbf{Anicic}\textbf{-L\'{e}ger}\\ \end{minipage}&\begin{minipage}{3.4cm} \footnotesize\vspace*{2mm} \textbf{Budiansky}\textbf{-Sanders}\\ \end{minipage}& \begin{minipage}{1.2cm} \footnotesize\vspace*{2mm} \textbf{Acharya}\\\end{minipage}& \begin{minipage}{2cm} \footnotesize\vspace*{2mm} \textbf{6-parameter}\\ \end{minipage}& \begin{minipage}{2.5cm}\footnotesize\ \\\textbf{generalized} \\ \textbf{Naghdi-type}\end{minipage}\\
			\hline
			\begin{minipage}{2cm}\footnotesize  \vspace*{2mm} nonlinear \\unconstrained\\ Cosserat \\\end{minipage}&  \begin{minipage}{7cm}\vspace*{2mm}$\mathcal{N} = n_0^T\, \big(\mbox{axl}(\overline{Q}_{e,s}^T\partial_{x_1}\overline{Q}_{e,s})\,|\, \mbox{axl}(\overline{Q}_{e,s}^T\partial_{x_2}\overline{Q}_{e,s}) \big)$ \end{minipage}& $\times$ & $\times$ & $\times$ &  $\times$&$\mathcal{N} $ &$\times$\\
			\hline
			\begin{minipage}{2cm}\footnotesize \vspace*{2mm}linear\\ unconstrained \\Cosserat \\\end{minipage}& \begin{minipage}{7cm}\vspace*{2mm}$\mathcal{N}^{\rm{lin}} \,= \, n_0^T\,(\partial_{x_1}\vartheta \,|\, \partial_{x_2}\vartheta)\,= \,n_0^T\,  (\nabla\vartheta)$ \end{minipage} & $\times$ & $\times$ & $\times$ &  $\times$ & $\mathcal{N}^{\rm{lin}} $&$\times$\\
			\hline
			\begin{minipage}{2cm}\footnotesize\vspace*{2mm} nonlinear\\
				modified\\ constrained \\Cosserat \\\end{minipage}&\begin{minipage}{7cm}\vspace*{2mm}$\mathcal{N}_\infty \coloneqq  n_0^T\, \big(\mbox{axl}({Q}_{ \infty }^T\partial_{x_1}{Q}_{ \infty })\,|\, \mbox{axl}({Q}_{ \infty }^T\partial_{x_2}{Q}_{ \infty }) \big)$ \end{minipage} & $\times$ & $\times$ & $\times$ &  $\times$ & $\mathcal{N}_\infty $&$\times$\\
			\hline
			\begin{minipage}{2cm}\footnotesize \vspace*{2mm}linear\\ constrained\\ Cosserat \\\end{minipage}& \begin{minipage}{7.2cm}\vspace*{2mm}$\mathcal{N}^{\rm{lin}}_\infty \,= \, n_0^T\,(\partial_{x_1}\vartheta \,|\, \partial_{x_2}\vartheta_\infty)= \,n_0^T\,  (\nabla\vartheta_\infty)$\\
				\hspace*{2.1mm}$\vartheta_\infty={\rm axl}(A_{\vartheta_\infty})$\\
				$A_{\vartheta_\infty}=-\skw( (\nabla v\,|\dd\sum_{\alpha=1,2}\bigl\langle n_0, \partial_{x_\alpha}v\bigr\rangle\, a^\alpha)[\nabla \Theta]^{-1})$ \end{minipage} & $\times$ & $\times$ & $\times$ &  $\times$ & $\mathcal{N}^{\rm{lin}}_\infty  $&$\times$\\
			\hline
	\end{tabular}}
	\caption{\footnotesize The drilling bending measures. }\label{tabeldrilling}
\end{table}

\section{Conclusions of this paper}
In this paper, we have highlighted that in the constrained Cosserat shell model, the change of metric tensor simplifies to $\sqrt{[\nabla\Theta ]^{-T}\,{\rm I}_m^{\flat }\,[\nabla\Theta ]^{-1}} - \sqrt{[\nabla\Theta ]^{-T}\,{\rm I}_{y_0}^{\flat }\,[\nabla\Theta ]^{-1}}$. Upon linearization of the constrained model, this change of metric tensor further reduces to the change of metric tensor $\mathcal{G}_{\rm{Koiter}}^{\rm{lin}}$, whose geometric interpretation is provided in \cite{Ciarlet00,mardare2008derivation,mardare2019nonlinear,ciarlet2018existence,ciarlet2019new}. The correspondence between the metric tensor in our Cosserat shell model and the change of metric tensor $\mathcal{G}_{\rm{Koiter}}^{\rm{lin}}$ employed in the linear Koiter model is clear and to be expected. It    is important to note that such a direct equivalence does not exist between the bending strain tensor $\mathcal{R}$ considered by us in the constrained Cosserat framework and the bending tensor or the change of curvature tensor in other theories. This disparity is not surprising because the tensor $\mathcal{R}_{\rm{Koiter}}^{\rm{lin}}$ in the Koiter linear theory cannot be simply labeled as "bending" or "change of curvature." Our bending strain tensor $\mathcal{R}$ extends and generalizes the linear Koiter-Sanders-Budiansky bending measure $\mathcal{R}_{\rm{KSB}}^{\rm{lin}}$ \cite{budiansky1963best,koiter1973foundations}, which vanishes during infinitesimal pure stretch deformations of a quadrant of a cylindrical surface \cite{acharya2000nonlinear}. This property is not shared by the classical "bending strain tensor" $\mathcal{R}_{\rm{Koiter}}^{\rm{lin}}$ in the Koiter model.

In our modelling framework we find that the bending strain tensor $\mathcal{R}$ plays a role through the elastic shell bending-curvature tensor $\mathcal{K}_{e,s}$. This tensor, $\mathcal{K}_{e,s}$, exerts influence on both the membrane-bending energy density and the bending-curvature energy density. In the context of the membrane-bending energy, its impact is incorporated along with the influence of the elastic shell tensor, resulting in the expression $\mathcal{E}_{m,s} {\rm B}_{y_0} + \mathrm{C}_{y_0} \mathcal{K}_{e,s}$. More specifically, it encompasses the effects of transverse shear deformations (represented by $\mathcal{T}$) and the changes in curvature (expressed by $ \mathcal{R}-\mathcal{G} \,{\rm L}_{y_0}$). The term "the change of curvature" is used to describe the nonsymmetric quantity $ \mathcal{R}$, combined with $\mathcal{G} \,{\rm L}_{y_0}$, and its justification is provided within the framework of the linearised  constrained theory.
				
				We posit that the influence of the tensor $\mathcal{E}_{m,s} {\rm B}_{y_0} + \mathrm{C}_{y_0} \mathcal{K}_{e,s}$, denoted as ``the change of curvature" tensor $ \mathcal{R}$, is not accounted for in other models. This assertion is further detailed in \cite[Section 6]{GhibaNeffPartI} by a comprehensive comparison with the general 6-parameter shell model. The natural inclusion of this tensor in the model, particularly after dimension reduction, and its intriguing geometric interpretation in its linearised form has been explored in this paper.

From the alternative formulations presented for both unconstrained and constrained models, we have seen that 	$\mathcal{K}_{e,s}$  influences the bending-curvature energy and that  the \textit{elastic shell bending-curvature tensor} $\mathcal{K}_{e,s}$ incorporates at the same time {\it bending effects} and {\it drilling bendings}.

We have to notice that by the names we have used for   $\mathcal{R}$ and  $ \mathcal{R}-\mathcal{G} \,{\rm L}_{y_0}$, it is indicated  that $\mathcal{R}$ measures rather  bending  while  $ \mathcal{R}-\mathcal{G} \,{\rm L}_{y_0}$ measures the changes of the curvature.

This line of thought, beside some other arguments presented in the linearised framework by 
Anicic and L\'eger \cite{anicic1999formulation}, see also \cite{anicic2001modele}, and more recently by {\v{S}}ilhav{\`y} \cite {vsilhavycurvature}, suggest  that the triple $\mathcal{G}_{ \infty }$, $\mathcal{R}_{ \infty }-2\,\mathcal{G}_{ \infty } \,{\rm L}_{y_0}$ and $\mathcal{K}_{ \infty }$ are  appropriate measures  to express the change of metric and of the curvatures ${\rm H}$ and ${\rm K}$, while the bending and drilling effects are both  additionally incorporated in the bending-curvature energy through the elastic shell bending-curvature tensor $\mathcal{K}_{\infty}$.

In the linearised  version of our Cosserat shell model we obtain all the  linear strain measures of  the theory of 6-parameter shells, due to the fact that the kinematical structure is equivalent, since there is  an explicit dependence of the internal energy density on the change of curvature measure and on the bending measure.  In order to make connections with existing works in the literature on      6-parameter shell models \cite{Eremeyev06,Pietraszkiewicz-book04,Pietraszkiewicz10}, see also  \cite[Section 6]{GhibaNeffPartI}, we conclude that  the influence of the change of curvature tensor $\mathcal{R}_{ \infty }-2\,\mathcal{G}_{ \infty } \,{\rm L}_{y_0}$, is omitted if a constrained Cosserat shell model would be derived from other available simpler 6-parameter shell models \cite{Eremeyev06,Pietraszkiewicz-book04,Pietraszkiewicz10,pietraszkiewicz2012exact,pietraszkiewicz2020thin}, even if the bending strain tensor $\mathcal{R}_{ \infty }$ is present (through the presence of the curvature energy).

\bigskip

\begin{footnotesize}
	\noindent{\bf Acknowledgements:}   Patrizio Neff is grateful for discussions with Sylvia Anicic, Universit\'e de Haute-Alsace, Mulhouse, France. 
	This research has been funded by the Deutsche Forschungsgemeinschaft (DFG, German Research Foundation) -- Project no. 415894848: NE 902/8-1.

	\bibliographystyle{plain} 

\appendix
\section{Appendix}
\subsection{Acharya's blog-entry  on iMechanica from 2007}
 https://imechanica.org/node/1408:  ``We know from strength of materials that non-uniform stretching of fibers along the cross section of a beam produces bending moments. But does this situation necessarily correspond to a 'bending' deformation? For that matter, what do we exactly mean kinematically when we talk about a bending deformation?
To make the question more concrete, consider a cylinder that expands uniformly along all radial rays. Does this deformation of the cylinder correspond to bending? I think it is fair to say that most would say that this is purely a stretching deformation with no bending. But then, what is precisely a bending deformation?
The most classical definition relates the change in the second fundamental form as a bending strain. By this is meant the following. Calculate the gradient of the unit normal fields (i.e. curvature tensors) on the deformed and undeformed shells. The difference of these curvatures, suitably adjusted for the fact that at each material point they are tensors on different tangent spaces that can be oriented very differently, is assigned to be the bending strain. But you see, a stretching of the shell changes the curvature and therefore the radially expanding cylinder would be described as undegoing a bending deformation, according to this classical measure.
So, something isn't quite right here. Starting with KOITER and then followed by SANDERS and BUDIANSKY, a bending strain measure was introduced for linear kinematics that does not have this shortcoming, up to the accuracy of the linear theory. Koiter and Budiansky later proposed nonlinear strain measures that predict vanishing bending strain in biaxial stretching of cylinders, up to second order in the radial and axial displacements.
In my opinion, the Koiter, Sanders, Budiansky developments were pioneering works towards clarifying what one might mean kinematically by bending. However, a clear physical definition leading to an exact mathematical statement of what constitutes bending deformation of a shell was lacking. For one thing, if such a measure was available then it would associate, without approximation, no bending strain to a biaxial stretching of a cylinder. 
In \cite{acharya2000nonlinear}, I tried to address this question of physical definition and corresponding mathematical form in the nonlinear setting with its connections to the linear KSB [Koiter-Sander-Budiansky] measure. The seemingly innocuous question became surprisingly subtle - at least for me - with things like the drill rotation spoiling the show to exact success. The physical definition naturally involves things like the polar decomposition of the deformation gradient on a manifold with almost the whole story revolving around being careful about the domain and range of the stretch and rotation tensors. 
For those of you who read it, I hope you enjoy it as much as I did working on the problem."

\subsection{The stretch $U_e$ that leaves tangent planes invariant}
\setcounter{equation}{0}
In this appendix we consider the specific form that a stretch $U_e$ must have, such that the stretch derives from a mapping $m$ that leaves normals invariant. Hence, the question: is every stretch $U_e$ for a mapping $m$ that leaves normals invariant of the form  $\U=\sqrt{F_e^TF_e}$ with $F_e=(\nabla m\,|\,n_0)(\nabla y_0\,|\,n_0)^{-1}$?

The answer is yes: First let us see which is such an $\U$ for a mapping $m$ that leaves normals invariant is of the form  $\U=\sqrt{F_e^TF_e}$ with $F_e=(\nabla m\,|\,n_0)(\nabla y_0\,|\,n_0)^{-1}$.
We compute 
\begin{align}\label{Ue1}
\U=\sqrt{F_e^TF_e}=\ \ &\sqrt{(\nabla y_0\,|\,n_0)^{-T}(\nabla m\,|\,n_0)^T(\nabla m\,|\,n_0)(\nabla y_0\,|\,n_0)^{-1}}\notag\\
\overset{n=n_0}{=}&\sqrt{(\nabla y_0\,|\,n_0)^{-T}(\nabla m\,|\,n)^T(\nabla m\,|\,n)(\nabla y_0\,|\,n_0)^{-1}}
=\sqrt{(\nabla y_0\,|\,n_0)^{-T}\begin{footnotesize}\left( \begin{array}{c|c}
	{\rm I}_m & \mathbf{0}\\\hline
	\raisebox{-2pt}{$ 0$} & \raisebox{-2pt}{$1$}
	\end{array} \right)\end{footnotesize}(\nabla y_0\,|\,n_0)^{-1}}.
\end{align}

We check that if  a mapping $m$ leaves normals invariant  then $\U$ given by $\eqref{Ue1}$ satisfies $\U \,n_0=n_0.$  We recall that 	 $a^1, a^2, a^3\,$ denote the rows of $[\nabla\Theta \,]^{-1}$, i.e. 
$
\nabla\Theta =(\nabla y_0|\,n_0)\,=\,(a_1|\,a_2|\,a_3),\  [\nabla\Theta \,]^{-1}=(a^1|\,a^2|\,a^3)^T,
$
and $ a^3=n_0\,$, $\langle a^1,a^3\rangle=0, \langle a^2,a^3\rangle=0$. 
We compute 
\begin{align}\label{Ue}
{\rm U}_e^2\, n_0
&=(\nabla y_0\,|\,n_0)^{-T}\begin{footnotesize}\left( \begin{array}{c|c}
	{\rm I}_m & \mathbf{0}\\\hline
	\raisebox{-2pt}{$ 0$} & \raisebox{-2pt}{$1$}
	\end{array} \right)\end{footnotesize}(\nabla y_0\,|\,n_0)^{-1}\, \, n_0=(\nabla y_0\,|\,n_0)^{-T}\begin{footnotesize}\left( \begin{array}{c|c}
	{\rm I}_m & \mathbf{0}\\\hline
	\raisebox{-2pt}{$ 0$} & \raisebox{-2pt}{$1$}
	\end{array} \right)\end{footnotesize}(a^1|\,a^2|\,n_0)^T\, n_0\\
	&=(\nabla y_0\,|\,n_0)^{-T}\begin{footnotesize}\left( \begin{array}{c|c}
	{\rm I}_m & \mathbf{0}\\\hline
	\raisebox{-2pt}{$ 0$} & \raisebox{-2pt}{$1$}
	\end{array} \right)\end{footnotesize}e_3=(\nabla y_0\,|\,n_0)^{-T}e_3=
	(a^1|\,a^2|\,n_0)e_3=n_0.\notag
\end{align}

The positive definite square root of ${\rm U}_e^2$ satisfies also ~${\U}\,n_0=n_0 $. This follows  from  the spectral decomposition of matrices and the uniqueness of the square root of a symmetric and positive definite matrix.

So, the remaining question is: is every stretch $U_{e}$ for a mapping $m$ that leaves normals invariant necessarily of the above form  \begin{align}\U=\sqrt{(\nabla y_0\,|\,n_0)^{-T}\begin{footnotesize}\left( \begin{array}{c|c}
	{\rm I}_m & \mathbf{0}\\\hline
	\raisebox{-2pt}{$ 0$} & \raisebox{-2pt}{$0$}
	\end{array} \right)\end{footnotesize}(\nabla y_0\,|\,n_0)^{-1}}\,\,\, ?
\end{align}

Consider a map $m$ such that the corresponding stretch leaves normals invariant. Then for this deformation  $m$ we can consider the polar decomposition \cite{neff2013grioli}
\begin{align}
(\nabla m\,|\,n)(\nabla y_0\,|\,n_0)^{-1}=R_e U_e,
\end{align}
where $U_e \, n_0=n_0$ and 	 ${U}_e(\xi)\in {\rm Sym}^+(3)$ is a linear mapping of the tangent plane to the initial surface at $\xi=y_0(x)$ into itself  and 
$
{R}_e(\xi)=R_0\,R_{n_0}(\xi),
$
with $R_0\in {\rm SO}(3)$ a uniform rotation, i.e., independent of position, and $R_{n_0}(\xi)\in{\rm SO}(3)$ belongs to the group of rotations about the unit normal $n_0(\xi)$ to the surface at $\xi=y_0(x)$ (pure drill).

Question: is $U_e=\U:=\sqrt{(\nabla y_0\,|\,n_0)^{-T}\begin{footnotesize}\left( \begin{array}{c|c}
	{\rm I}_m & \mathbf{0}\\\hline
	\raisebox{-2pt}{$ 0$} & \raisebox{-2pt}{$1$}
	\end{array} \right)\end{footnotesize}(\nabla y_0\,|\,n_0)^{-1}}\in {\rm Sym}^+(3)$? 
Indeed, since
\begin{align}
U_e^2=U_e U_e=U_e^T U_e&=(\nabla y_0\,|\,n_0)^{-T}(\nabla m\,|\,n)^TR_eR_e^T(\nabla m\,|\,n)(\nabla y_0\,|\,n_0)^{-1}\\
&=(\nabla y_0\,|\,n_0)^{-T}(\nabla m\,|\,n)^T(\nabla m\,|\,n)(\nabla y_0\,|\,n_0)^{-1}=(\nabla y_0\,|\,n_0)^{-T}\begin{footnotesize}\left( \begin{array}{c|c}
	{\rm I}_m & \mathbf{0}\\\hline
	\raisebox{-2pt}{$ 0$} & \raisebox{-2pt}{$1$}
	\end{array} \right)\end{footnotesize}(\nabla y_0\,|\,n_0)^{-1},\notag
\end{align}
the answer is yes.

Altogether, we have shown that 
every stretch $U_{e}$ for a mapping $m$ that leaves normals invariant is of the form  \begin{align}\U=\sqrt{(\nabla y_0\,|\,n_0)^{-T}\begin{footnotesize}\left( \begin{array}{c|c}
	{\rm I}_m & \mathbf{0}\\\hline
	\raisebox{-2pt}{$ 0$} & \raisebox{-2pt}{$1$}
	\end{array} \right)\end{footnotesize}(\nabla y_0\,|\,n_0)^{-1}}.
\end{align}
\end{footnotesize}

\end{document}